\documentclass[floatfix,twocolumn,showpacs,preprintnumbers,amsmath,amssymb,pra,superscriptaddress,longbibliography]{revtex4-1}
\usepackage{color}
\usepackage[usenames,dvipsnames,svgnames,table]{xcolor}
\usepackage[colorlinks=true,linkcolor=blue,urlcolor=blue,citecolor=blue]{hyperref}
\usepackage{mathtools}
\usepackage{graphicx}
\usepackage{dcolumn}
\usepackage{array}
\usepackage{lipsum}
\usepackage{bm}
\usepackage{subfigure}
\usepackage{amssymb}
\usepackage{multirow}
\usepackage{tabularx}
\usepackage{amsmath}
\usepackage{braket}
\graphicspath{{plots/}}
 \usepackage{lipsum}
\usepackage{mathrsfs}

\newcommand{\beq}{\begin{equation}}
\newcommand{\eeq}{\end{equation}}
\newcommand{\bea}{\begin{eqnarray}}
\newcommand{\eea}{\end{eqnarray}}

\begin{document}

\title{Temperature analysis of X-ray Thomson scattering data}

\author{Tobias Dornheim}
\email{t.dornheim@hzdr.de}

\affiliation{Center for Advanced Systems Understanding (CASUS), D-02826 G\"orlitz, Germany}
\affiliation{Helmholtz-Zentrum Dresden-Rossendorf (HZDR), D-01328 Dresden, Germany}

\author{Maximilian P.~B\"ohme}

\affiliation{Center for Advanced Systems Understanding (CASUS), D-02826 G\"orlitz, Germany}

\affiliation{Helmholtz-Zentrum Dresden-Rossendorf (HZDR), D-01328 Dresden, Germany}

\affiliation{Technische  Universit\"at  Dresden,  D-01062  Dresden,  Germany}

\author{David A.~Chapman}
\affiliation{First Light Fusion, Yarnton, Oxfordshire, United Kingdom}

\author{Dominik Kraus}
\affiliation{Institut f\"ur Physik, Universit\"at Rostock, D-18057 Rostock, Germany}
\affiliation{Helmholtz-Zentrum Dresden-Rossendorf (HZDR), D-01328 Dresden, Germany}

\author{Thomas R.~Preston}
\affiliation{European XFEL, D-22869 Schenefeld, Germany}

\author{Zhandos A.~Moldabekov}

\affiliation{Center for Advanced Systems Understanding (CASUS), D-02826 G\"orlitz, Germany}
\affiliation{Helmholtz-Zentrum Dresden-Rossendorf (HZDR), D-01328 Dresden, Germany}

\author{Niclas Schl\"unzen}

\affiliation{Center for Advanced Systems Understanding (CASUS), D-02826 G\"orlitz, Germany}
\affiliation{Helmholtz-Zentrum Dresden-Rossendorf (HZDR), D-01328 Dresden, Germany}

\author{Attila Cangi}

\affiliation{Center for Advanced Systems Understanding (CASUS), D-02826 G\"orlitz, Germany}
\affiliation{Helmholtz-Zentrum Dresden-Rossendorf (HZDR), D-01328 Dresden, Germany}

\author{Tilo D\"oppner}
\affiliation{Lawrence Livermore National Laboratory (LLNL), California 94550 Livermore, USA}

\author{Jan Vorberger}
\affiliation{Helmholtz-Zentrum Dresden-Rossendorf (HZDR), D-01328 Dresden, Germany}

\begin{abstract}
The accurate interpretation of experiments with matter at extreme densities and pressures is a notoriously difficult challenge. In a recent work [T.~Dornheim \emph{et al.}, \emph{Nature Comm.}~(in print), arXiv:2206.12805], we have introduced a formally exact methodology that allows extracting the temperature of arbitrarily complex materials without any model assumptions or simulations. Here, we provide a more detailed introduction to this approach and analyze the impact of experimental noise on the extracted temperatures. In particular, we extensively apply our method both to synthetic scattering data and to previous experimental measurements over a broad range of temperatures and wave numbers. We expect that our approach will be of high interest to a gamut of applications, including inertial confinement fusion, laboratory astrophysics, and the compilation of highly accurate equation-of-state databases.
\end{abstract}

\maketitle

\section{Introduction\label{sec:introduction}}
Over the last decades, there has been a surge of interest in the properties of matter at extreme conditions~\cite{fortov_review}. The phase space representing temperatures of $T\sim10^4-10^8\,$K and pressures of $P\sim1-10^4\,$Mbar is called \emph{warm dense matter} (WDM), which is ubiquitous throughout our Universe and occurs in a variety of astrophysical objects such as giant planet interiors~\cite{Brygoo2021,Kraus_Science_2022,He2022,Benuzzi_Mounaix_2014,Militzer_2008} and brown dwarfs~\cite{saumon1,becker}. In addition, WDM plays an important role in a number of cutting-edge technological applications. For example, the fuel capsule in an inertial confinement fusion experiment~\cite{Betti2016,Zylstra2022} has to traverse the WDM regime on its pathway towards nuclear fusion~\cite{hu_ICF}. Other practical applications include the discovery of novel materials~\cite{Lazicki2021,Kraus2016,Kraus2017} and hot-electron chemistry~\cite{Brongersma2015}.

In the laboratory, WDM is generated at large research facilities using a number of techniques, see, e.g., the topical overview by Falk~\cite{falk_wdm}. However, the central obstacle is the rigorous interpretation of the experiment, because basic parameters such as the temperature cannot be directly measured. In this situation, the X-ray Thomson scattering (XRTS) approach~\cite{siegfried_review} has emerged as a highly useful method. More specifically, it has become common practice to fit an experimentally observed XRTS signal with a theoretical model to infer system parameters such as the temperature~\cite{kraus_xrts,Gregori_PRE_2003,Falk_PRL_2014}.
Unfortunately, the rigorous theoretical description of WDM is notoriously difficult~\cite{wdm_book,new_POP,Dornheim_review}. In practice, one, therefore, has to rely on uncontrolled approximations such as the artificial decomposition into \emph{bound} and \emph{free} electrons within the widely used Chihara decomposition~\cite{Gregori_PRE_2003,Chihara_1987}.
Consequently, the actual interpretation of an experiment might strongly depend on a particular model, which limits the accuracy of equation of state (EOS) tables~\cite{Falk_PRL_2014} and other observations.

To overcome this unsatisfactory situation, we have recently introduced a new methodology~\cite{Dornheim_T_2022} that allows extracting the temperature from a given XRTS signal directly without the need for any theoretical models or simulations. In particular, we have proposed to compute the two-sided Laplace transform [Eq.~(\ref{eq:F}) below] of the measured intensity, which has a number of key advantages: 1) we can completely remove the impact of the instrument function without the need for a numerically unstable explicit deconvolution, 2) our method is very robust with respect to noise in the experimental data, and 3) we can actually \emph{measure} the temperature of arbitrarily complex materials without any model assumptions. The high practical value of this new approach has been demonstrated in Ref.~\cite{Dornheim_T_2022} by re-evaluating the XRTS measurements of warm dense dense beryllium by Glenzer \emph{et al.}~\cite{Glenzer_PRL_2007}, aluminium by Sperling \emph{et al.}~\cite{Sperling_PRL_2015}, and graphite by Kraus \emph{et al.}~\cite{kraus_xrts}.

In the present work, we provide a more detailed introduction to this method, including a comprehensive discussion of the underlying theoretical framework. In addition, we present an extensive analysis of synthetic XRTS data over a broad range of temperatures and wave numbers. This allows us to clearly delineate the limitations of this approach, and to rigorously predict the required experimental specifications to resolve a given $T$. Finally, we systematically investigate the impact of random noise in the experimentally measured intensity and show how it can be used to quantify the uncertainty in the extracted temperature.

In addition to its direct value for WDM diagnostics, we note that the Laplace domain of the dynamic structure factor has a clear physical interpretation as an imaginary-time correlation function~\cite{dornheim2022physical,Dornheim_PTR_2022,Dornheim_review}. The latter naturally emerges in Feynman's path integral formulation of statistical mechanics~\cite{HagenKleinertPI,Berne_JCP_1983} and contains the same information as the usual frequency representation. In fact, both representations are complementary and tend to emphasize different aspects of the same information about a given system~\cite{dornheim2022physical}. Therefore, our approach has the potential to give novel insights beyond the temperature, such as the excitation energy of quasi particles or physical effects like the exchange--correlation induced alignment of pairs of electrons at metallic densities~\cite{Dornheim_Nature_2022}.

This paper is organized as follows: In Sec.~\ref{sec:theory}, we introduce the relevant theoretical background, including a brief discussion of XRTS (\ref{sec:XRTS}), the extraction of the temperature in the Laplace domain (\ref{sec:ITCF}), its connection to imaginary-time correlation functions~\cite{Dornheim_JCP_ITCF_2021,Berne_JCP_1982}, and some practical remarks on the convergence with respect to the experimentally observed frequency range (\ref{sec:convergence}). Sec.~\ref{sec:results_synthetic} is devoted to the analysis of synthetic data and is followed by a new framework for the study of the impact of random noise provided in Sec.~\ref{sec:noise}. In Sec.~\ref{sec:experiment}, we re-analyse the aforementioned experiments by Kraus \emph{et al.}~\cite{kraus_xrts} and Glenzer \emph{et al.}~\cite{Glenzer_PRL_2007} and, thereby, complement the earlier analysis in Ref.~\cite{Dornheim_T_2022} by quantifying the given uncertainties in different properties. The paper is concluded with a summary and an outlook in Sec.~\ref{sec:summary}.

\section{Theory\label{sec:theory}}

\subsection{Characteristic parameters\label{sec:characteristic}}
We consider Hartree atomic units, i.e., $\hbar=k_\textnormal{B}=m_e=1$ 
throughout unless specified otherwise.
From a theoretical perspective, the WDM regime is conveniently characterised by two parameters that are both of the order of unity~\cite{Ott2018}: 1) the Wigner-Seitz radius $r_s=d/a_\textnormal{B}=\left(3/4\pi n_e\right)^{1/3}$, where $d$, $a_\textnormal{B}$ and $n_e$ are the average inter-particle distance, the Bohr radius, and the electron number density, and 2) the degeneracy temperature $\Theta=k_\textnormal{B}T/E_\textnormal{F}$ with $E_\textnormal{F}$ denoting the Fermi energy. It is connected to the Fermi wave number
\begin{eqnarray}
q_\textnormal{F} = \left(\frac{9\pi}{4}\right)^{1/3} \frac{1}{r_s}
\end{eqnarray}
via $E_\textnormal{F}=q_\textnormal{F}^2/2$.
For completeness, we note that we restrict ourselves to the fully unpolarized case with the same number of spin-up and spin-down electrons, $n_e^\uparrow = n_e^\downarrow = n_e/2$.

\subsection{X-ray Thomson scattering experiments\label{sec:XRTS}}

\begin{figure}
    \centering
    \includegraphics[width=0.49\textwidth]{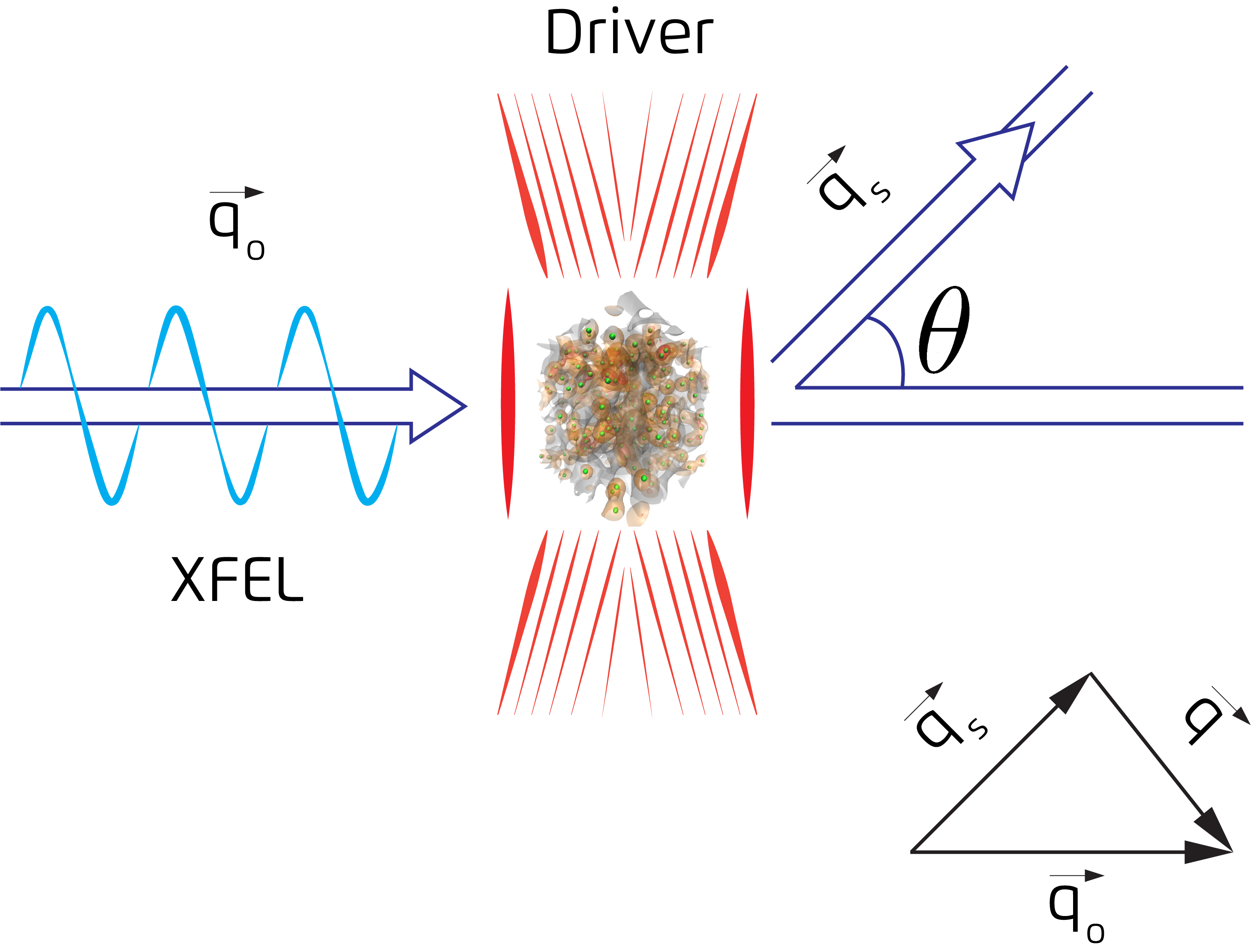}
    \caption{
        Sketch of a typical WDM experimental setup. The sample is compressed using a powerful long-pulse laser (\lq{Driver}\rq~from top and bottom). The diagnostics is provided by a highly brilliant x-ray beam (\lq{XFEL}~\rq from the left) with a variable delay time relative to the drive laser. A number of detectors is placed at different scattering angles $\theta$ behind the target to record the scattering signal.
    }
    \label{fig:XRTS}
\end{figure}

In spectrally resolved XRTS experiments~\cite{siegfried_review} (Fig.~\ref{fig:XRTS}) a detector instrument (usually a crystal spectrometer coupled to a CCD or microchannel plate) placed at some orientation relative to an incident monochromatic source of probing x-ray photons measures in each pixel an energy $E_{\text{pixel}} \in \{E_{\text{s}}, E_{\text{s}}+\Delta E\}$ over the source duration:
\begin{align}
    \label{eq:energy_per_pixel}
    E_{\text{pixel}}
    = &\,
    \int_{0}^{t_{\text{probe}}} \mathrm{d}t \, 
    \int_{E_{\text{s}}/\hbar}^{(E_{\text{s}}+\Delta E)/\hbar} \mathrm{d}\omega_{s} \, 
    \frac{ \partial P_{\text{s}} }{ \partial\omega_{\text{s}} }
    \nonumber \\
    \approx &\,
    \frac{\Delta E}{\hbar}
    \int \text{d}t\,
    \left. \frac{ \partial P_{\text{s}} }{ \partial\omega_{\text{s}} } \right|_{\hbar\omega_{\text{s}} = E_{\text{s}} + \Delta E /2}
    \,.
\end{align}
In Eq.\;\eqref{eq:energy_per_pixel}, $E_{\text{s}} = \hbar\omega_{\text{s}}$ is the energy of the scattered x-rays, $\Delta E$ is the energy range associated with the pixel as determined by the properties (crystal orientation and dispersion relation, etc.) of the experimental apparatus, and $\partial P_{\text{s}}/\partial \omega_{\text{s}}$ is the scattered power per unit frequency as seen by the detector. The meaning of the approximation in Eq.\;\eqref{eq:energy_per_pixel} is that the differential scattered power is treated as constant over each pixel and, thus, is evaluated at the mid-point frequency of the energy interval.

\subsubsection{Differential scattered power spectrum}

As discussed by Fortman et al.~\cite{Fortmann_HEDP_2006}, the differential scattered power spectrum can be written in terms of a more general, higher-order differential quantity
\begin{align}
    \label{eq:power_spectrum}
    \frac{ \partial P_{\text{s}} }{ \partial\omega_{\text{s}} }
    = &\,
    \int \text{d}\mathcal{V}
    \int \text{d}\Omega\,
    \frac{ \partial^{3}P_{\text{s}} }
    { \partial\mathcal{V} \partial\Omega \partial\omega_{\text{s}}}
    \nonumber\\
    = &\,
    \int \text{d}\mathcal{V}
    \int \text{d}\Omega\,
    \left(
    I_{0}\,G(\theta,\phi) n_{e} \frac{ \partial^{2} \sigma }{ \partial\Omega\partial\omega_{\text{s}} }
    \right)
    \,.
\end{align}
In the above $I_{0}$ is the intensity of the incident x-rays, the geometrical term $G(\theta,\phi) = (\hat{\mathrm{e}}_{0}\cdot\hat{\mathrm{e}}_{\text{s}})^{2}$ gives the projection of the unit vectors for the incident and scattered x-ray polarizations \cite{siegfried_review}, $n_{e}$ is the mean electron number density in the volume element $\text{d}\mathcal{V}$, and $\partial^{2}\sigma/\partial\Omega\partial\omega_{\text{s}}$ is the double-differential cross section per unit solid angle, per unit frequency. For an interacting many-electron system the latter is \cite{Crowley_NewJPhys_2013}
\begin{align}
    \label{eq:double_dif_cross_section}
    \frac{ \partial^{2} \sigma }{ \partial\Omega\partial\omega_{\text{s}} }
    = &\,
    \left( \frac{\omega_{\text{s}}}{\omega_{0}} \right)^{n}
    \sigma_{\text{T}}\, 
    S(\mathbf{q},\omega)
    \,,
\end{align}
in terms of the Thomson cross section for a single electron, $\sigma_{\text{T}} \approx 6.65\times10^{-29}\;\mathrm{m^{-2}}$, and the total electron-electron dynamic structure factor (DSF), $S(\mathbf{q},\omega)$, which contains all the information on spatio-temporal correlations between the electrons in the target \cite{kremp2006quantum}. The power on the first term in Eq.\;\eqref{eq:double_dif_cross_section} is $n = 1,2$, depending on whether the detector is 'quantum' (counting the number of incident photons) or 'classical' (sensitive to the total incident energy) \cite{Crowley_NewJPhys_2013}.

\subsubsection{Realistic restrictions for data analysis}

If the x-ray source has close-to-uniform spatial and temporal intensity profiles, the volume of plasma probed by the x-rays is sufficiently small (relative to its distance from both the source and detector) and is also reasonably homogeneous, then the volume integration in Eq.\;\eqref{eq:power_spectrum} can be ignored and the solid angle integration can be approximated by multiplying by the subtended solid angle element $\text{d}\Omega$. There are numerous approximate treatments of incorporating spatial inhomogeneity within the target \cite{Fortmann_HEDP_2009, Chapman_POP_2014, Kraus_PhysRevE_2016, Kozlowski_SciRep_2016}, and the incorporation of such effects into the present analysis framework remains an important task for future works. The same is true for the effect of $k$-blurring, which may be important for large sample volumes in close proximity to a divergent x-ray source. Fortunately both of these considerations are usually negligible for XFEL experiments. A dedicated discussion related to these restrictions is required and, thus, shall not be addressed further here. 

Assuming the foregoing conditions are indeed fulfilled, then the power spectrum reduces to
\begin{align}
    \label{eq:power_spectrum_2}
    \frac{ \partial P_{\text{s}} }{ \partial\omega_{\text{s}} }
    \approx &\,
    I_{0} \, 
    G(\theta,\phi) \, 
    \mathrm{d}\Omega \,
    \sigma_{\text{T}} \, 
    n_{e} \mathcal{V} \,
    \left(\frac{\omega_{\text{s}}}{\omega_{0}}\right)^{n}
    S(\mathbf{q},\omega)
    \,.
\end{align}
The shifts in the incident frequency and wave vector are given straightforwardly by energy and momentum conservation, i.e., $\omega = \omega_{0} - \omega_{\text{s}}$ and $\mathbf{q} = \mathbf{q}_{0} - \mathbf{q}_{\text{s}}$ (see the inset of vector triangle in Fig.\,\ref{fig:XRTS}). For any isotropic distribution function, it is well known that the DSF depends only on the magnitude of the wave vector shift, $q = |\mathbf{q}|$, which can be obtained by straightforward application of the cosine formula
\begin{align}
    \label{eq:full_q}
    q
    = &\,
    \sqrt{q_{0}^{2} + q_{\text{s}}^{2} - 2q_{0}q_{\text{s}}\cos\left(\theta\right)}
    \,.
\end{align}
Ignoring dispersion of the probing radiation in the target (a condition universally satisfied for multi-keV x-rays), then one may reduce Eq.~(\ref{eq:full_q}) to the commonly used approximate form (see Appendix \ref{app:A})
\begin{align}
    \label{eq:approx_q}
    q
    \approx &\,
    2q_{0}\sin(\theta/2)
\end{align}
and correspondingly ignore the factor $(\omega_{\text{s}}/\omega_{0})^{n} \approx 1$ in Eq.\,(\ref{eq:power_spectrum_2}), thereby reducing the scattered power spectrum to a single dynamic term (the DSF). The robustness of this approximation plays an important role in the basis of the novel diagnostic method discussed in this paper. 

\begin{figure}
    \centering
    \includegraphics[width=0.475\textwidth]{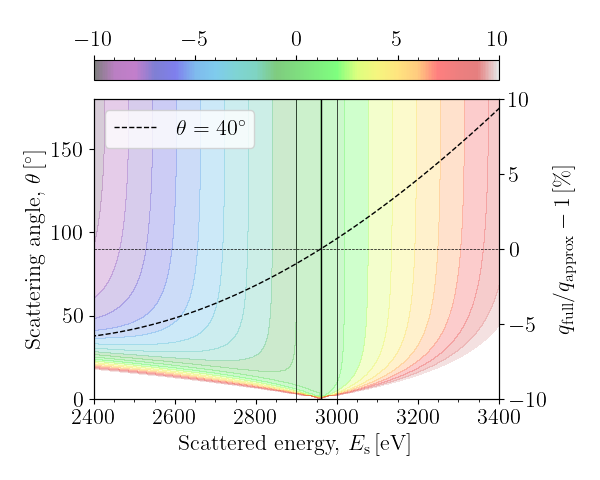}
    \caption{
        Contour map of the percentage difference of the full and approximate forms of the wave number shift as a function of scattered energy and scattering angle, shown for the example of the plasmon scattering data of Glenzer et al. \cite{Glenzer_PRL_2007} using x-rays produced by Cl Ly-$\alpha$ line emission at $2.96\,\mathrm{keV}$. The heavy dashed black curve related to the right-hand axis shows a slice through the surface at the nominal angle of $\theta = 40^{\circ}$ used in the experiment. The two thin black vertical lines denote the dynamic range of the scattering data, from which it is clear that the approximate form holds to better than $1\%$.
    }
    \label{fig:mom_cons_importance_Be_2007_Glenzer}
\end{figure} 

To emphasize this point, Fig.\,\ref{fig:mom_cons_importance_Be_2007_Glenzer} shows the percentage difference between the full \eqref{eq:full_q} and approximate \eqref{eq:approx_q} forms of the wave number $q$ as a function of scattered energy $E_{\text{s}}$ and scattering angle $\theta$ for the example of the collective scattering data taken by Glenzer et al. \cite{Glenzer_PRL_2007}. Clearly, the approximate expression \eqref{eq:approx_q} is well fulfilled over the entire dynamic range of the experiment (the central region bounded by the vertical thin black lines). Similar results are found for all other cases considered. Despite the nonlinearity of the physics governing the response of the plasma to the probing radiation (e.g., the Landau damping rate of plasmons \cite{kremp2006quantum}) with respect to $q$, forward modelling shows that the small differences between the full and approximate forms of $q$ have negligible impact on the shape of the scattered power spectrum. It should, however, be noted that this approximation often does not hold for low-energy probes (particularly for the visible-light lasers used in optical Thomson scattering to probe samples with the plasmon frequency being only slightly lower than the laser frequency; see, e.g., Refs.\,\cite{Froula_2006, Ross_PhysRevLett_2010, Palastro_PhysRevE_2010}), meaning that our diagnostic is currently limited to analyzing XRTS experiments; the extension to low-energy probes will be investigated in detail in future works.

\subsubsection{X-ray source profile and detector response}

Finally, the spectral bandwidth and features of the x-ray source; be it either a relatively narrow, single sharp peak, such as those produced by XFELs \cite{Tschentscher_2017,LCLS_2016,SACLA_2011}, or the more structured emission typical of thermal line emission from hot plasmas~\cite{MacDonald_RSI_2018,MacDonald_POP_2021}, can be accommodated by convolving the scattered power spectrum with both the source function $\Sigma(\omega)$ and the response function of the detector $\Delta(\omega)$. In practice, only information of the convolution $R(\omega) = \Sigma(\omega)\circledast \Delta(\omega)$ [see also Eq.~(\ref{eq:intensity}) below] can be known, e.g., by performing a source characterization experiment. With the foregoing considerations, and in the absence of an absolutely calibrated detector, one may dispense with all the contributions to $\partial P_{\text{s}}/\partial\omega_{\text{s}}$ other than the remaining dynamic terms, working instead with a reduced intensity

\begin{eqnarray}\label{eq:intensity}
I(\mathbf{q},\omega) &=& S(\mathbf{q},\omega) \circledast R(\omega) \\\nonumber &=& \int_{-\infty}^\infty \textnormal{d}\omega'\ S(\mathbf{q},\omega') R\left(\omega'-\omega\right)\ .
\end{eqnarray}

Consequently, in a realistic XRTS experiment, wherein the probe has some finite bandwidth and/or spectral structure, the energy received by the detector on each pixel can be approximately reduced from Eq.\;(\ref{eq:energy_per_pixel}) to 
\begin{align}
    \label{eq:energy_per_pixel_2}
    E_{\text{pixel}}
    \approx &\,
    A\frac{\Delta E}{\hbar}
    \int_{0}^{t_{\text{pulse}}} \text{d}t\,
    \left. 
    I(\mathbf{q},\omega)
    \right|_{\hbar\omega_{\text{s}} = E_{\text{s}} + \Delta E /2}
    \,,
\end{align}
where the factor $A$ accommodates all the modulation separate from the reduced intensity. Finally, if the state of the plasma under study evolves slowly compared to the duration of the x-ray probe, then the integration over time reduces to multiplying by $t_{\text{probe}}$ and one finds
\begin{align}
    \label{eq:energy_per_pixel_3}
    E_{\text{pixel}}
    \propto &\,
    I(\mathbf{q},\omega_{0} - \omega_{\text{s}}')
    \,,
\end{align}
where $\omega_{\text{s}}'$ is frequency associated with the characteristic energy discriminated by the pixel. From Eq.\;\eqref{eq:energy_per_pixel_3}, and since it is extremely rare for the detectors used in XRTS experiments to be absolutely calibrated, it suffices to describe the measured intensity spectrum solely in terms of the reduced intensity Eq.\;\eqref{eq:intensity}.

\subsection{Dynamic structure factor}

A very general and convenient definition of the DSF is obtained from the Fourier transform of the microscopic density-density autocorrelation function
\begin{eqnarray}\label{eq:DSF}
S(\mathbf{q},\omega) &=& \mathcal{F}\left[F(\mathbf{q},t)\right]\nonumber\\
&=& \int_{-\infty}^\infty \textnormal{d}t\ e^{i\omega t} F(\mathbf{q},t)\ .
\end{eqnarray}
The latter is also known as intermediate scattering function (ISF) in the literature~\cite{siegfried_review}, and is given by
\begin{eqnarray}\label{eq:ISF}
F(\mathbf{q},t) = \braket{\hat{n}(\mathbf{q},t)\hat{n}(-\mathbf{q},0)}\, ,
\end{eqnarray}
with $\hat{n}(\mathbf{q},t)$ being the density operator (expressed here in reciprocal space) at time $t$. We note that $\braket{\dots}$ denotes a thermodynamic expectation value.

In thermodynamic equilibrium, the Hamiltonian $\hat{H}$ does not explicitly depend on $t$, and the DSF obeys the detailed balance between positive and negative frequencies~\cite{quantum_theory},
\begin{eqnarray}\label{eq:detailed_balance}
S(\mathbf{q},-\omega) = e^{-\beta\omega} S(\mathbf{q},\omega)\ ,
\end{eqnarray}
where we have defined the inverse temperature $\beta = 1/k_\textnormal{B}T$.
Eq.~(\ref{eq:detailed_balance}) implies that a scattered photon can either loose [$S(\mathbf{q},\omega)$] or gain [$S(\mathbf{q},-\omega)$] the amount of energy $\Delta E(\omega)=\hbar\omega$, and the ratio of the respective probabilities is given by the statistical factor of $e^{-\beta\omega}$. In the ground state, an energy gain is not possible leading to $S(\mathbf{q},\omega<0)\equiv0$ as $\beta\to\infty$.
In principle, Eq.~(\ref{eq:detailed_balance}) would directly allow to extract the temperature from a given $S(\mathbf{q},\omega)$ if it includes both positive and negative frequencies. In particular, it has the considerable advantage that no theoretical model for $S(\mathbf{q},\omega)$ is required, and the corresponding analysis would be exact.
Yet, the convolved XRTS intensity [Eq.~(\ref{eq:intensity})] 
does not fulfill Eq.~(\ref{eq:detailed_balance}). 
Moreover, the deconvolution of the intensity to obtain the actual DSF is, in general, rendered highly unstable by the noise in the experimental signal and, therefore, does not constitute a possibility in most cases.
Although an approximate utilization of the detailed balance relation might still be possible in some cases~\cite{DOPPNER2009182}, its key advantages -- being a) exact and model-free and b) generally applicable -- cannot be leveraged.

Therefore, an alternative for obtaining the temperature (as well as a host of other system parameters such as the number density $n$ or ionization degree $Z$) has become common practice~\cite{Gregori_PRE_2003,siegfried_review,kraus_xrts}:  1) construct a suitable model $S_\textnormal{model}[T](\mathbf{q},\omega)$ for the DSF, 2) convolve it with the instrument function $R(\omega)$, and 3) compare it to the experimentally measured intensity $I(\mathbf{q},\omega)$. In this way, the originally unknown parameters such as the temperature $T$ are effectively reconstructed from a fit to the XRTS signal.
Naturally, this approach strongly relies on the utilized model description for $S(\mathbf{q},\omega)$, which can substantially affect the obtained free parameters. For example, Gregori \emph{et al.}~\cite{Gregori_PRE_2003} have suggested using the Chihara decomposition~\cite{Chihara_1987}, where the total DSF is split into separate contributions from bound electrons, free electrons, and transitions between the two. Yet, the validity of this \emph{chemical picture} is particularly questionable in the WDM regime, where electrons can be weakly localized around the ions~\cite{Baczewski_PRL_2016}.

The state of the art is using the Kubo-Greenwood (KG) formalism based on eigenvalues and occupations of Kohn-Sham density functional theory~\cite{PhysRev.140.A1133,Holst_2011,Preising_2020} for obtaining the dielectric function in the optical limit. It can subsequently be extended to all wavenumbers in terms of the Mermin dielectric function, with the required collision frequencies calculated from the KG dielectric function~\cite{Plagemann_2012}.

A more sophisticated alternative is the use of time-dependent density functional theory (TD-DFT) simulations~\cite{Baczewski_PRL_2016,Mo_PRL_2018,Ramakrishna_PRB_2021}, an in-principle exact method for determining the quantum dynamics of electrons under external time-dependent perturbations. TD-DFT neither presupposes an artificial decomposition nor a continuation from the optical limit. On the other hand, present implementations of TD-DFT rely on approximations that might limit their utility under WDM conditions. The development of more accurate exchange-correlation approximations beyond the adiabatic approximation is an active area of research~\cite{GrKo1986,Do1994,ViKo1996,ViUl1997,DoBu1997,MaBu2002,GrBa2004,To2007,My2008,Panholzer_PRL_2018,leblanc}. Moreover, the considerable computational cost of TD-DFT calculations makes them impracticable as a method for optimizing over a wide range of parameters required for reproducing XRTS signals. Currently, this rules out TD-DFT for on-the-fly interpretation of experiments. 

Finally, we note that, despite impressive recent progress, the reliable modelling of $S(\mathbf{q},\omega)$ using potentially more accurate methods such as non-equilibrium Green functions~\cite{kwong_prl-00,Kas_PRL_2017} or even exact path integral Monte Carlo methods~\cite{dornheim_dynamic,dynamic_folgepaper,Dornheim_PRE_2020} is presently not feasible for realistic WDM applications. Moreover, the inevitable systematic errors of less accurate methods such as the Chihara decomposition are expected to become more pronounced for complex materials, such as the ablator coating of an ICF fuel capsule~\cite{Betti2016} or complex mixtures of elements that occur in planetary interiors~\cite{Georg2007}.

\subsection{Temperature extraction in the Laplace domain\label{sec:ITCF}}
Let us next consider the two-sided Laplace transform of the dynamic structure factor
\begin{eqnarray}\label{eq:F}
\mathcal{L}\left[S(\mathbf{q},\omega)\right] &=& \int_{-\infty}^\infty \textnormal{d}\omega\ e^{-\tau\omega}\ S(\mathbf{q},\omega) \\\nonumber
&=& F(\mathbf{q},\tau)\ .
\end{eqnarray}
In fact, Eq.~(\ref{eq:F}) directly corresponds to the intermediate scattering function [Eq.~(\ref{eq:ISF})], but evaluated at an \emph{imaginary time} $t=-i\hbar\tau$, with $\tau\in[0,\beta]$,
\begin{eqnarray}
F(\mathbf{q},\tau) = \braket{\hat{n}(\mathbf{q},\tau)\hat{n}(-\mathbf{q},0)}\ .
\end{eqnarray}
Such imaginary-time correlation functions~\cite{Dornheim_JCP_ITCF_2021,dornheim2022physical} naturally emerge within Feynman's path integral picture of statistical mechanics~\cite{HagenKleinertPI} and give access to a wealth of linear~\cite{dornheim_dynamic,dornheim_ML,Dornheim_SciRep_2022} and nonlinear~\cite{Dornheim_JCP_ITCF_2021} response properties of a given system. In particular, Eq.~(\ref{eq:F}) often constitutes the starting point for an \emph{analytic continuation}~\cite{JARRELL1996133}, where one tries to reconstruct $S(\mathbf{q},\omega)$ based on highly accurate path integral Monte Carlo data for the imaginary-time ISF $F(\mathbf{q},\tau)$. 

In the context of the present work, the main utility of $F(\mathbf{q},\tau)$ is its symmetry with respect to $\tau=\beta/2$, which directly follows from inserting the detailed balance relation [Eq.~(\ref{eq:detailed_balance})] into Eq.~(\ref{eq:F}) and using symmetry to reduce the integration interval to only positive frequencies,
\begin{eqnarray}\label{eq:symmetry}
F(\mathbf{q},\tau) &=& \int_0^\infty \textnormal{d}\omega\ S(\mathbf{q},\omega)\left\{ e^{-\tau\omega} + e^{-\omega(\beta-\tau)} \right\}\\\nonumber
 &=& F(\mathbf{q},\beta-\tau)\ .
\end{eqnarray}
In practice, $F(\mathbf{q},\tau)$ has a minimum at $\tau=\beta/2$ [cf.~Fig.~\ref{fig:Synthetic_DSF_x} b) below], and thus directly determines the temperature. 

The final obstacle on the way towards the exact, model-free extraction of the temperature from an XRTS measurement is then the convolution with the instrument function $R(\omega)$, Eq.~(\ref{eq:intensity}). It is easy to see that the two-sided Laplace transform of the intensity, $\mathcal{L}\left[S(\mathbf{q},\omega)\circledast R(\omega)\right]$, does not obey the symmetry of Eq.~(\ref{eq:symmetry}). Conveniently, this problem can be fully overcome by making use of the well-known convolution theorem
\begin{eqnarray}\label{eq:convolution_theorem}
\mathcal{L}\left[S(\mathbf{q},\omega)\right] = \frac{ \mathcal{L}\left[S(\mathbf{q},\omega)\circledast R(\omega)\right] }{ \mathcal{L}\left[ R(\omega)\right] }\ .
\end{eqnarray}
Naively, the direct evaluation of the LHS of Eq.~(\ref{eq:convolution_theorem}) gives straightforward access to the temperature of any given system. But at a closer look, it does require the explicit deconvolution of Eq.~(\ref{eq:intensity}), which is rendered numerically highly unstable by the noise in the experimental signal. The evaluation of the enumerator of the RHS, on the other hand, is very robust with respect to noise due to its definition as an integral over the entire relevant frequency range. 
Moreover, the impact of the instrument function is completely removed by the denominator $\mathcal{L}\left[R(\omega)\right]$, which, too, can be evaluated without any problems. In other words, Eq.~(\ref{eq:convolution_theorem}) directly implies that it is possible to extract the exact temperature of a given system from an XRTS measurement without the need for an explicit deconvolution, and, at the same time, without the bias due the broadening by the instrument function.

\subsection{Integration range and convergence\label{sec:convergence}}
It is easy to see that the symmetry relation Eq.~(\ref{eq:symmetry}) holds for any symmetric integration range, i.e., $F_{ab}(\mathbf{q},\tau)=F_{ab}(\mathbf{q},\beta-\tau)$
with
\begin{eqnarray}\nonumber
F_{ab}(\mathbf{q},\tau) &=& \int_{-b}^{-a} S(\mathbf{q},\omega) e^{-\tau\omega} \textnormal{d}\omega\ + \int_{a}^{b} S(\mathbf{q},\omega) e^{-\tau\omega} \textnormal{d}\omega \\
&=& \int_a^b S(\mathbf{q},\omega) \left\{ e^{-\tau\omega} + e^{-\omega(\beta-\tau)} \right\}\ .
\label{eq:cut_out}\end{eqnarray}
Therefore, deconvolved data for $S(\mathbf{q},\omega)$ would only be required on a finite frequency interval for both positive and negative values of $\omega$.
On the other hand, the proof of the convolution theorem of the two-sided Laplace transform explicitly requires the infinite integration range. Yet, in an XRTS experiment the intensity is measured in a finite frequency range $\omega\in[\omega_\textnormal{min},\omega_\textnormal{max}]$ with reasonable accuracy.

Hence, we define the symmetrically truncated Laplace transform of the XRTS signal as
\begin{eqnarray}
\mathcal{L}_x\left[S(\mathbf{q},\omega)\circledast R(\omega)\right] &=& \int_{-x}^x \textnormal{d}\omega\ e^{-\tau\omega} \left\{S(\mathbf{q},\omega)\circledast R(\omega)\right\}\ , \nonumber \\ & & \label{eq:Laplace_truncated}
\end{eqnarray}
with the corresponding truncated imaginary-time ISF
\begin{eqnarray}\label{eq:F_truncated}
F_x(\mathbf{q},\tau) = \frac{\mathcal{L}_x\left[S(\mathbf{q},\omega)\circledast R(\omega)\right] }{\mathcal{L}\left[R(\omega)\right]}\ .
\end{eqnarray}
Clearly, it holds
\begin{eqnarray}\label{eq:lim1}
\lim_{x\to\infty} F_x(\mathbf{q},\tau) = F(\mathbf{q},\tau)\ ,
\end{eqnarray}
and the convergence with respect to $x$ can simply be checked in practice.

\section{Results: Synthetic data\label{sec:results_synthetic}}

\subsection{Imaginary time intermediate scattering function\label{sec:ITCF_synthetic}}
\begin{figure*}\centering
\includegraphics[width=0.45\textwidth]{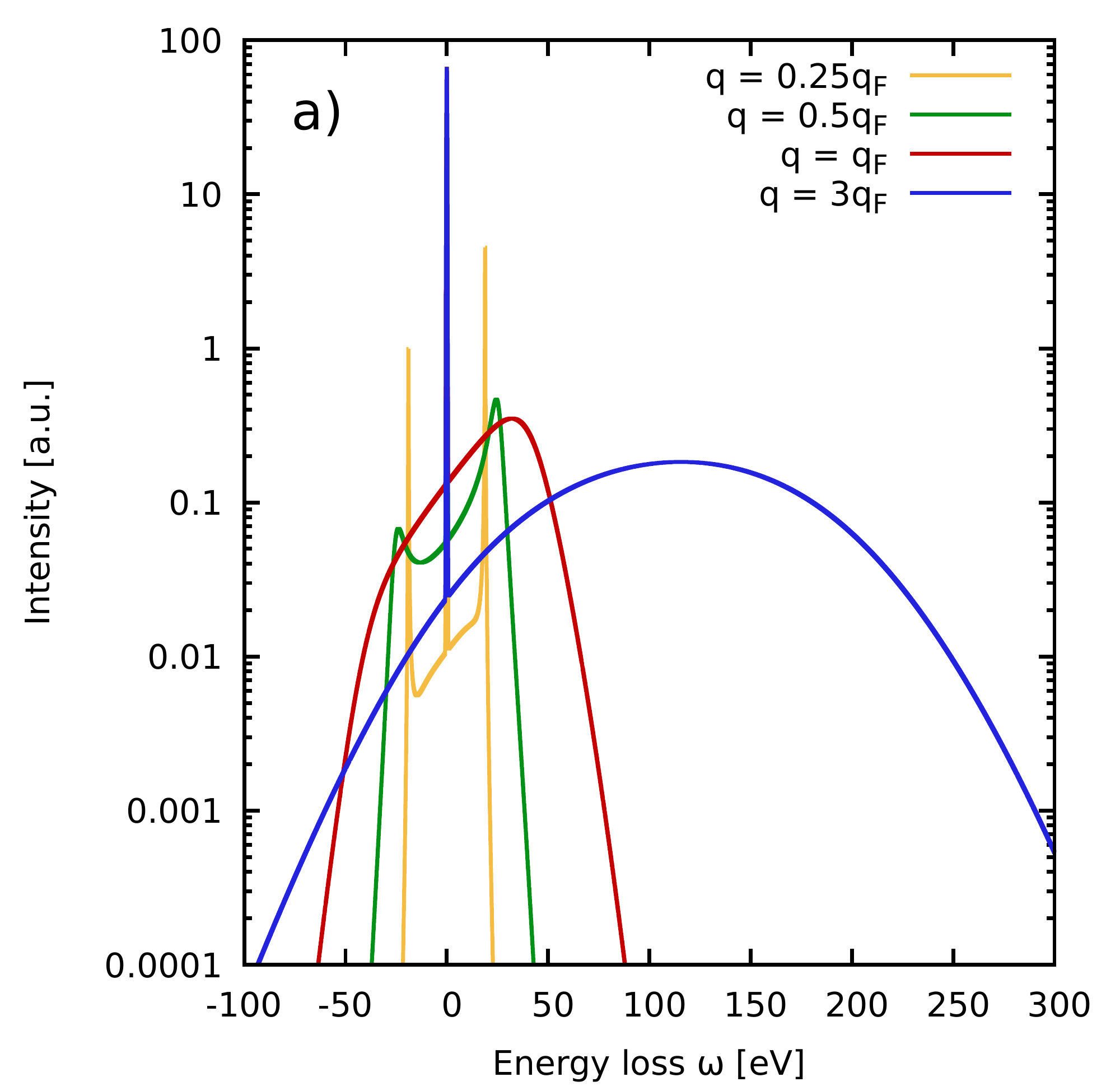}\includegraphics[width=0.45\textwidth]{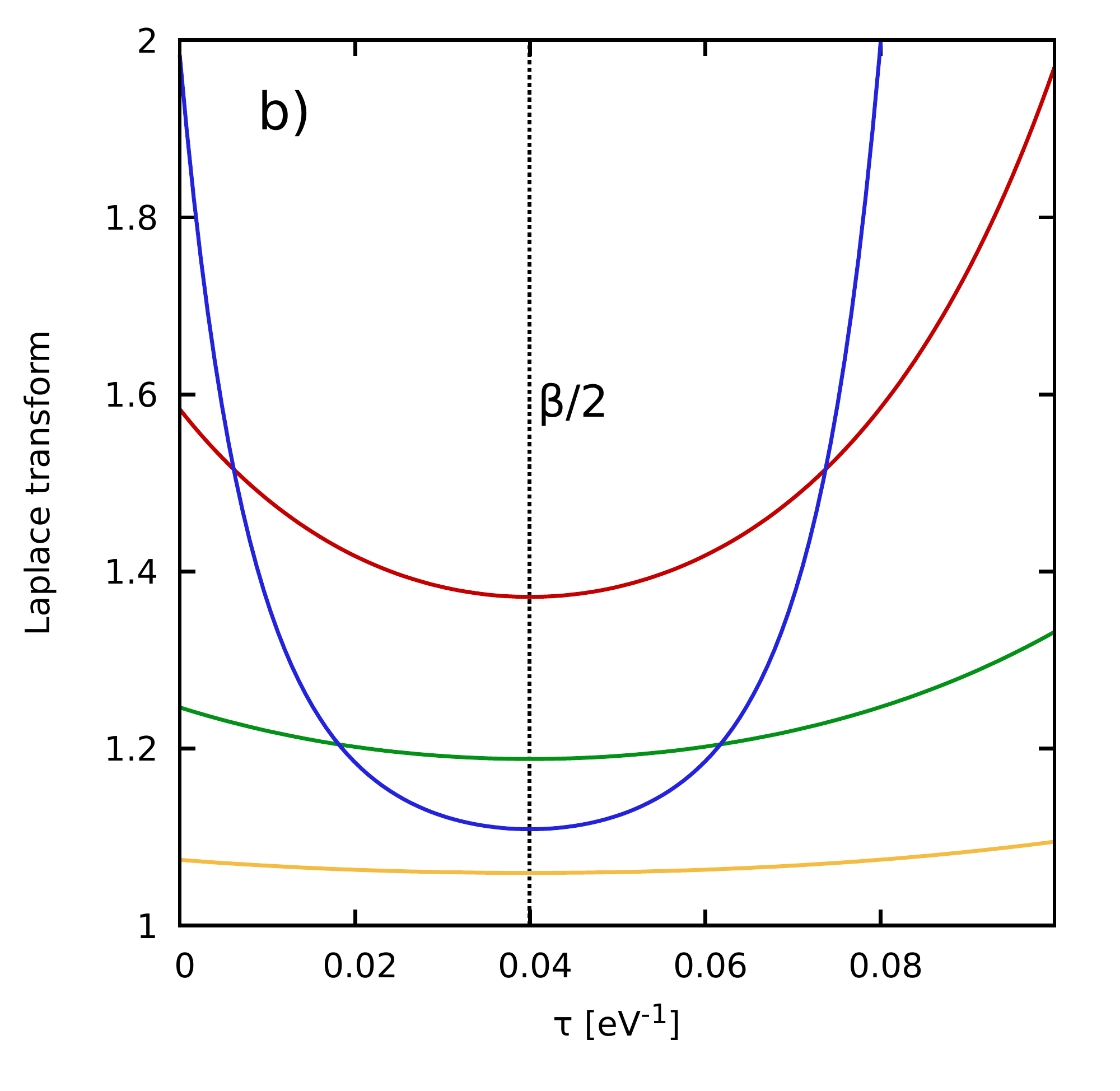}
\caption{\label{fig:Synthetic_DSF_x} a) Synthetic data of the (unconvolved) dynamic structure factor $S(\mathbf{q},\omega)$ at $r_s=2$ and $\Theta=1$ for different values of the wave number $q=|\mathbf{q}|$. b) Corresponding imaginary-time intermediate scattering functions $F(\mathbf{q},\tau)=\mathcal{L}\left[S(\mathbf{q},\omega)\right]$. We note the perfect symmetry around $\tau=\beta/2$ (vertical dotted black line) for all $q$.
}
\end{figure*} 

We illustrate our new methodology by generating synthetic scattering intensities based on a uniform electron gas model~\cite{dornheim_dynamic,dynamic_folgepaper,review}. Specifically, we combine a DSF in terms of a neural-network representation of the static local field correction of the uniform electron gas~\cite{dornheim_ML} with a sharp elastic feature around $\omega=0$,
\begin{eqnarray}\label{eq:DSF_synthetic}
S(\mathbf{q},\omega) = S_\textnormal{UEG}(\mathbf{q},\omega)  + \frac{e^{-\omega^2/2\eta^2}}{\sqrt{2\pi\eta^2}}\,,
\end{eqnarray}
with $\eta$ being the standard deviation.  In particular, the last term on the RHS becomes a delta distribution in the limit of $\eta\to0$. In practice, we use $\eta = \omega_{\text{p}e}/100$ for numerical convenience, with $\omega_{\text{p}e}=\sqrt{3/r_s^3}$ denoting the usual plasma frequency~\cite{Ott2018}.

In Fig.~\ref{fig:Synthetic_DSF_x} a), we show the corresponding dynamic structure factors for relevant values of the wave number $q$ at the electronic Fermi temperature $\Theta=1$ ($T=12.53$eV) and a metallic density of $r_s=2$. By design, all DSFs exhibit the same sharp elastic feature around $\omega=0$. The yellow curve corresponding to a quarter of the Fermi wave number $q=0.25q_\textnormal{F}$ exhibits a sharp plasmon peak around $\omega=25$eV. Upon increasing $q$, the plasmon is first broadened (green curve, $q=0.5q_\textnormal{F}$), then disappears in a single broad inelastic curve at $q=q_\textnormal{F}$ (red). Finally, the blue curve computed for a large wave number $q=3q_\textnormal{F}$ in the non-collective, single-particle regime exhibits a broad Gaussian form, and its peak position increases parabolically with $q$. 
Fig.~\ref{fig:Synthetic_DSF_x} b) shows the corresponding imaginary time intermediate scattering function $F(\mathbf{q},\tau)=\mathcal{L}\left[S(\mathbf{q},\omega)\right]$, i.e., the two-sided Laplace transform of the DSF defined in Eq.~(\ref{eq:F}). Evidently, the different curves substantially depend on the wave number, thereby reflecting the transition from the collective regime $q\ll q_\textnormal{F}$ to the single-particle regime $q\gg q_\textnormal{F}$. This has been analyzed in detail in the recent Ref.~\cite{Dornheim_PTR_2022}.
At the same time, all curves are perfectly symmetric around the same value of $\tau=\beta/2$. In other words, knowledge of the DSF allows for a straightforward extraction of the temperature for any value of the wave vector $\mathbf{q}$ without any physical assumptions or models.

\begin{figure}\centering
\includegraphics[width=0.45\textwidth]{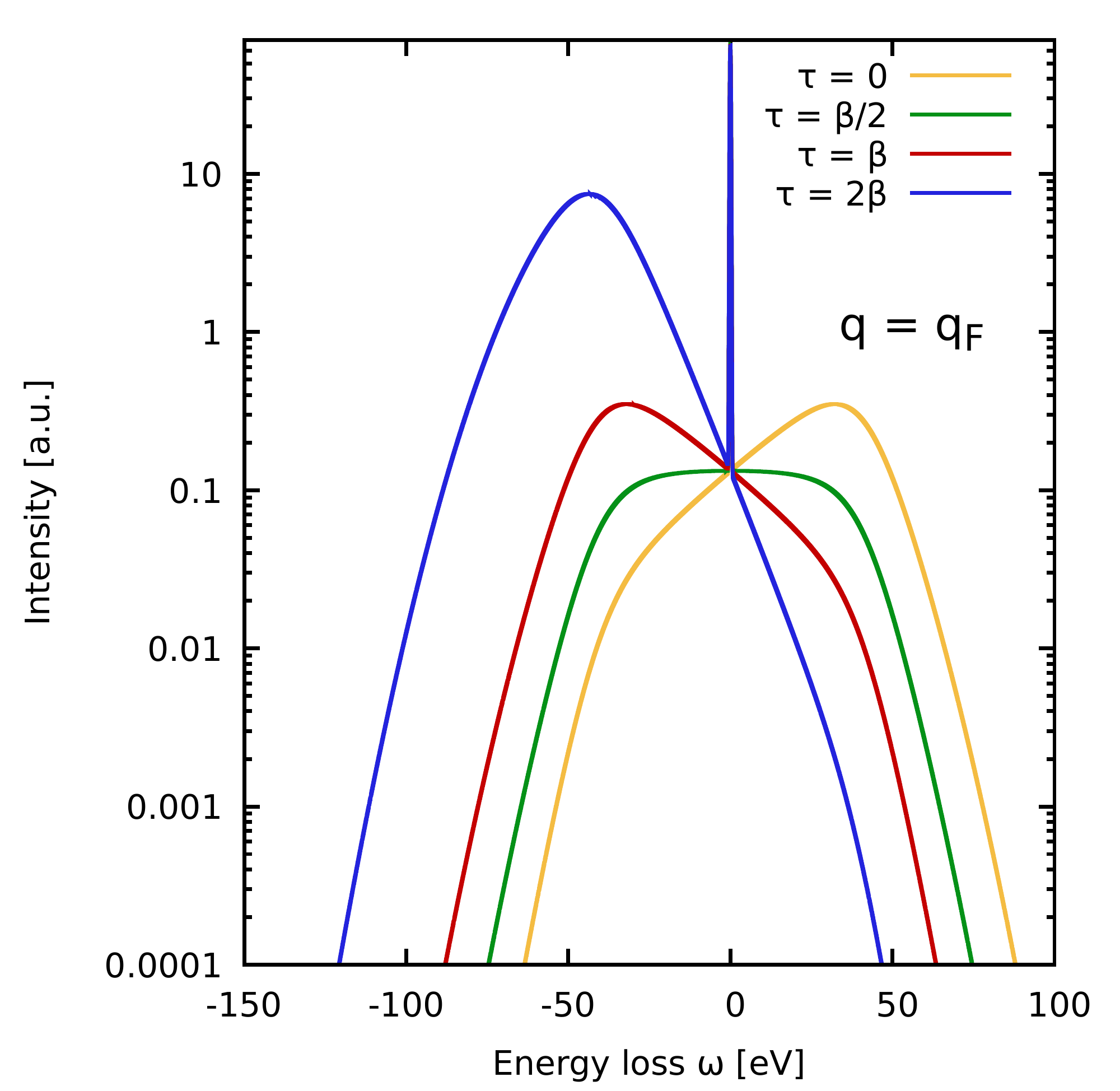}
\caption{\label{fig:Synthetic_DSF_contribution} Contribution to the two-sided Laplace transform $\mathcal{L}\left[S(\mathbf{q},\omega)\right]$, $S(\mathbf{q},\omega)e^{-\tau\omega}$, for synthetic data at $r_s=2$, $\Theta=1$, and $q=q_\textnormal{F}$ for selected values of the imaginary time $\tau$.
}
\end{figure}

\begin{figure*}\centering
\includegraphics[width=0.45\textwidth]{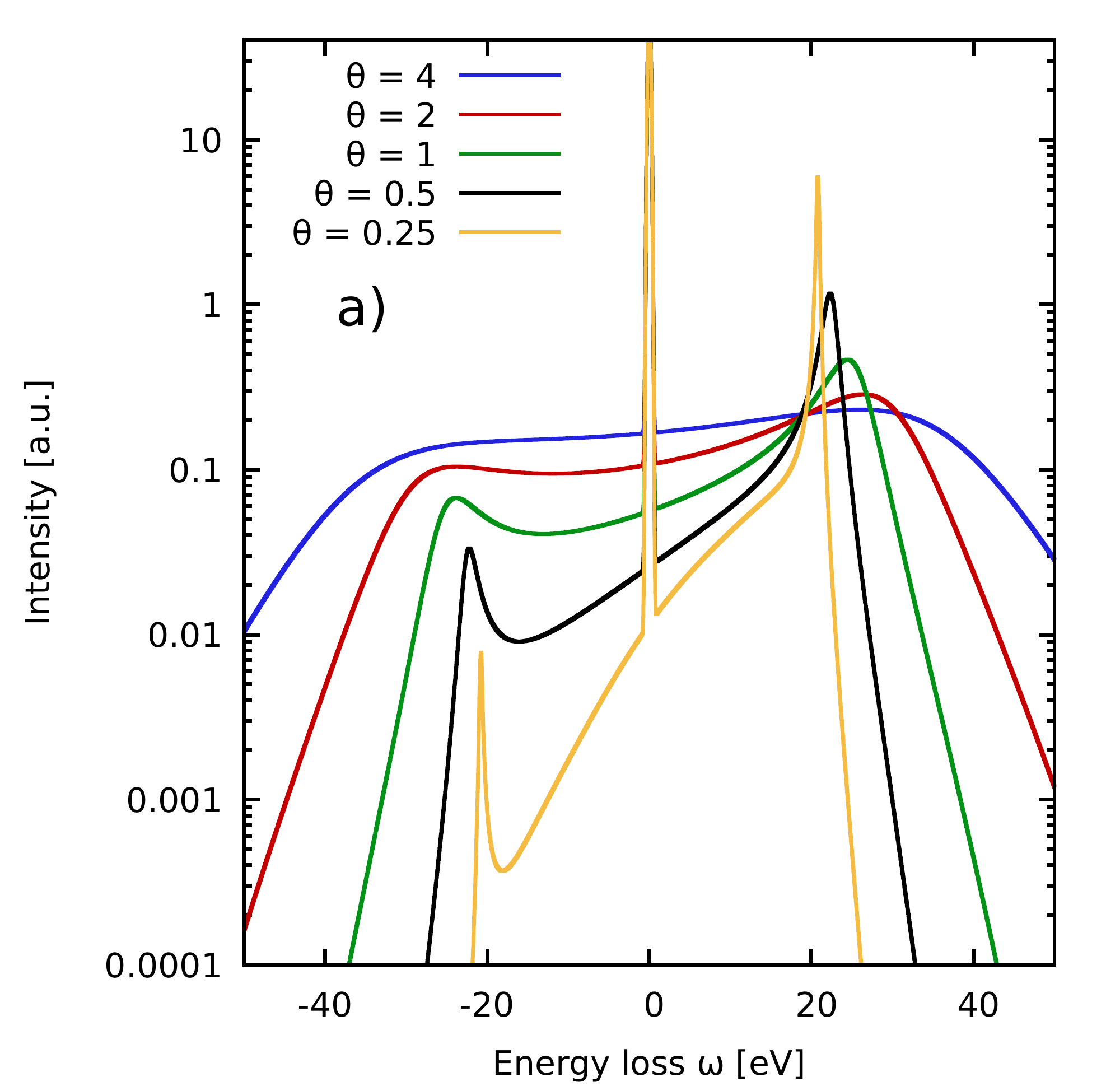}\includegraphics[width=0.45\textwidth]{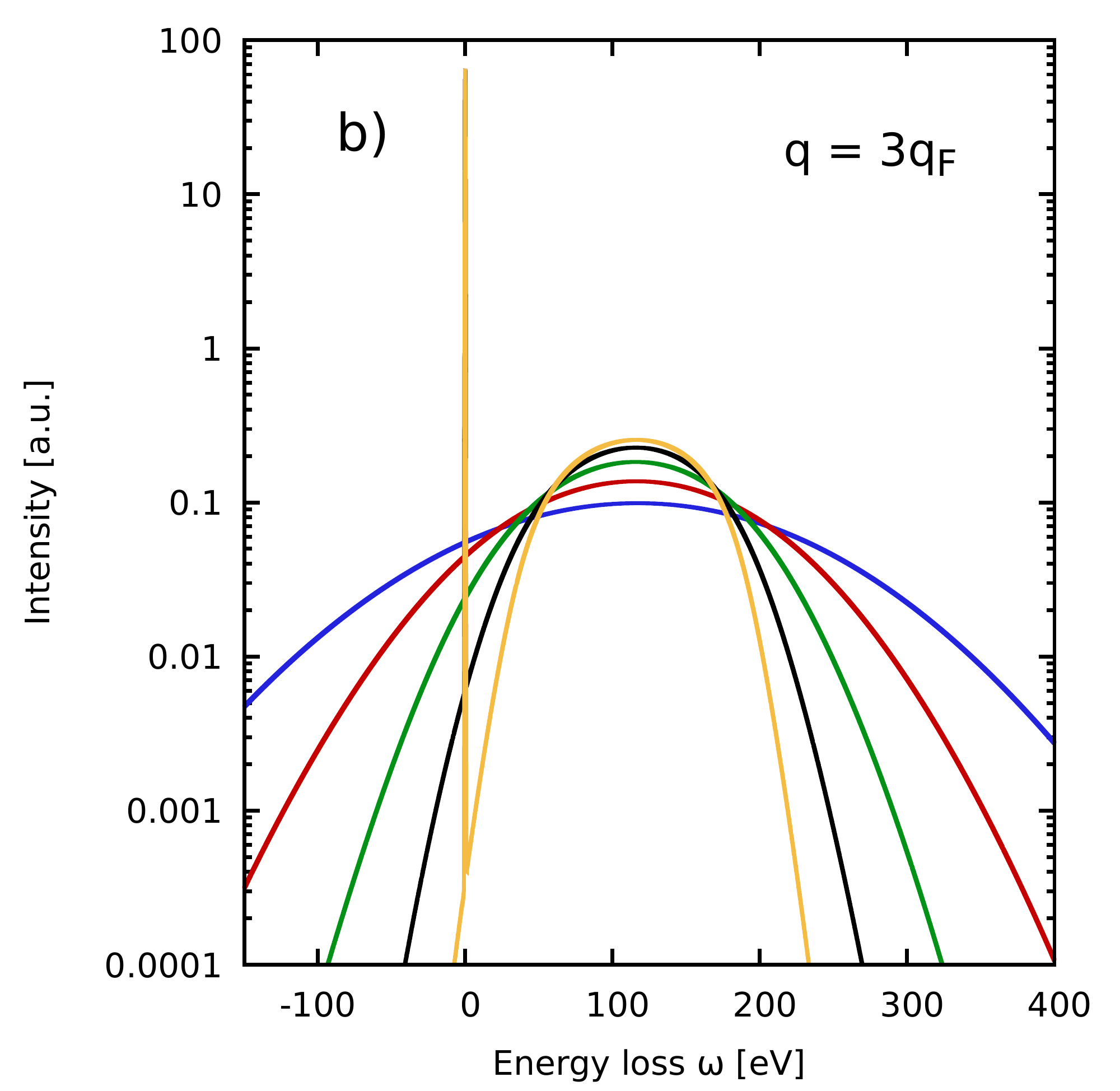}\\\includegraphics[width=0.45\textwidth]{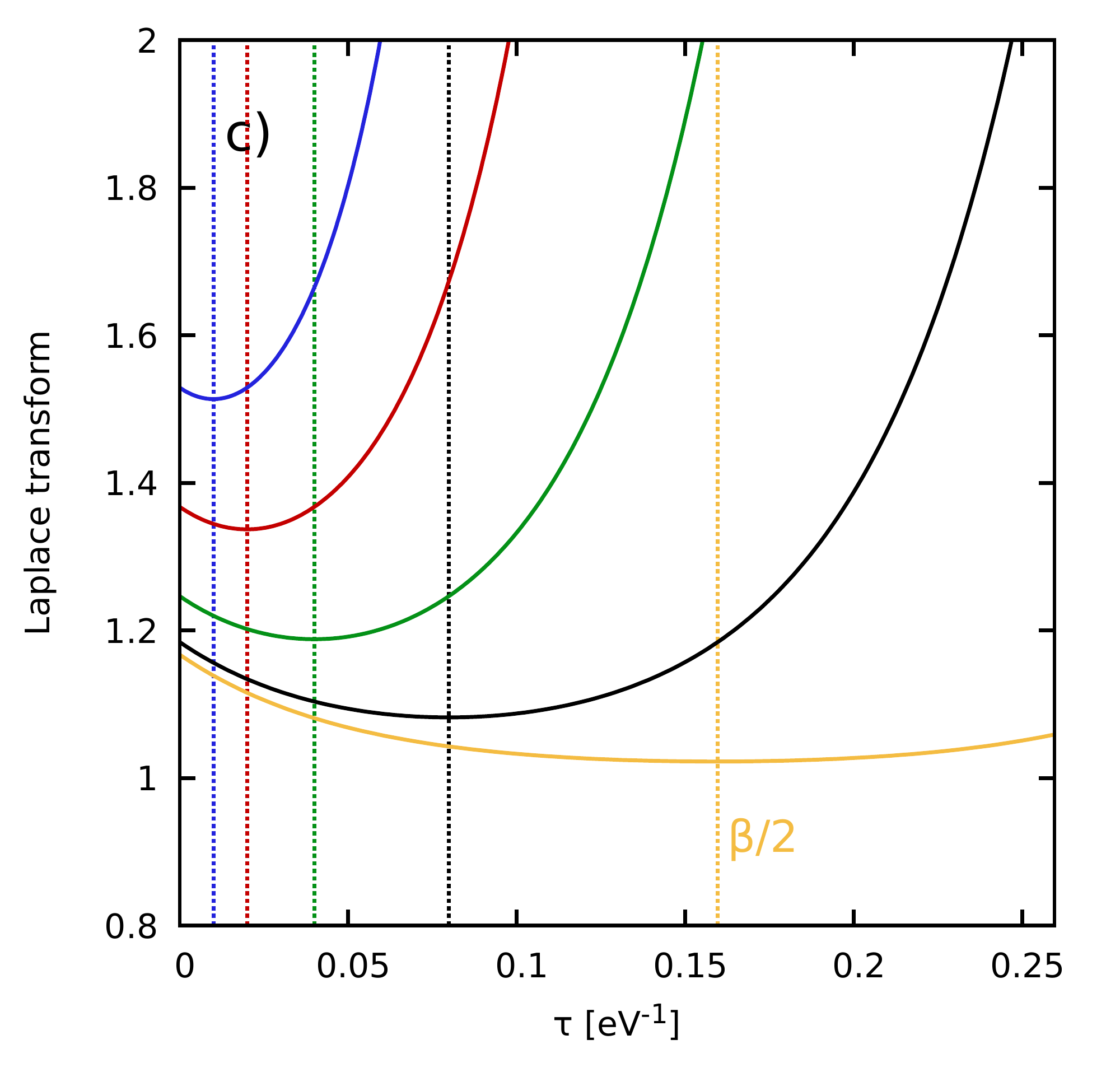}\includegraphics[width=0.45\textwidth]{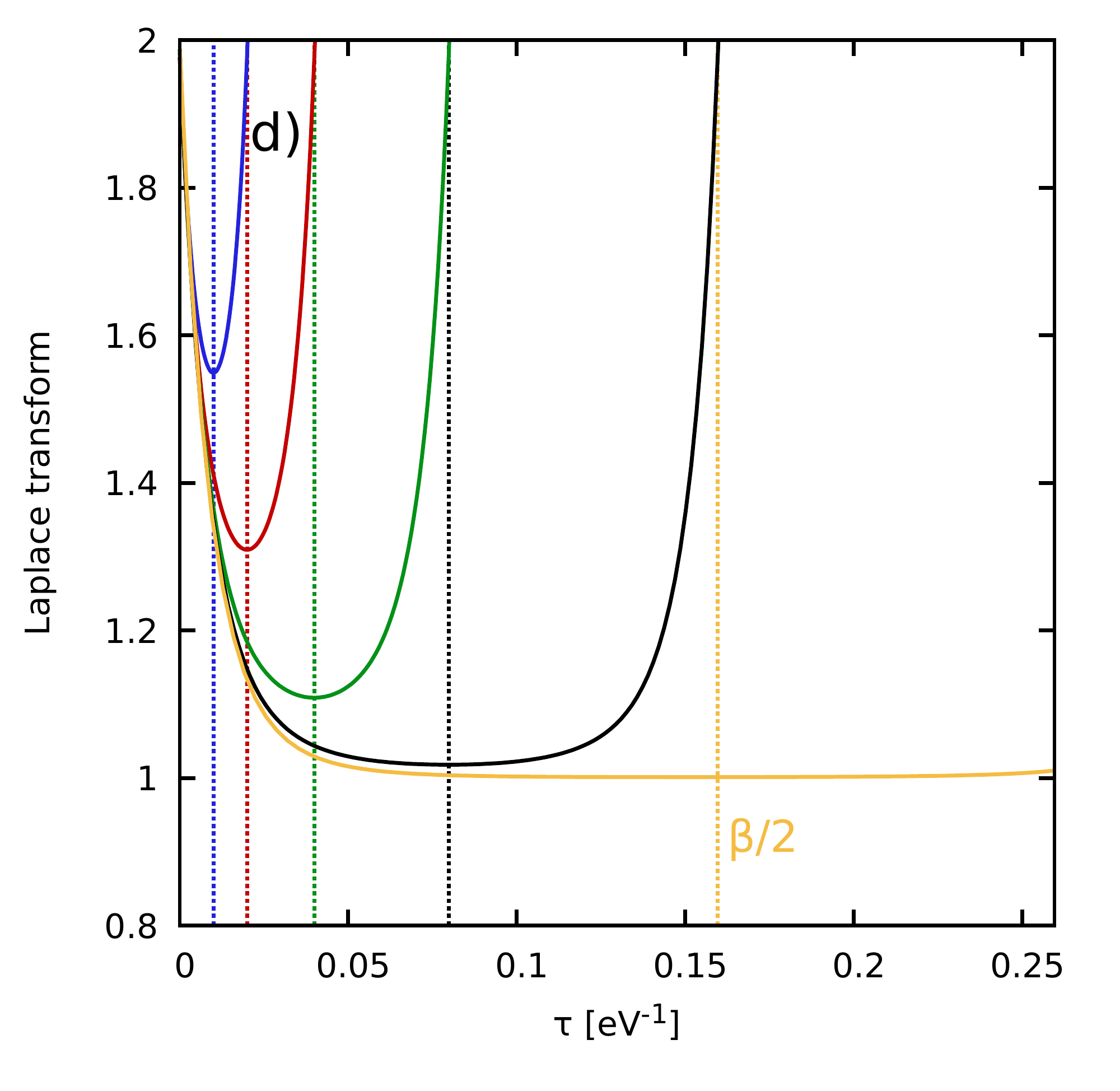}
\caption{\label{fig:Snythetic_DSF_theta} Top row: Dynamic structure factor $S(\mathbf{q},\omega)$ computed from a UEG model [Eq.~(\ref{eq:DSF_synthetic})] for $r_s=2$ and different values of the temperature parameter $\Theta$ at a) $q=0.5q_\textnormal{F}$ (collective) and b) $q=3q_\textnormal{F}$ (single-particle). Bottom row: Corresponding imaginary time intermediate scattering functions $F(\mathbf{q},\tau)=\mathcal{L}\left[S(\mathbf{q},\omega)\right]$. The respective minima at $\tau=\beta/2$ are indicated by the vertical dotted lines.
}
\end{figure*}

We further illustrate the origin of this behavior by analyzing the frequency-resolved contribution to $F(\mathbf{q},\tau)$ for $q=q_\textnormal{F}$ in Fig.~\ref{fig:Synthetic_DSF_contribution}. More specifically, we plot
$S(\mathbf{q},\omega)e^{-\tau\omega}$ for three different values of the imaginary time $\tau$. The yellow curve corresponds to $\tau=0$, i.e., to the original DSF that is also shown in Fig.~\ref{fig:Synthetic_DSF_x} a). The green curve has been obtained for $\tau=\beta/2$, where $F(\mathbf{q},\tau)$ attains its minimum value. Evidently, the corresponding curve is symmetric around $\omega=0$. This is a general property of the dynamic structure factor and can directly be seen by inserting the detailed balance relation Eq.~(\ref{eq:detailed_balance}) into the modified quantity $C(\mathbf{q},\omega)=S(\mathbf{q},\omega)e^{-\beta\omega/2}$,
\begin{eqnarray}
C(\mathbf{q},\omega) &=& e^{-\beta\omega/2} S(\mathbf{q},\omega) \\ \nonumber
&=& S(\mathbf{q},-\omega) e^{\beta\omega} e^{-\beta\omega/2} \\ \nonumber
&=& S(\mathbf{q},-\omega) e^{\beta\omega/2} \\
\nonumber &=& C(\mathbf{q},-\omega)\ .
\end{eqnarray}
The red curve has been obtained for $\tau=\beta$, and corresponds to the original DSF, but mirrored around the $y$-axis. For completeness, we also include a curve for $\tau=2\beta$, which has no physical equivalent in Feynman's imaginary-time path integral picture, but can be easily computed from the two-sided Laplace transform Eq.~(\ref{eq:F}). In this case, the negative frequency range gets substantially enhanced by the exponential factor, whereas, conversely, the positive frequency range gets damped. In practice, the evaluation of Eq.~(\ref{eq:F}) at such large values of $\tau$ would require high-quality data of the DSF at very low frequencies, which is unrealistic at present. At the same time, we note that it is not needed to locate the minimum and hence extract the temperature of the system.

Let us conclude this analysis of the unconvolved DSF by directly considering the impact of the temperature. This is shown in Fig.~\ref{fig:Snythetic_DSF_theta} in the collective regime ($q=0.5q_\textnormal{F}$, left column) and in the single-particle regime ($q=3q_\textnormal{F}$, right column). In particular, Fig.~\ref{fig:Snythetic_DSF_theta} a) shows the DSF evaluated from the usual UEG model at different values of the temperature; beware that the elastic peak of the depicted synthetic model data does not depend on $\Theta$. The yellow curve has been obtained for $\Theta=0.25$ ($T=3.13\,$eV) and exhibits sharp plasmon peaks around $\omega=\pm 20\,$eV. Increasing the temperature by a factor of two ($\Theta=0.5$, $T=6.26\,$eV) yields the black curve where the impact of increasing thermal effects is two-fold: firstly, the DSF is broadened overall and decays more slowly for large $|\omega|$; secondly, the plasmon is damped and shifted to significantly larger frequencies both in the positive and negative frequency domain. We note that approximate models for this \emph{plasmon shift}~\cite{Fortmann_PRE_2010} have been used to determine the temperature in previous XRTS experiments~\cite{Preston_APL_2019}.
Increasing the temperature further to $\Theta=1$ ($T=12.53\,$eV, green) enhances both the broadening and the plasmon shift, until the plasmon is eventually damped out for $\Theta=2$ ($T=25.06\,$eV, red) and $\Theta=4$ ($T=50.12\,$eV).

\begin{figure*}\centering
\includegraphics[width=0.315\textwidth]{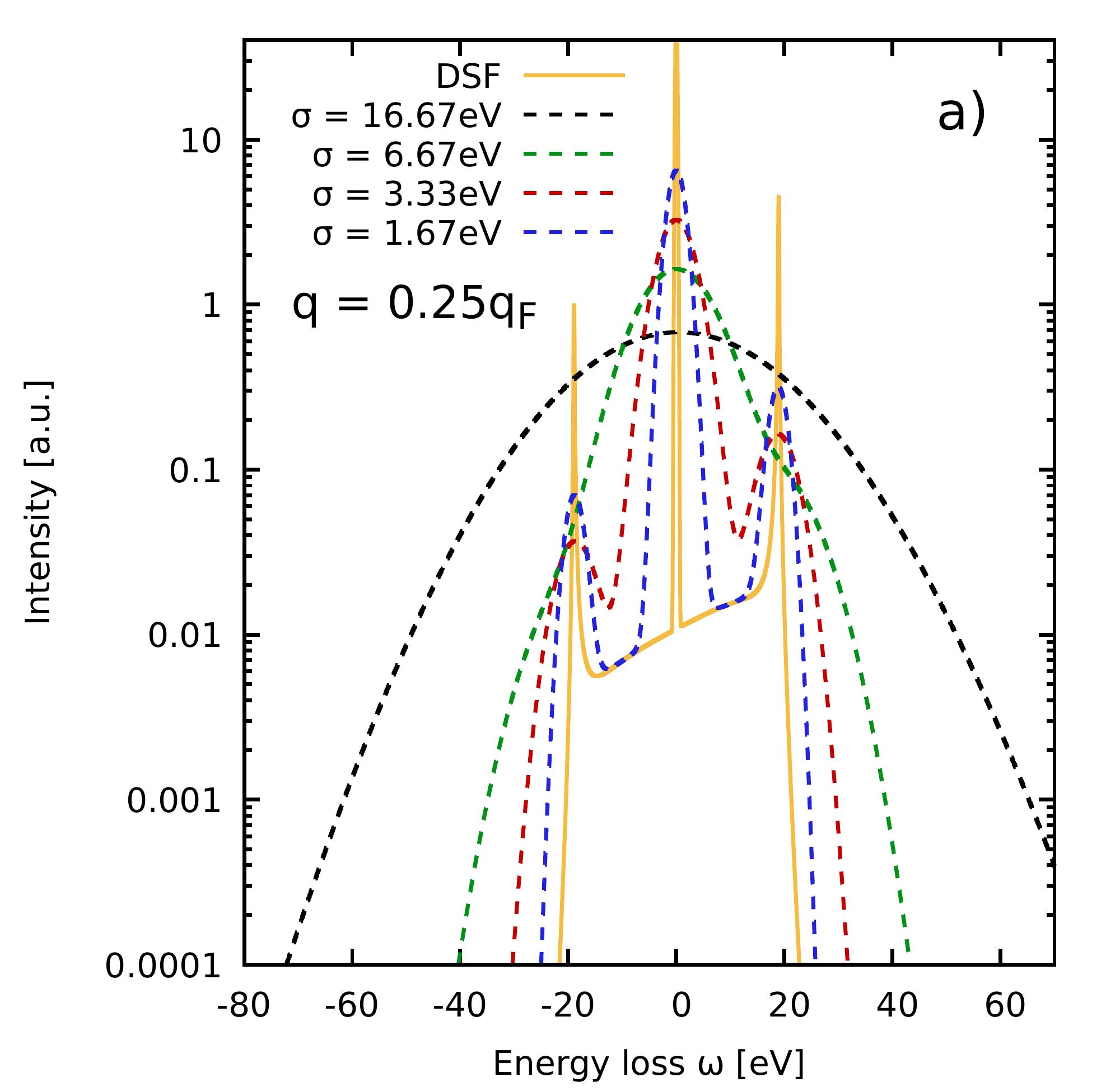}\includegraphics[width=0.315\textwidth]{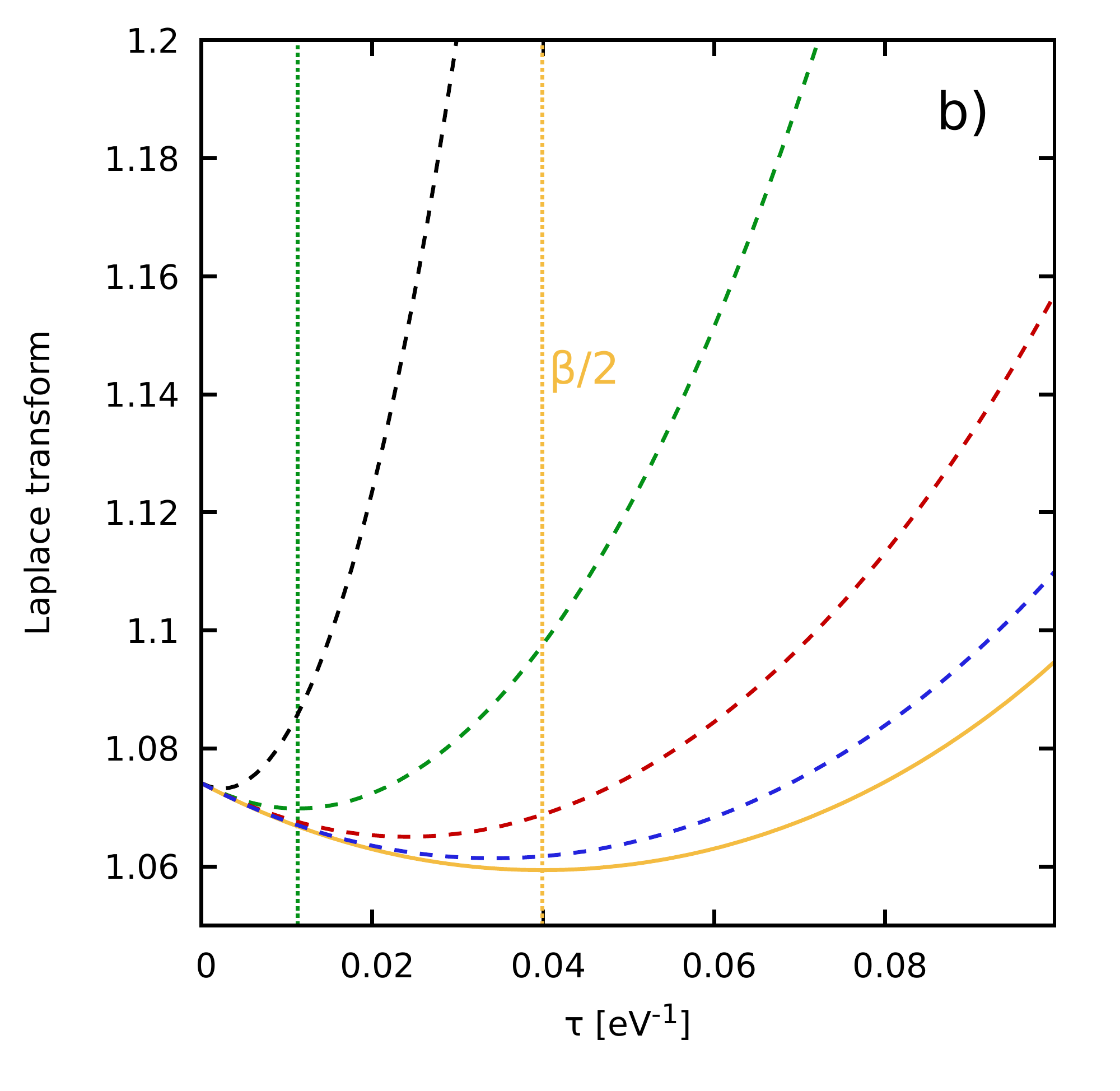}\includegraphics[width=0.315\textwidth]{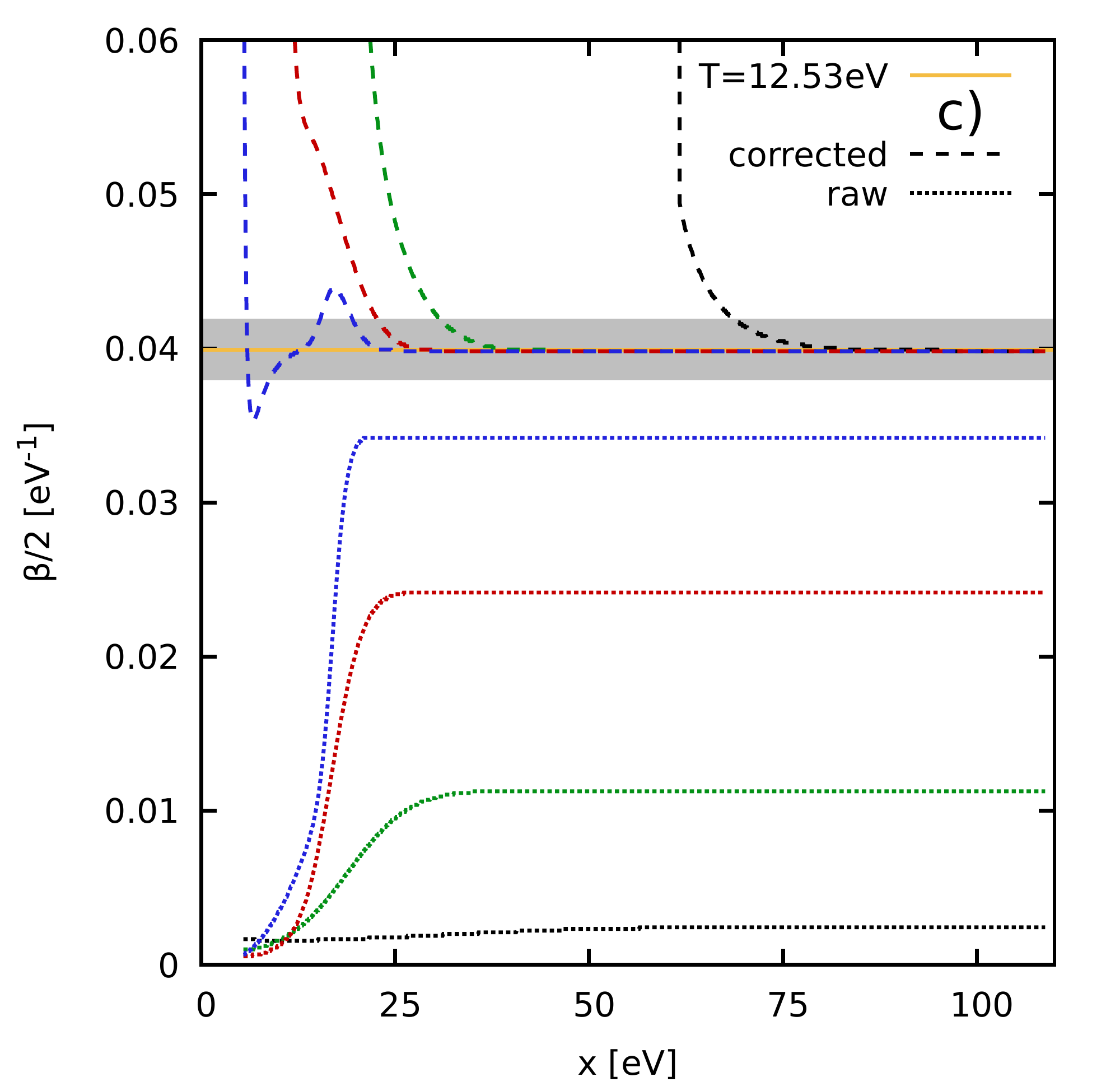}\\
\includegraphics[width=0.315\textwidth]{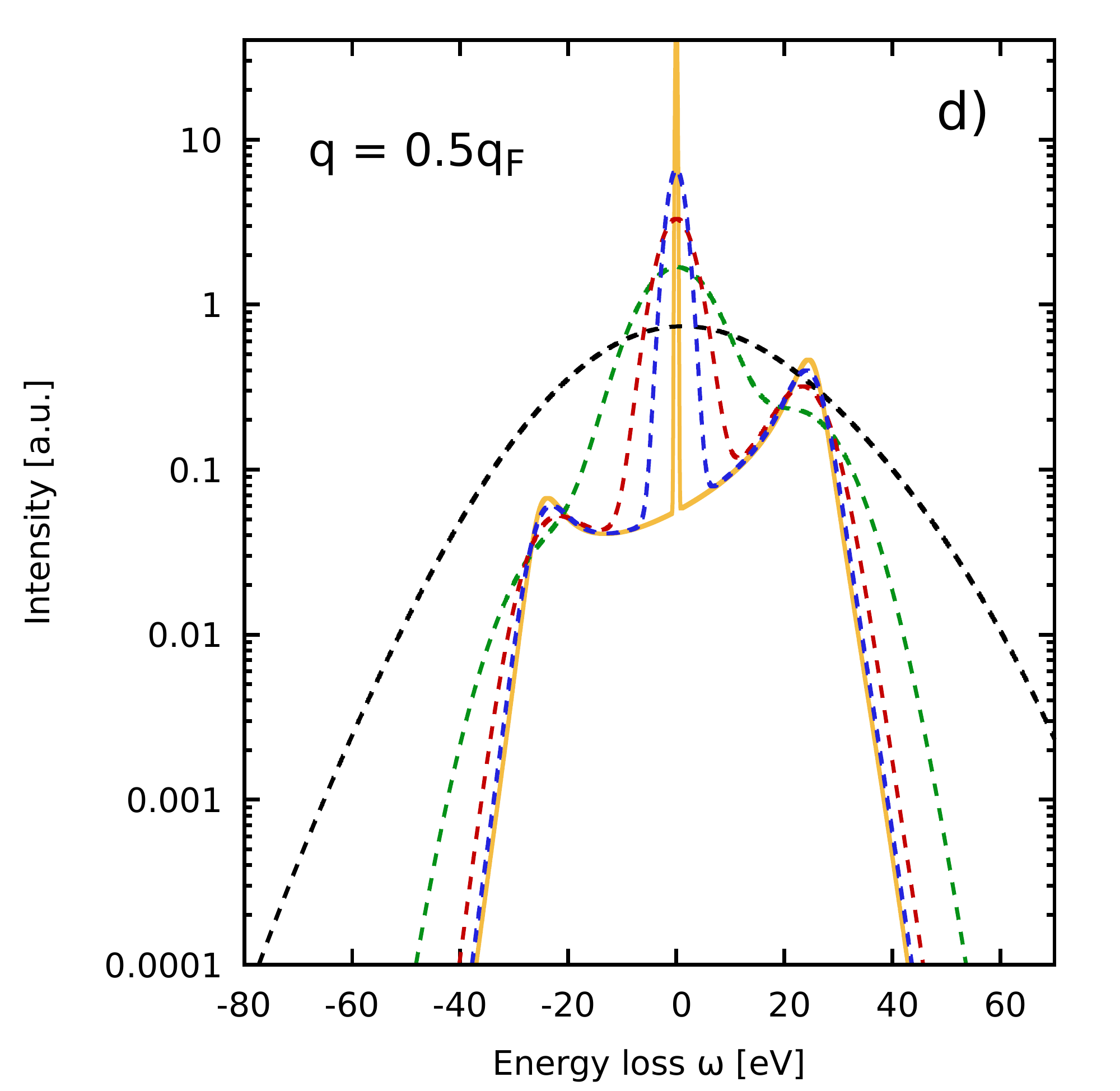}\includegraphics[width=0.315\textwidth]{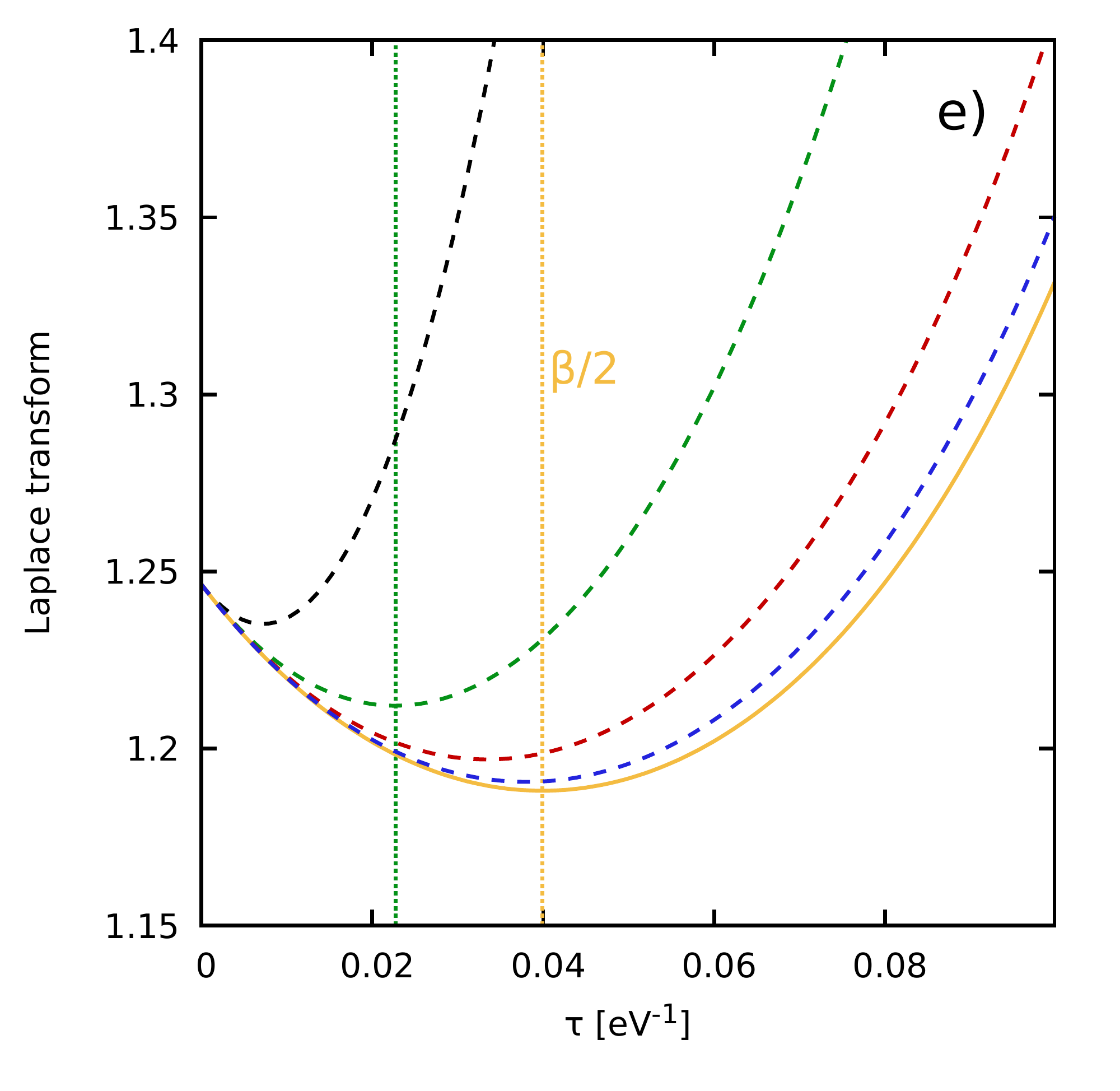}\includegraphics[width=0.315\textwidth]{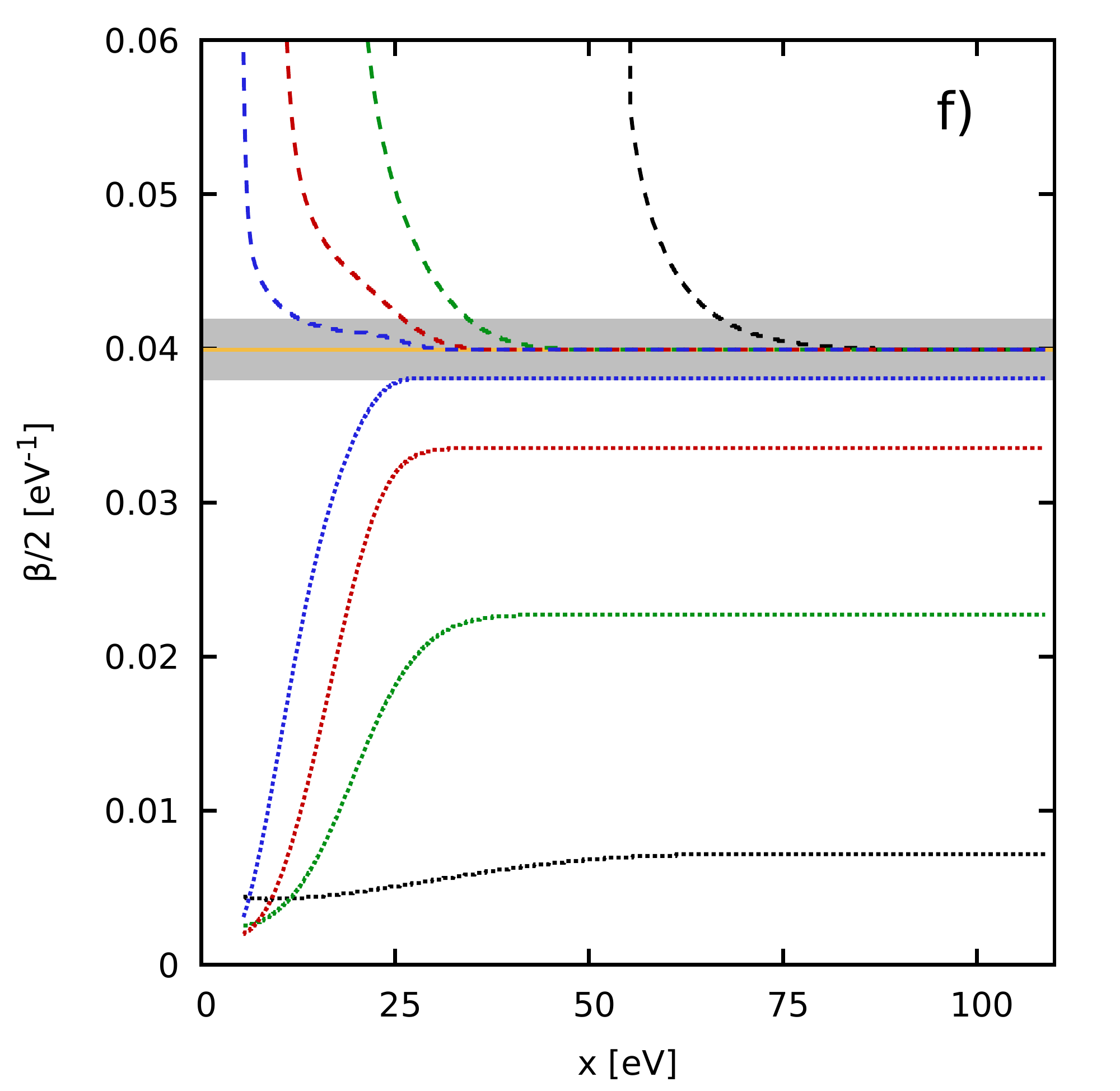}\\\includegraphics[width=0.315\textwidth]{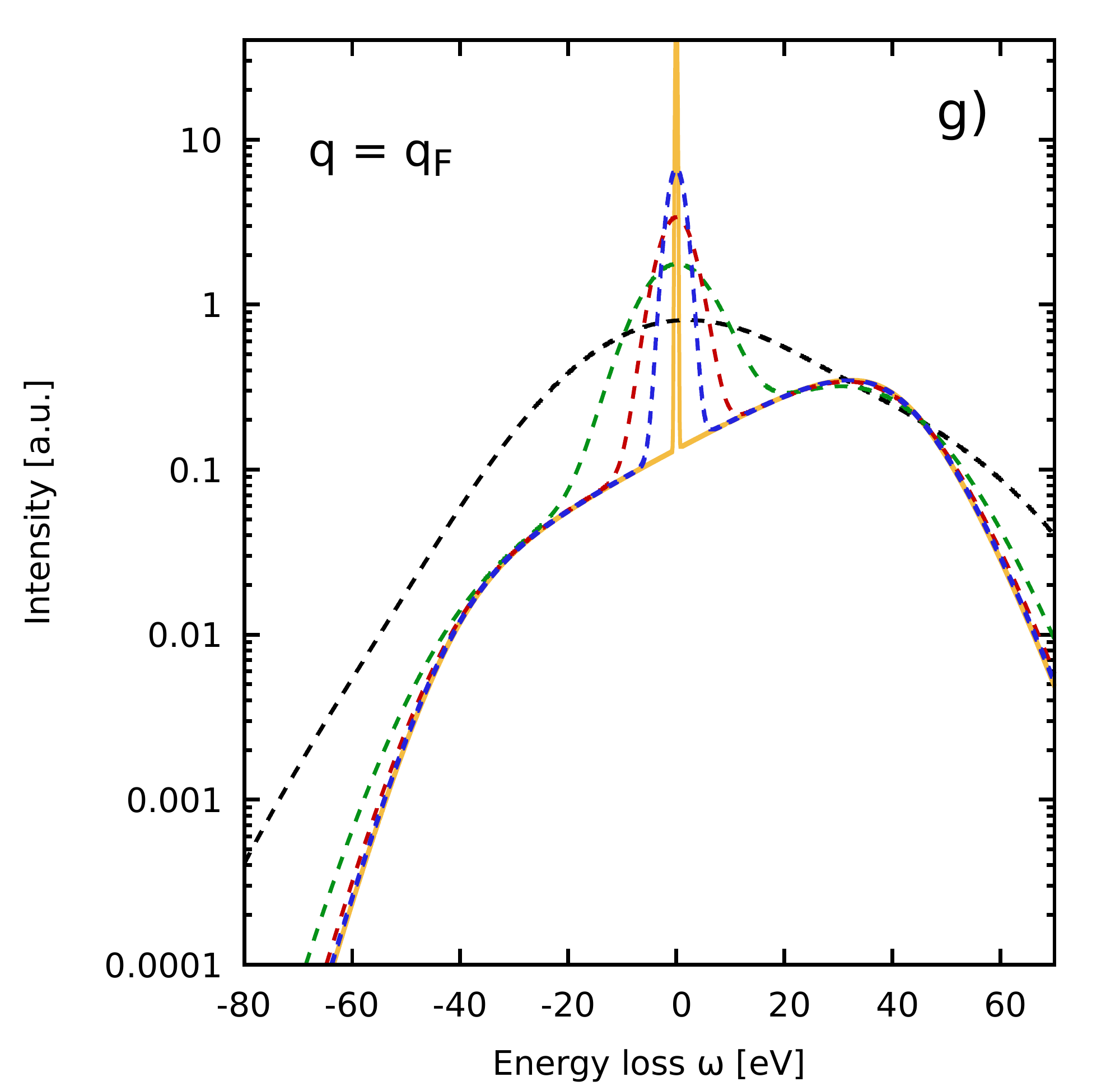}\includegraphics[width=0.315\textwidth]{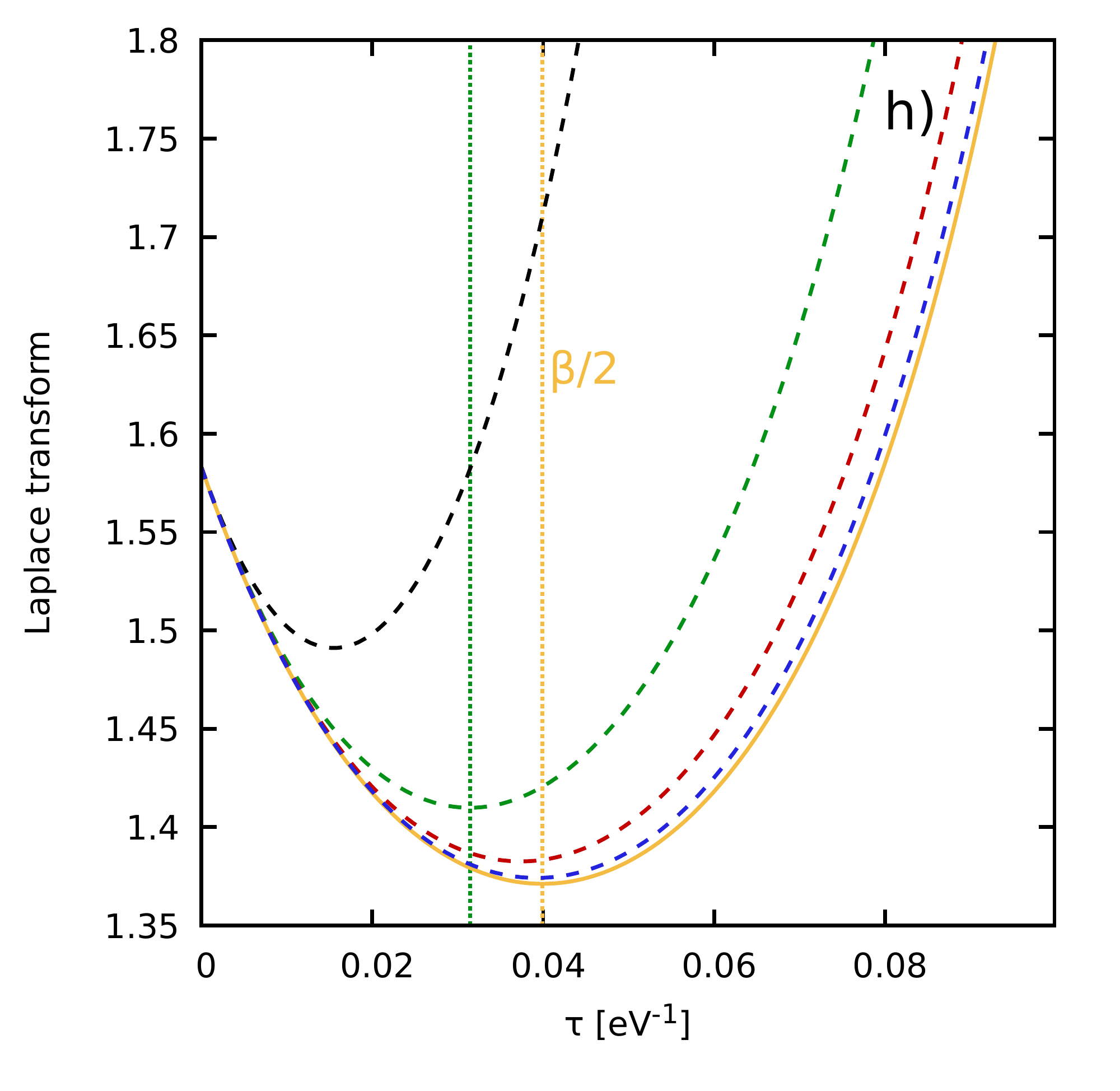}\includegraphics[width=0.315\textwidth]{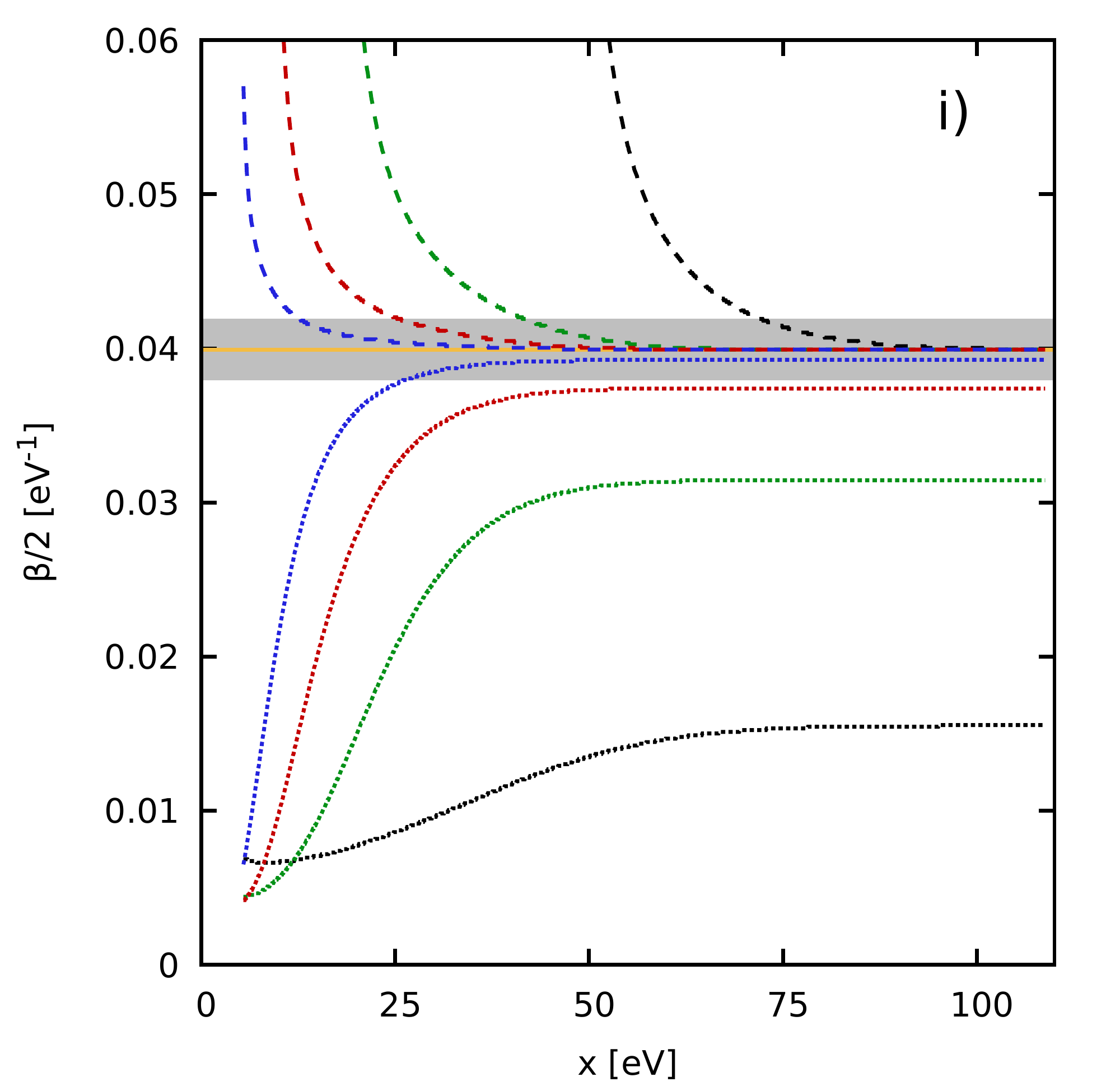}\\\includegraphics[width=0.315\textwidth]{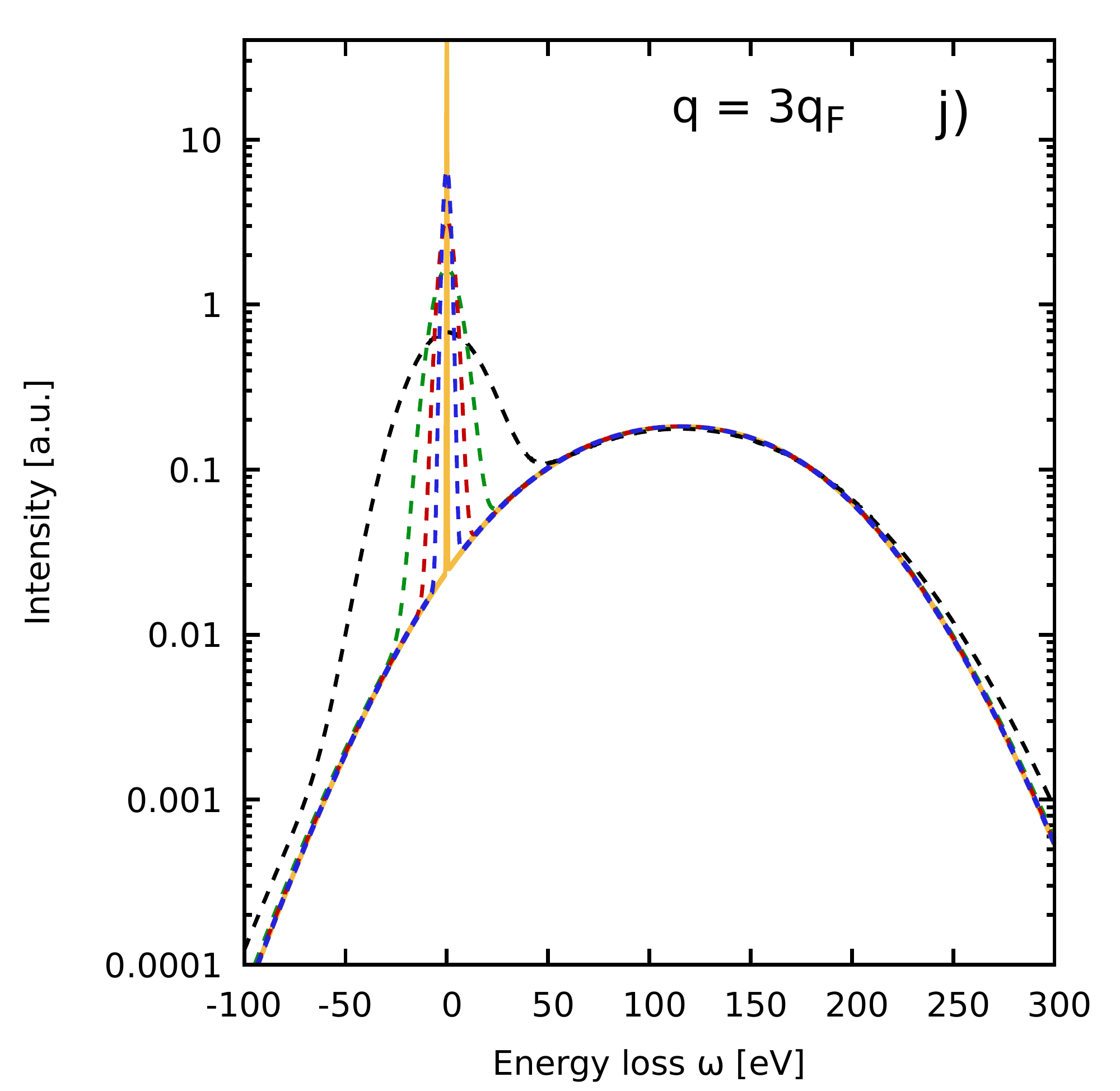}\includegraphics[width=0.315\textwidth]{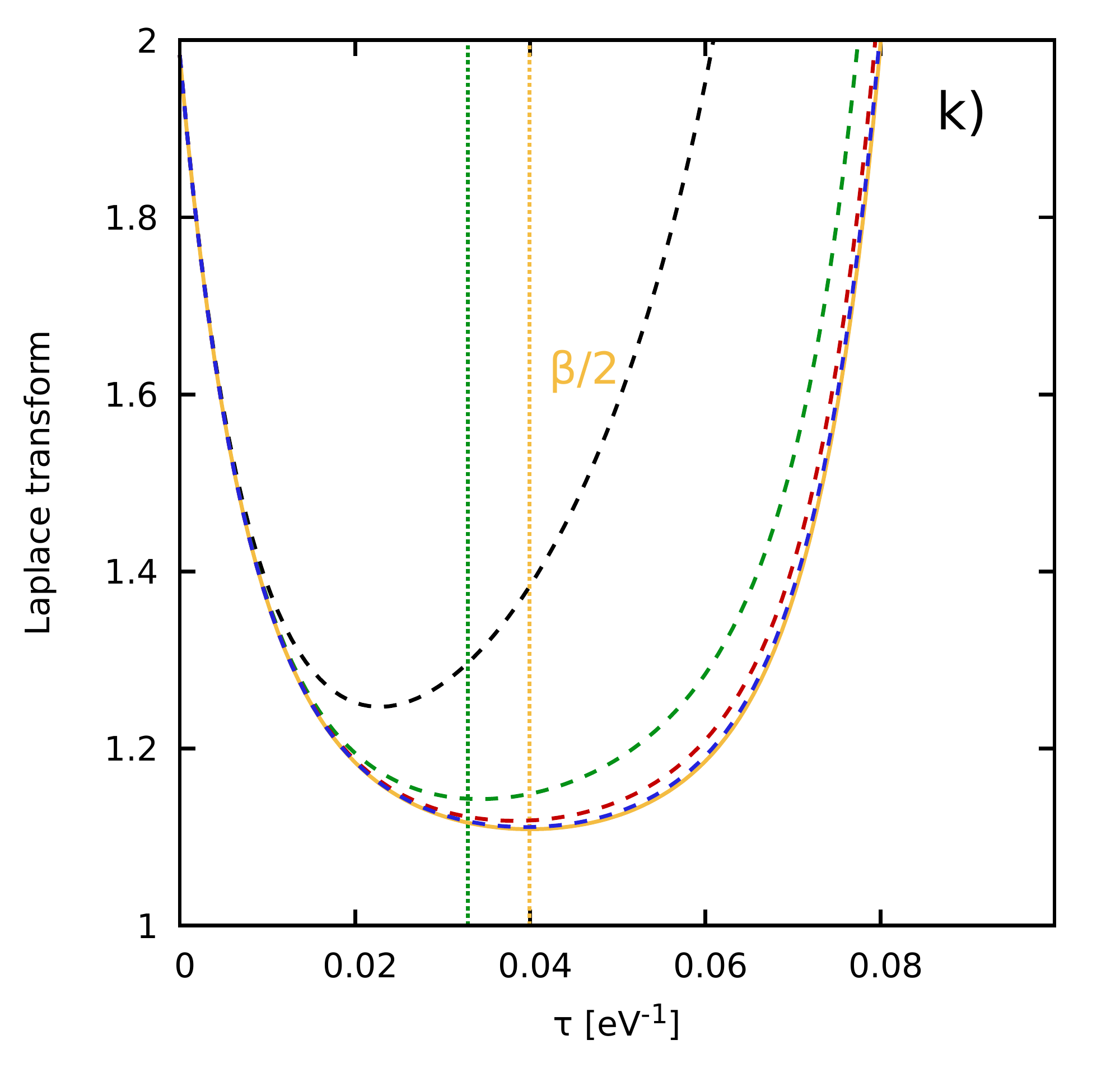}\includegraphics[width=0.315\textwidth]{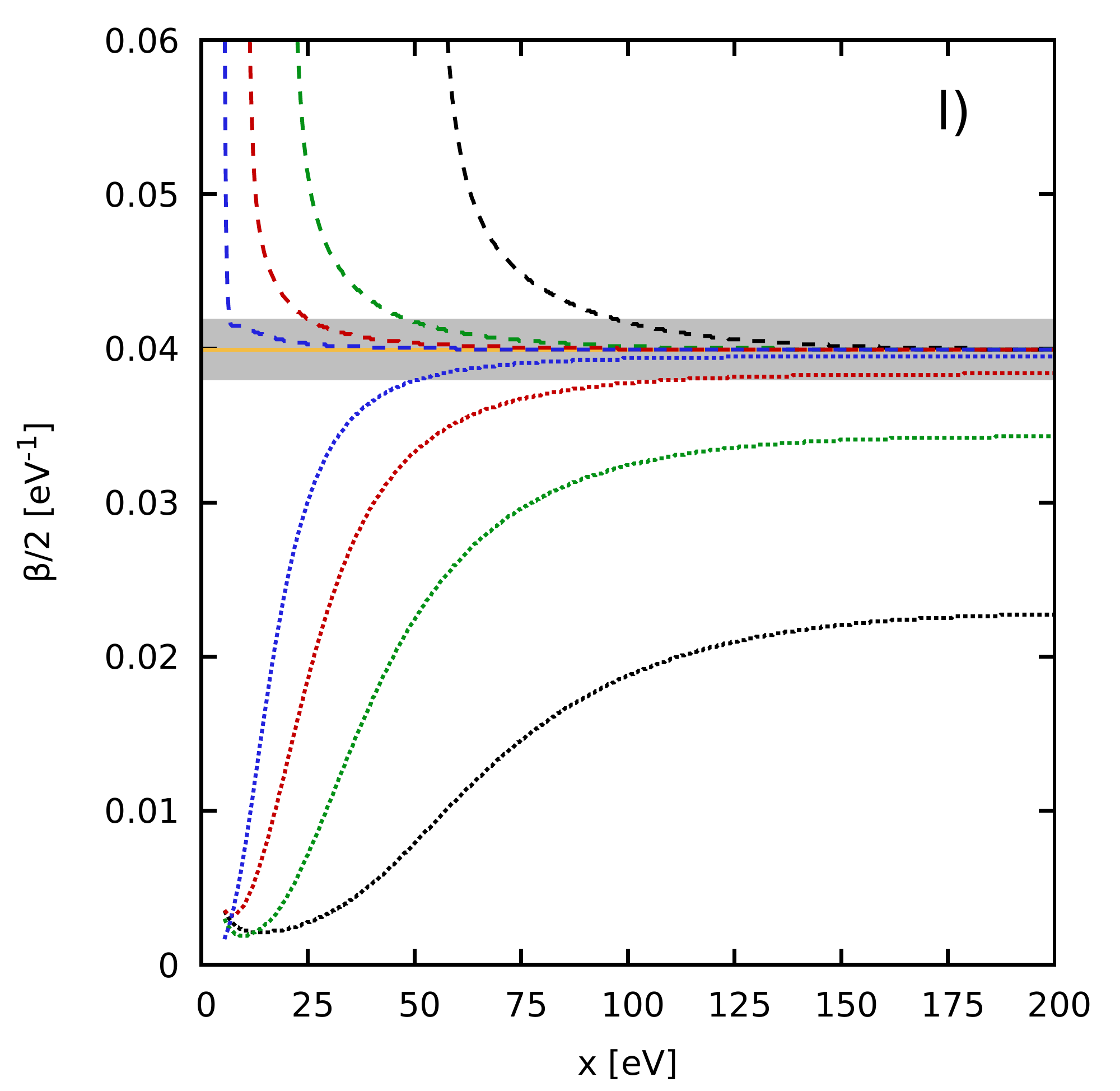}
\caption{\label{fig:Synthetic_convolution} 
Temperature extraction from synthetic UEG data for the convolved intensity $I(\mathbf{q},\omega) = S(\mathbf{q},\omega)\circledast R(\omega)$ [Eq.~(\ref{eq:intensity})] at $r_s=2$ and $\Theta=1$. Left column: original DSF (solid yellow), and intensities obtain from convolutions with Gaussian $R(\omega)$ with different widths $\sigma$ (dashed lines). Middle column: Corresponding two-sided Laplace transforms of original (yellow) DSF and the convolved curves \emph{without} the correction by $\mathcal{L}\left[R(\omega)\right]$. Right column: convergence of the temperature-extraction from the truncated Laplace transform $\mathcal{L}_x\left[I(\mathbf{q},\omega)\right]$, Eq.~(\ref{eq:Laplace_truncated}), with respect to the integration boundary $x$. The dashed (dotted) curves have been obtained with (without) the correction due to $R(\omega)$, and the shaded grey areas indicate a $5\%$ interval around the exact (yellow) temperature.
}
\end{figure*} 

The corresponding extraction of the temperature is illustrated in Fig.~\ref{fig:Snythetic_DSF_theta} c), where we show the respective data for the imaginary time intermediate scattering function $F(\mathbf{q},\tau)=\mathcal{L}\left[S(\mathbf{q},\omega)\right]$. 
All curves are symmetric with respect to the corresponding value of $\tau=\beta/2$. For the lowest temperature, $F(\mathbf{q},\tau)$ is very flat around the minimum, which might make the precise location of the latter more difficult in the case of noisy input data. Yet, this does not pose a fundamental obstacle and can easily be mitigated by considering the relation
\begin{eqnarray}\label{eq:origin}
F(\mathbf{q},0) = F(\mathbf{q},\beta)\ .
\end{eqnarray}
In other words, we can look where the two-sided Laplace transform of $S(\mathbf{q},\omega)$ attains the same value as for $\tau=0$ as an alternative way to determine $\beta$ to circumvent potential problems associated with the occurrence of a shallow minimum in $F(\mathbf{q},\tau)$.

Let us next consider the temperature dependence of the DSF in the non-collective regime, i.e., at $q=3q_\textnormal{F}$ depicted in Fig.~\ref{fig:Snythetic_DSF_theta} b). In this regime, all curves exhibit qualitatively similar broad peaks around $\omega=120\,$eV. The main impact of the temperature is given by the substantially more slowly vanishing tails for large $\omega$ for larger values of $\Theta$ and the less pronounced intensities of the DSF at negative frequencies at low $\Theta$ due to the detailed balance relation. 
In Fig.~\ref{fig:Snythetic_DSF_theta} d), we show the corresponding curves for $F(\mathbf{q},\tau)$, which give the same correct values for the (inverse) temperature as in Fig.~\ref{fig:Snythetic_DSF_theta} c). Notably, the minimum in $F(\mathbf{q},\tau)$ at $\Theta=0.25$ is even more shallow than at $q=0.5q_\textnormal{F}$, which makes the usage of Eq.~(\ref{eq:origin}) even more essential.

\subsection{Convolution with the instrument function\label{sec:convolution}}
In the previous section, we have conclusively demonstrated that knowledge of the dynamic structure factor $S(\mathbf{q},\omega)$ allows a straightforward extraction of the temperature independent of the wave-number regime (collective vs.~single-particle) and without the need for any physical models or simulations. Yet, in a real scattering experiment, we do not have direct access to the DSF, because the measured intensity $I(\mathbf{q},\omega)$ is convolved with the instrument function $R(\omega)$ as stated in Eq.~(\ref{eq:intensity}). We, therefore, analyze in detail the impact of the convolution on extracting the temperature across the relevant range of wave numbers $q$ in Fig.~\ref{fig:Synthetic_convolution}. 

The top row corresponds to the collective regime, where the inelastic part of the deconvolved DSF [solid yellow, Fig.~\ref{fig:Synthetic_convolution} a)] exhibits a sharp plasmon peak around $\pm20\,$eV. 
The dashed lines have been obtained by convolving the yellow curve with Gaussian model instrument functions of different widths $\sigma$. Evidently, the main effect of the convolution is a substantial broadening of the sharp features in the original DSF, which becomes more pronounced with increasing $\sigma$. Indeed, the convolved intensity appears to consist of a single broad elastic peak for $\sigma=\omega_p=16.67\,$eV, and no trace of the plasmon peaks can be recognized with the bare eye. 
In Fig.~\ref{fig:Synthetic_convolution} b), we show the corresponding results for the two-sided Laplace transform of the intensity. 
As usual, the solid yellow line corresponds to the exact $F(\mathbf{q},\tau)=\mathcal{L}\left[S(\mathbf{q},\omega)\right]$, with a minimum around $\tau=\beta/2$ (vertical line). In addition, the dashed curves show results for the Laplace transform of the convolved curves $\mathcal{L}\left[I(\mathbf{q},\omega)\right]$ for different $\sigma$. Evidently, the minimum in the Laplace transforms shifts to smaller $\tau$ with increasing width of the instrument function. In other words, the broadening from the convolution makes the thus extracted temperatures too large~\cite{Dornheim_T_2022}.
Given accurate knowledge of the instrument function $R(\omega)$, it might seem natural to attempt an explicit deconvolution of Eq.~(\ref{eq:intensity}) to reconstruct the original DSF $S(\mathbf{q},\omega)$. This, in turn, would allow one to subsequently obtain $F(\mathbf{q},\tau)=\mathcal{L}\left[S(\mathbf{q},\omega)\right]$ and thus to extract the unbiased minimum of $\tau=\beta/2$.
In practice, such a deconvolution is notoriously unstable with respect to the noise in the input data, which usually prevents the explicit extraction of $S(\mathbf{q},\omega)$. Yet, this obstacle is completely circumvented within our methodology due to the convolution theorem in Eq.~(\ref{eq:convolution_theorem}). Particularly, the instrument function and the DSF can be separated in a straightforward way in the Laplace domain. Consequently, we can completely remove the impact of the artificial broadening by dividing the dashed curves by the Laplace transform of the instrument function $\mathcal{L}\left[R(\omega)\right]$, which gives the original solid yellow curve in all cases.

\begin{figure}
\centering
\includegraphics[width=0.45\textwidth]{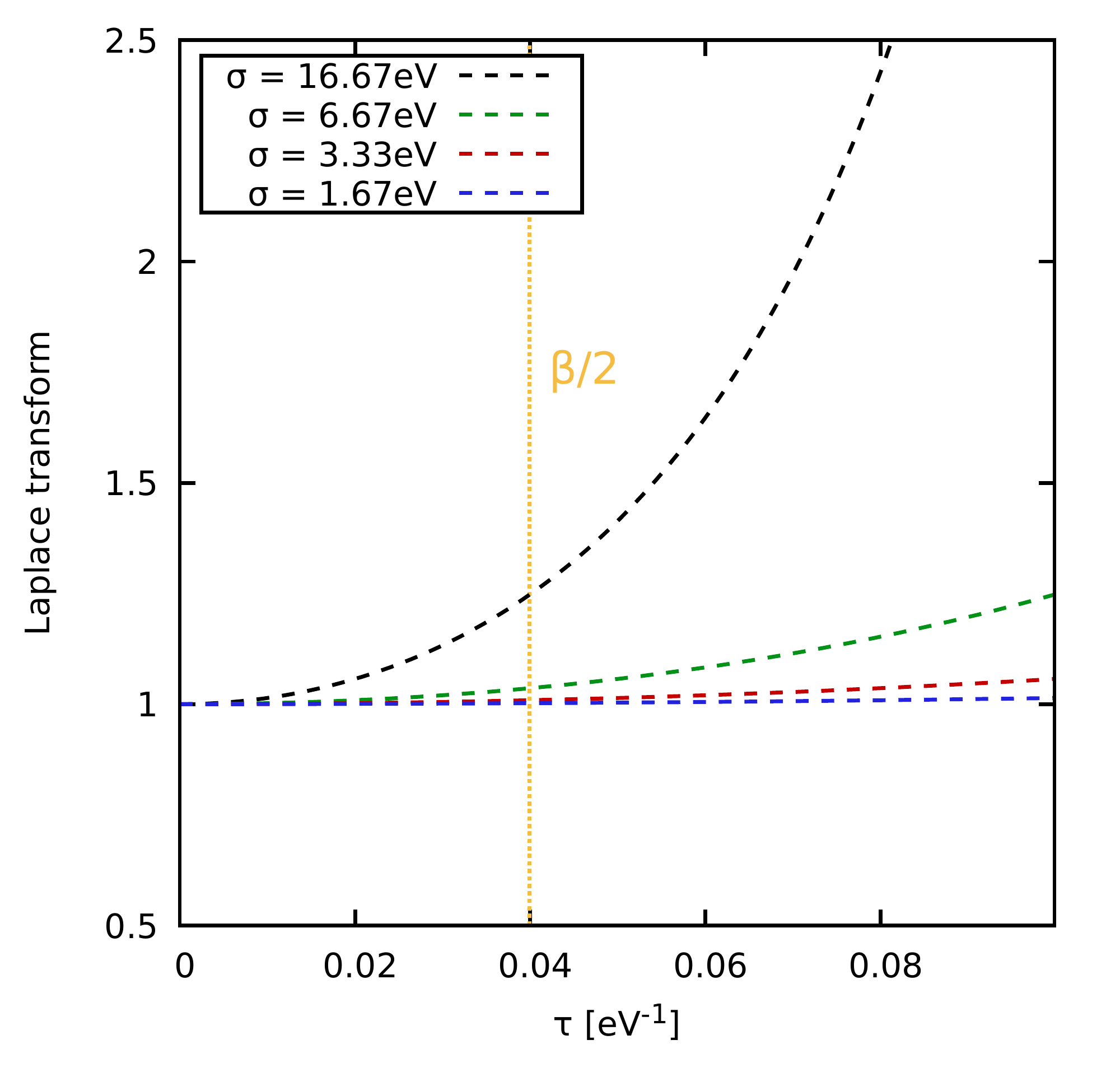}
\caption{\label{fig:Laplace_probe} Two-sided Laplace transform of the Gaussian instrument function $R_\sigma(\omega)$, see Eq.~(\ref{eq:Laplace_probe}), for different relevant values of the width $\sigma$. The dotted yellow vertical line indicates $\tau=\beta/2$ for $T=12.53\,$eV, cf.~Fig.~\ref{fig:Synthetic_convolution}, and has been included as a reference.
}
\end{figure} 

For a Gaussian probe function of width $\sigma$ and centered around $\omega=0$, $R_\sigma(\omega)$, the two-sided Laplace transform can be carried out analytically, 
\begin{eqnarray}\label{eq:Laplace_probe}
\mathcal{L}\left[R_\sigma(\omega)\right] = e^{\sigma^2\tau^2/2}\ .
\end{eqnarray}
The results are shown as the dashed lines in Fig.~\ref{fig:Laplace_probe} for the same values of $\sigma$ as in Fig.~\ref{fig:Synthetic_convolution}. The vertical dotted yellow line indicates $\tau=\beta/2$ for $T=12.53\,$eV, i.e., $\Theta=1$ at $r_s=2$, and has been included as a reference. For the most narrow instrument function with $\sigma=1.67\,$eV, Eq.~(\ref{eq:Laplace_probe}) attains a nearly constant value of one over the entire relevant $\tau$-range. Consequently, the impact of the instrument function on $\mathcal{L}\left[I(\mathbf{q},\omega)\right]$ is small, and the dashed blue line in Fig.~\ref{fig:Synthetic_convolution} b) is very close to the exact result for $F(\mathbf{q},\tau)=\mathcal{L}\left[S(\mathbf{q},\omega)\right]$.
With increasing $\sigma$, $\mathcal{L}\left[R(\omega)\right]$ starts to increasingly deviate from unity, which manifests as a shift of the minimum in $\mathcal{L}\left[I(\mathbf{q},\omega)\right]$ towards smaller values of $\tau$.

In addition, we find that the particular value of $x$ for which convergence is reached strongly increases with the width of the instrument function $R_\sigma(\omega)$. In other words, the integral boundaries for which the exact convolution theorem Eq.~(\ref{eq:convolution_theorem}) is recovered scale with $\sigma$. 

\begin{figure}\centering
\includegraphics[width=0.45\textwidth]{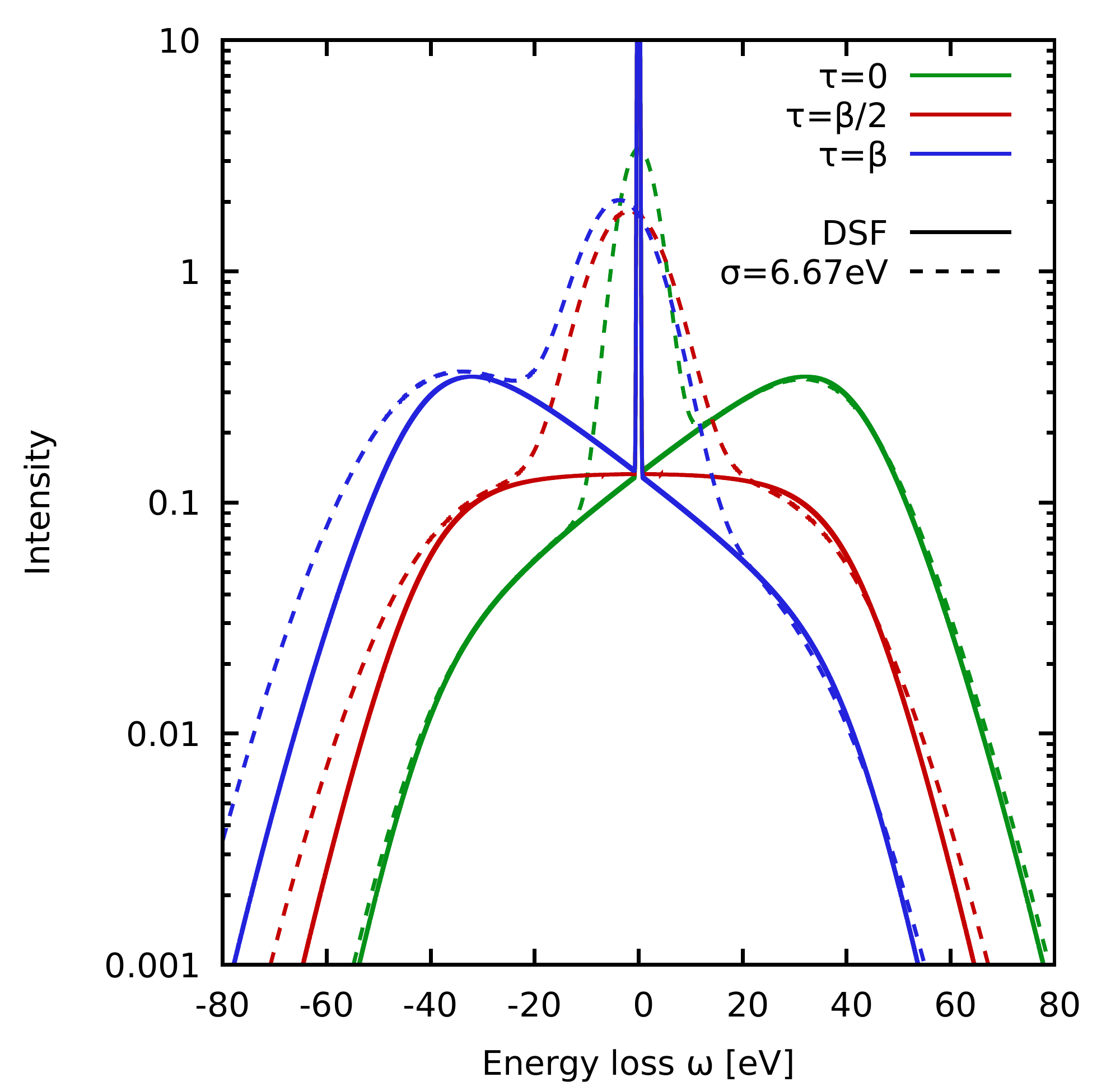}
\caption{\label{fig:Synthetic_contribution} Contribution to the two-sided Laplace transform $\mathcal{L}\left[S(\mathbf{q},\omega)\right]$ (solid) and $\mathcal{L}\left[I(\mathbf{q},\omega)\right]$ (dashed) as a function of the frequency $\omega$ at $r_s=2$, $\Theta=1$, and $q=q_\textnormal{F}$ for selected values of the imaginary time $\tau$.%
}
\end{figure} 

\begin{figure}\centering
\includegraphics[width=0.45\textwidth]{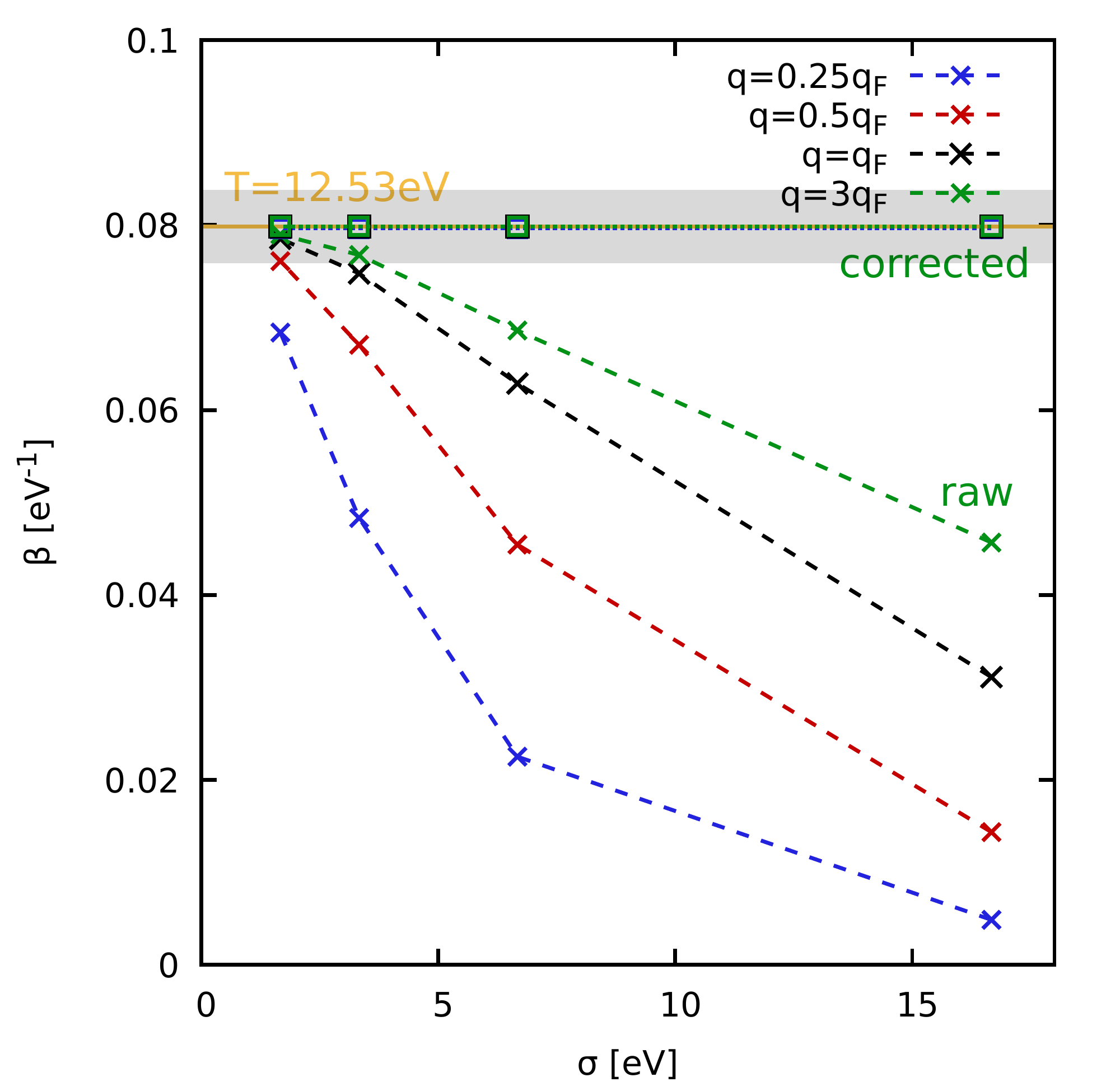}
\caption{\label{fig:Synthetic_meta} 
Temperature extraction at $r_s=2$ and $\Theta=1$ as a function of the width $\sigma$ of the Gaussian instrument function for different wave numbers $q$. Squares (crosses): corrected (uncorrected) for influence of $\mathcal{L}\left[R(\omega)\right]$. Horizontal yellow: exact inverse temperature. Shaded grey area: interval of $\pm5\%$, included as a reference. }
\end{figure} 

A further interesting point of this analysis is the required accuracy of the intensity needed to extract the exact value of $\tau=\beta/2$. For example, at $\sigma=3.33\,$eV convergence is reached around $x=25\,$eV. In this case, the intensity [see Fig.~\ref{fig:Synthetic_convolution} a)] at $\omega=-25\,$eV is reduced by a single order of magnitude compared to the size of the plasmon peak at $\omega=20\,$eV. Resolving the inelastic intensity over such a range in a scattering experiment is feasible in modern laser facilities~\cite{SACLA_2011,LCLS_2016,Tschentscher_2017}. For the broadest instrument function with $\sigma=16.67\,$eV, the extracted temperature converges around $x=75\,$eV. Yet, here the convolved intensity has already decayed by more than three orders of magnitude and therefore will be difficult to resolve in an actual experiment. This clearly illustrates the importance of a narrow probe function for the accurate and practical analysis of experimental scattering data.

Let us conclude this discussion of Fig.~\ref{fig:Synthetic_convolution} c) by considering the dotted lines, which have been obtained by determining the minimum in $\mathcal{L}\left[I(\mathbf{q},\omega)\right]$ without the correction by $\mathcal{L}\left[R(\omega)\right]$. We find that the finite width of the instrument function then substantially influences (in fact, decreases) the extracted values of $\tau=\beta/2$ even in the case of the relatively narrow Gaussian with $\sigma=1.67\,$eV.

\begin{figure*}\centering
\includegraphics[width=0.45\textwidth]{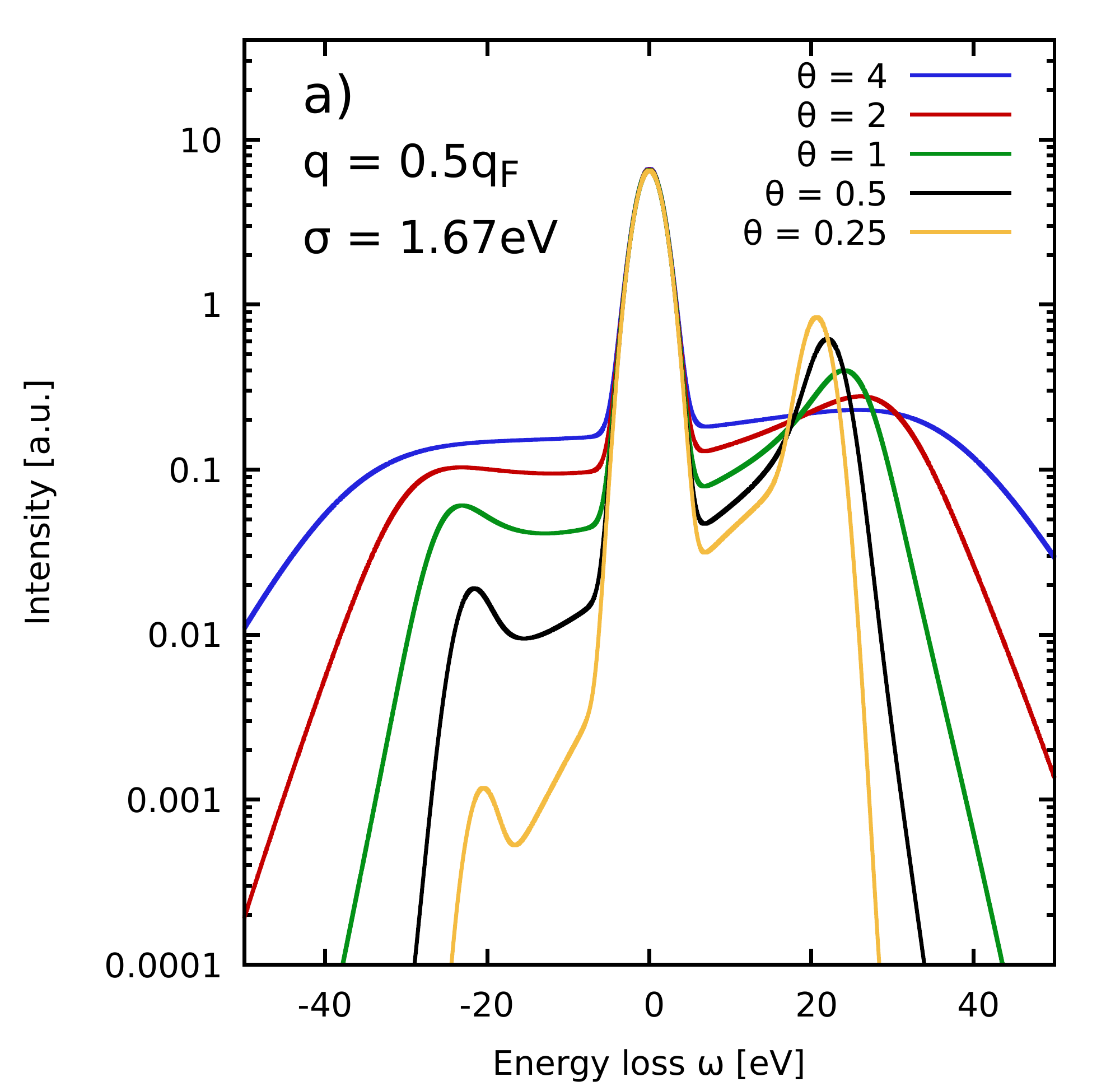}\includegraphics[width=0.45\textwidth]{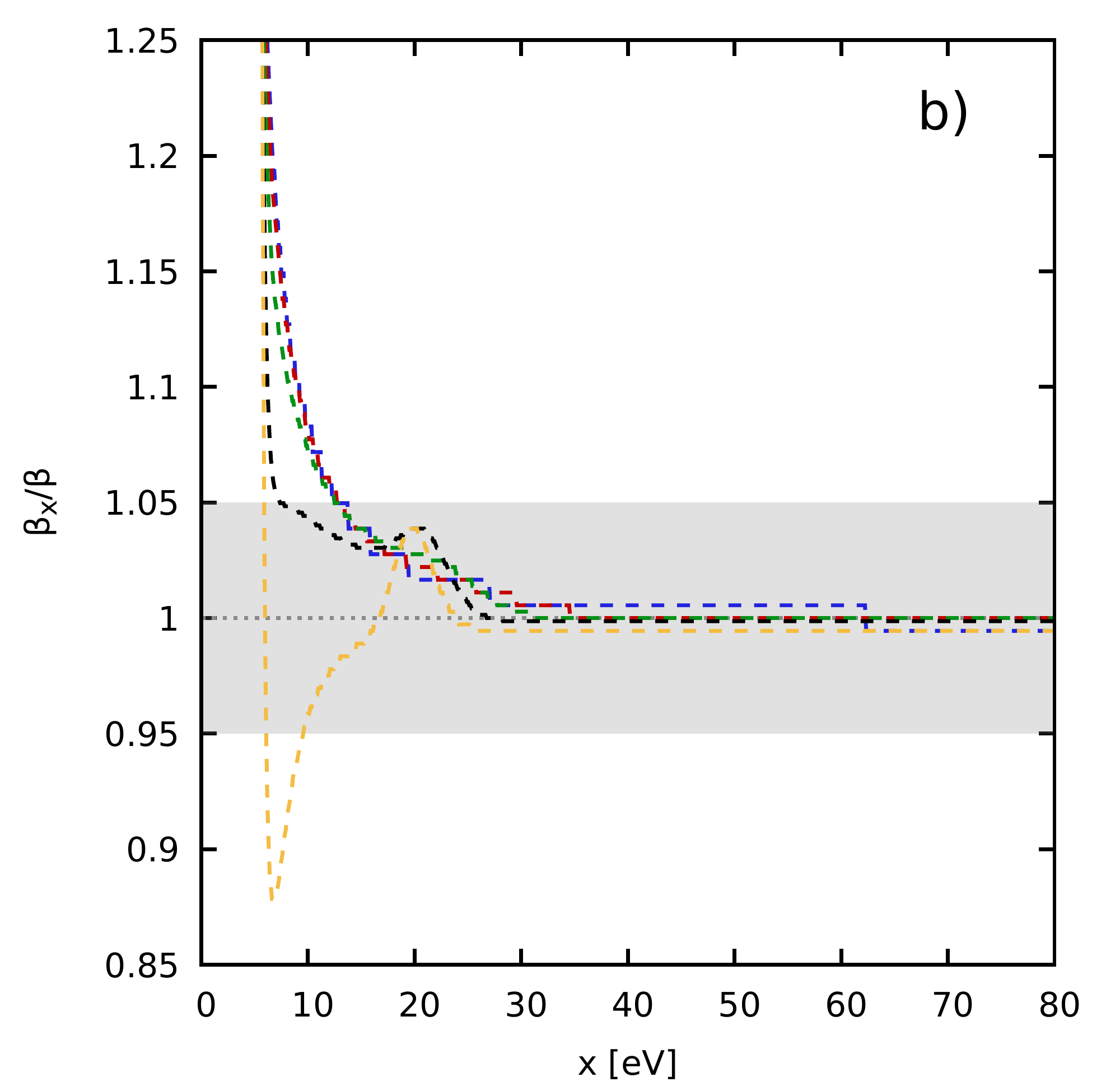}\\\includegraphics[width=0.45\textwidth]{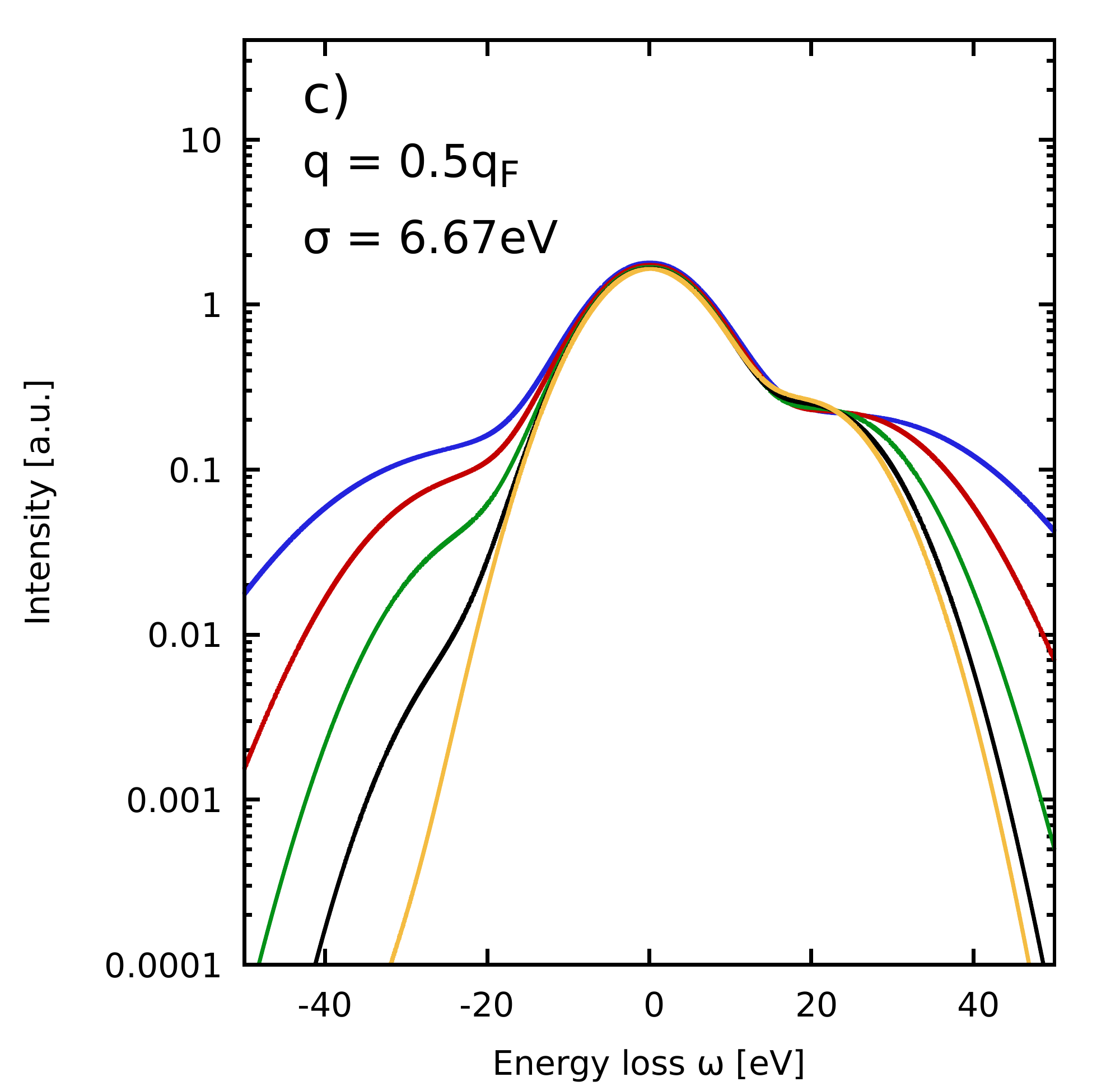}\includegraphics[width=0.45\textwidth]{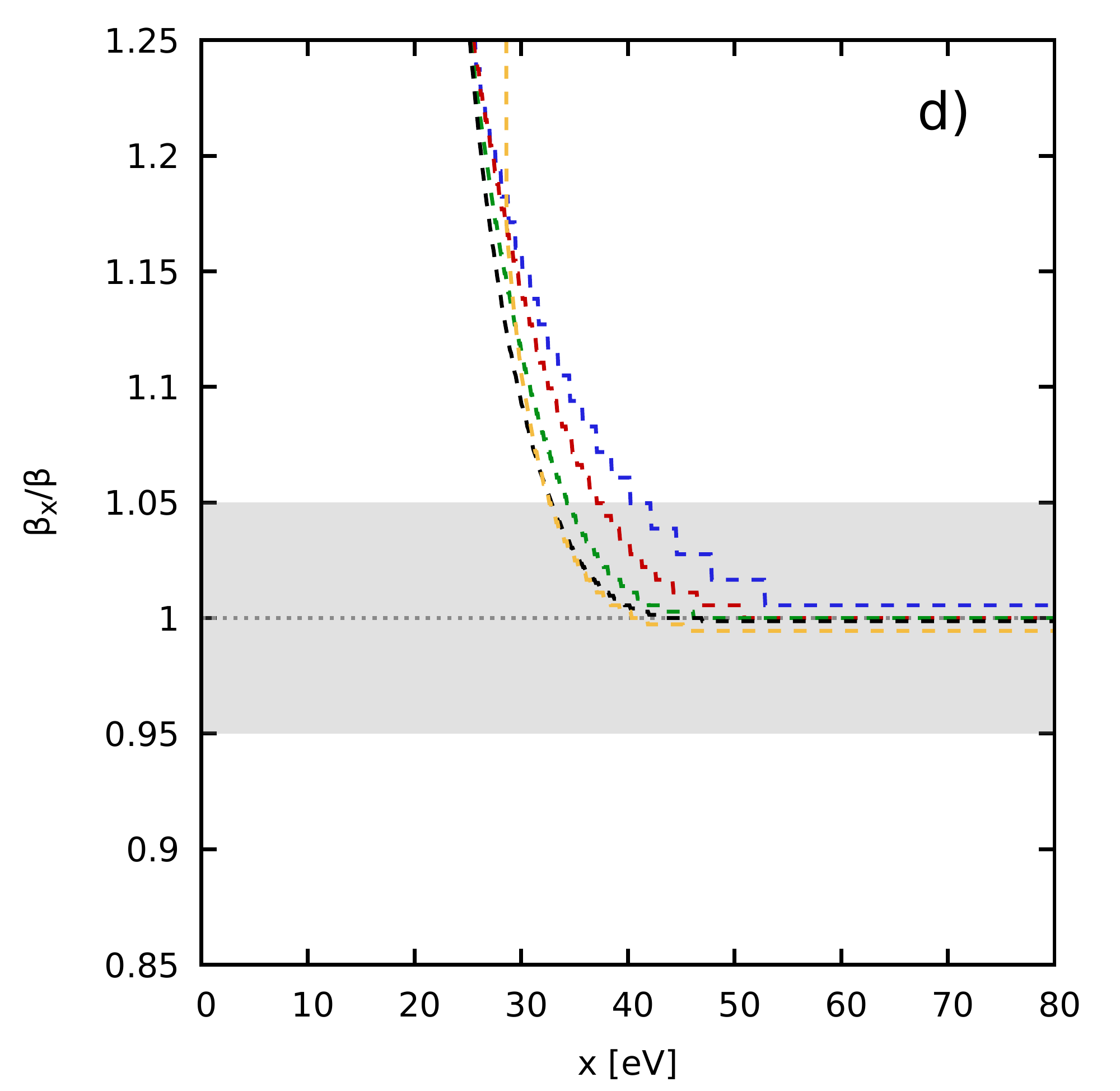}
\caption{\label{fig:STRIPLE_lim} Influence of the temperature parameter $\Theta$ on the extraction of the temperature at $r_s=2$ and $q=0.5q_\textnormal{F}$. Top (bottom) row: width of the instrument function $\sigma=1.67\,$eV ($\sigma=6.67\,$eV). Left: convolved scattering intensity. Right: convergence of the extracted inverse temperature with the integration boundary $x$, rescaled by the corresponding exact value of $\beta$. The shaded grey area indicates an interval of $\pm5\%$ and has been included as a reference. 
}
\end{figure*} 

The bottom three rows of Fig.~\ref{fig:Synthetic_convolution} contain the same analysis, but for increasing values of the wave number $q$. We, therefore, restrict ourselves here to a concise discussion of the main differences between the different regimes. First, we reiterate our earlier point about the increasing width of the unconvolved DSF with increasing $q$. This, in turn, means that the impact of the Gaussian instrument function becomes less pronounced for large $q$. Indeed, the uncorrected curves for both $\sigma=1.67\,$eV and $\sigma=3.33\,$eV are within $5\%$ of the correct temperature in the single-particle regime (see Fig.~\ref{fig:Synthetic_convolution} l)). For the narrowest instrument function this even holds at the Fermi wave number $q=q_\textnormal{F}$ (see Fig.~\ref{fig:Synthetic_convolution} i)).
As a second observation, we find that the convergence of the extracted temperature with the integration boundary $x$ is shifted to somewhat larger frequencies. This is completely unproblematic for $\sigma\in\left\{1.67,3.33,6.67\right\}\,$eV, as the width of the actual intensity increases similarly. Therefore, the intensity does not have to be resolved over substantially more than one order of magnitude. For $\sigma=16.67\,$eV, on the other hand, reaching convergence in practice will be difficult.

We further illustrate the impact of the instrument function on the two-sided Laplace transform of the intensity by showing both $I(\mathbf{q},\omega)e^{-\tau\omega}$ (dashed) and $S(\mathbf{q},\omega)e^{-\tau\omega}$ (solid) for $q=q_\textnormal{F}$ in Fig.~\ref{fig:Synthetic_contribution} for three relevant values of the imaginary time $\tau$. The green curves have been obtained for $\tau=0$ and thus show the original intensity and DSF, respectively. The red curves correspond to $\tau=\beta/2$, where $F(\mathbf{q},\tau)$ attains its minimum. In this case, the contribution to $\mathcal{L}\left[S(\mathbf{q},\omega)\right]$ is symmetric around $\omega=0$, whereas the convolution with $R(\omega)$ noticeably skews the corresponding curve to lower frequencies. This trend is even more pronounced for $\tau=\beta$ (blue curves), where $S(\mathbf{q},\omega)e^{-\tau\omega}$ is equal to the solid green curve mirrored around $\omega=0$, whereas this clearly does not hold for the corresponding dashed curve.

Let us conclude with a more systematic analysis of the impact of the width of the instrument function $\sigma$ on the extraction of the temperature, which is shown in Fig.~\ref{fig:Synthetic_meta}. We plot the obtained inverse temperature $\beta$ as a function of $\sigma$ for the four wave numbers from Fig.~\ref{fig:Synthetic_convolution}. The squares show the values where we have corrected for the impact of $\mathcal{L}\left[R(\omega)\right]$, and we find a perfect agreement with the exact temperature for all combinations of $\sigma$ and $q$. The crosses show the extracted \emph{raw} temperatures without this correction. Overall, all four curves exhibit the same qualitative trend: the error in the uncorrected temperature monotonically decreases with decreasing $\sigma$, as is expected. Moreover, the curves are strictly ordered with $q$, as large wave numbers correspond to broader DSFs, where the impact of the convolution is less pronounced. The shaded grey area shows an interval of $\pm5\%$ around the exact inverse temperature, which can be reached without the correction either for a very narrow instrument function or in the single particle regime ($q\gg q_\textnormal{F}$).
This directly implies that large scattering angles as they can be realized in backscattering experiments make the method more robust against possible uncertainties in the characterization of the instrument function $R(\omega)$.

\begin{figure}\centering
\includegraphics[width=0.45\textwidth]{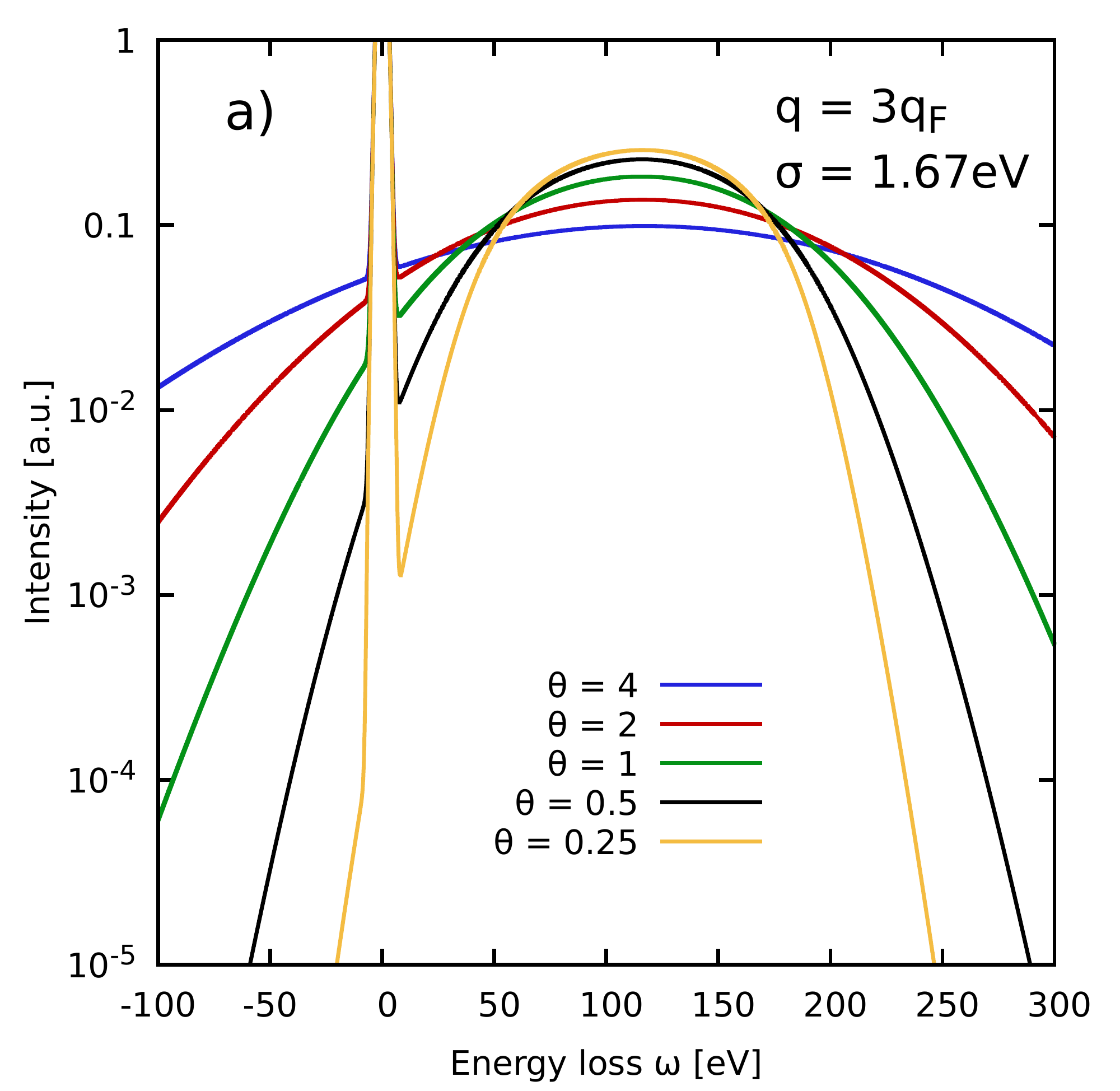}\\\includegraphics[width=0.45\textwidth]{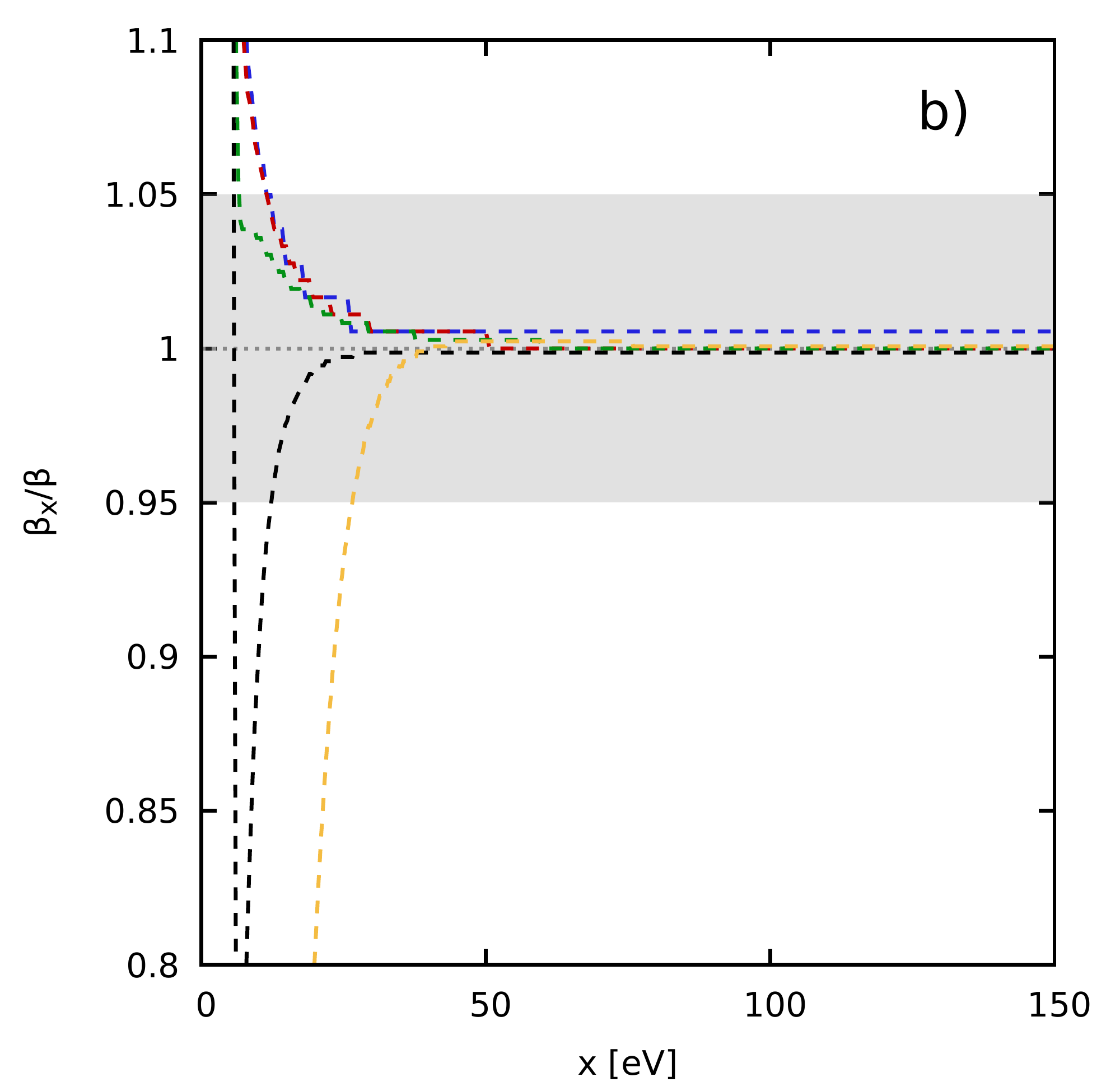}
\caption{\label{fig:STRIPLE_lim_x3} Influence of the temperature parameter $\Theta$ on the extraction of the temperature at $r_s=2$ and $q=3q_\textnormal{F}$ with the width of the instrument function $\sigma=1.67\,$eV. Top: convolved scattering intensity. Bottom: convergence of the extracted inverse temperature with the integration boundary $x$, rescaled by the corresponding exact value of $\beta$. The shaded grey area indicates an interval of $\pm5\%$ and has been included as a reference.
}
\end{figure} 

\subsection{Different temperatures}
In the previous section, we analyzed in detail the impact of the wave number and the width of the instrument function on the extracted temperature from a convolved scattering intensity signal. 
In Fig.~\ref{fig:STRIPLE_lim}, we extend these considerations by analyzing different values of the temperature $\Theta$. 
Fig.~\ref{fig:STRIPLE_lim} a) shows results for $I(\mathbf{q},\omega)$ at $r_s=2$ and $q=0.5q_\textnormal{F}$ for a narrow instrument function with $\sigma=1.67\,$eV. A comparison with the corresponding deconvolved results for the DSF (see Fig.~\ref{fig:Snythetic_DSF_theta} above) reveals the substantial broadening of the plasmon peak, in particular at low temperatures. 
The convergence of the extracted temperature with the integration boundary $x$ is shown in Fig.~\ref{fig:STRIPLE_lim} b). The curves have been rescaled by the respective value of $\beta$ to allow for a more straightforward comparison. As usual, the shaded grey area indicates an interval of $\pm5\%$ and has been included as a reference.

First, we find that the extracted temperature converges towards the exact value for all values of $\Theta$, as is expected; the small deviations from one at large $x$ are a direct consequence of the finite $\tau$-resolution in our numerical implementation, which could be increased if necessary. 
In addition, the values of $x$ for which convergence is reached appear to be nearly independent of $\Theta$. The accurate extraction of the temperature is thus substantially more challenging at low temperatures, where the scattering intensity at negative frequencies can be orders of magnitude smaller than in the positive $\omega$ range. For example, the negative plasmon is reduced by three orders of magnitude at $\Theta=0.25$, whereas it is not even reduced by a full order of magnitude for $\Theta=1$. From a practical perspective, this means that the accurate measurement of the intensity at $\omega<0$ is of prime importance and decisively determines the quality of the extracted temperature for $\Theta \ll 1$, as $\omega>0$ and $\omega<0$ equally contribute to $F(\mathbf{q},\beta/2)$, cf.~Fig.~\ref{fig:Synthetic_contribution}.

The bottom row of Fig.~\ref{fig:STRIPLE_lim} shows the same analysis for a broader instrument function with $\sigma=6.67\,$eV. Overall, the conclusions are similar to the previous case, although we do find a more pronounced dependence of the convergence with $x$ on $\Theta$. Still, the importance of the negative frequency range remains the same.

Let us conclude this analysis of synthetic data by analyzing the effect of the temperature parameter on the convolved intensity in the single-particle regime. The corresponding results are shown in Fig.~\ref{fig:STRIPLE_lim_x3} for a narrow probe function with $\sigma=1.67\,$eV. 
The main difference regarding the extraction of the temperature compared to the smaller wave number shown in Fig.~\ref{fig:STRIPLE_lim}
is that we had to use the relation Eq.~(\ref{eq:origin}) at $\Theta=0.25$, as the minimum in $F(\mathbf{q},\tau)$ is extremely shallow. Still, we resolve the correct temperature for all temperatures, and the effect of $\Theta$ on the value of $x$ for which convergence is reached is small.

\section{Error analysis and the role of noise\label{sec:noise}}
In a real scattering experiment, the measured intensity $I_\textnormal{exp}(\mathbf{q},\omega)$ is always afflicted with some form of random noise $\Delta I(\mathbf{q},\omega)$, such that
\begin{eqnarray}
I_\textnormal{exp}(\mathbf{q},\omega) = I(\mathbf{q},\omega) + \Delta I(\mathbf{q},\omega)\ ,
\end{eqnarray}
with $I(\mathbf{q},\omega)$ being the true convolution of $S(\mathbf{q},\omega)$ and the instrument function $R(\omega)$. Throughout this analysis, we will assume perfect knowledge of the latter; the role of uncertainty in the instrument function will be investigated in more detail in a separate publication.
In the following, we will systematically investigate the impact of the noise on the symmetrically truncated Laplace transform of the experimental scattering signal, which is given by
\begin{eqnarray}\label{eq:Laplace_noise}
\mathcal{L}_x\left[I_\textnormal{exp}(\mathbf{q},\omega)\right] &=& \mathcal{L}_x\left[I(\mathbf{q},\omega) + \Delta I(\mathbf{q},\omega)\right] \\\nonumber &=& \mathcal{L}_x\left[I(\mathbf{q},\omega)\right] + \mathcal{L}_x\left[\Delta I(\mathbf{q},\omega)\right]\ .
\end{eqnarray}
In a counting-based scattering experiment, the error distribution is given by~\cite{sheffield2010plasma}
\begin{eqnarray}\label{eq:scattering_error}
\Delta I(\mathbf{q},\omega) = \xi_{\sigma_\Delta}(\omega)\sqrt{I(\mathbf{q},\omega)}\ ,
\end{eqnarray}
where $\xi_{\sigma_\Delta}(\omega)$ is a Gaussian random variable (centered around zero) with a variance $\sigma_\Delta$.

Eqs.~(\ref{eq:Laplace_noise}) and (\ref{eq:scattering_error}) imply that it is sufficient to analyze the symmetrically truncated Laplace transform of the product of Gaussian random noise of unit variance with the square root of the actual intensity, $\mathcal{L}_x\left[\xi_1(\omega)\sqrt{I(\mathbf{q},\omega)}\right]$, as any particular noise level $\sigma_\Delta$ can simply be included as a pre-factor,
\begin{eqnarray}\nonumber
\mathcal{L}_x\left[I_\textnormal{exp}(\mathbf{q},\omega)\right] &=& \mathcal{L}_x\left[I(\mathbf{q},\omega)\right] + \sigma_\Delta \mathcal{L}_x\left[\xi_1(\omega)\sqrt{I(\mathbf{q},\omega)}\right]\ .\\  \label{eq:unit_noise}
\end{eqnarray}

\begin{figure}\centering
\includegraphics[width=0.45\textwidth]{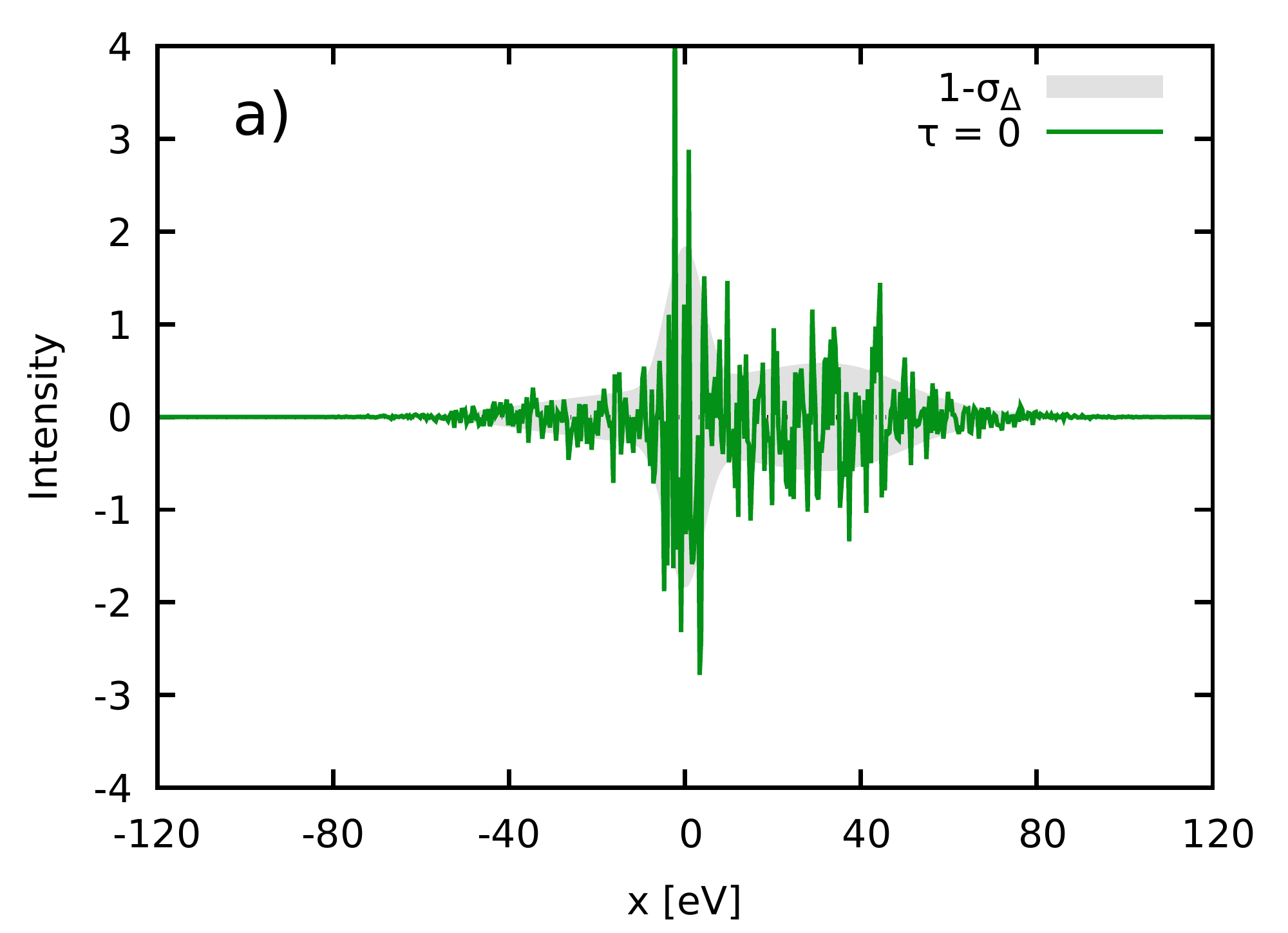}\\\vspace{-0.8cm}\includegraphics[width=0.45\textwidth]{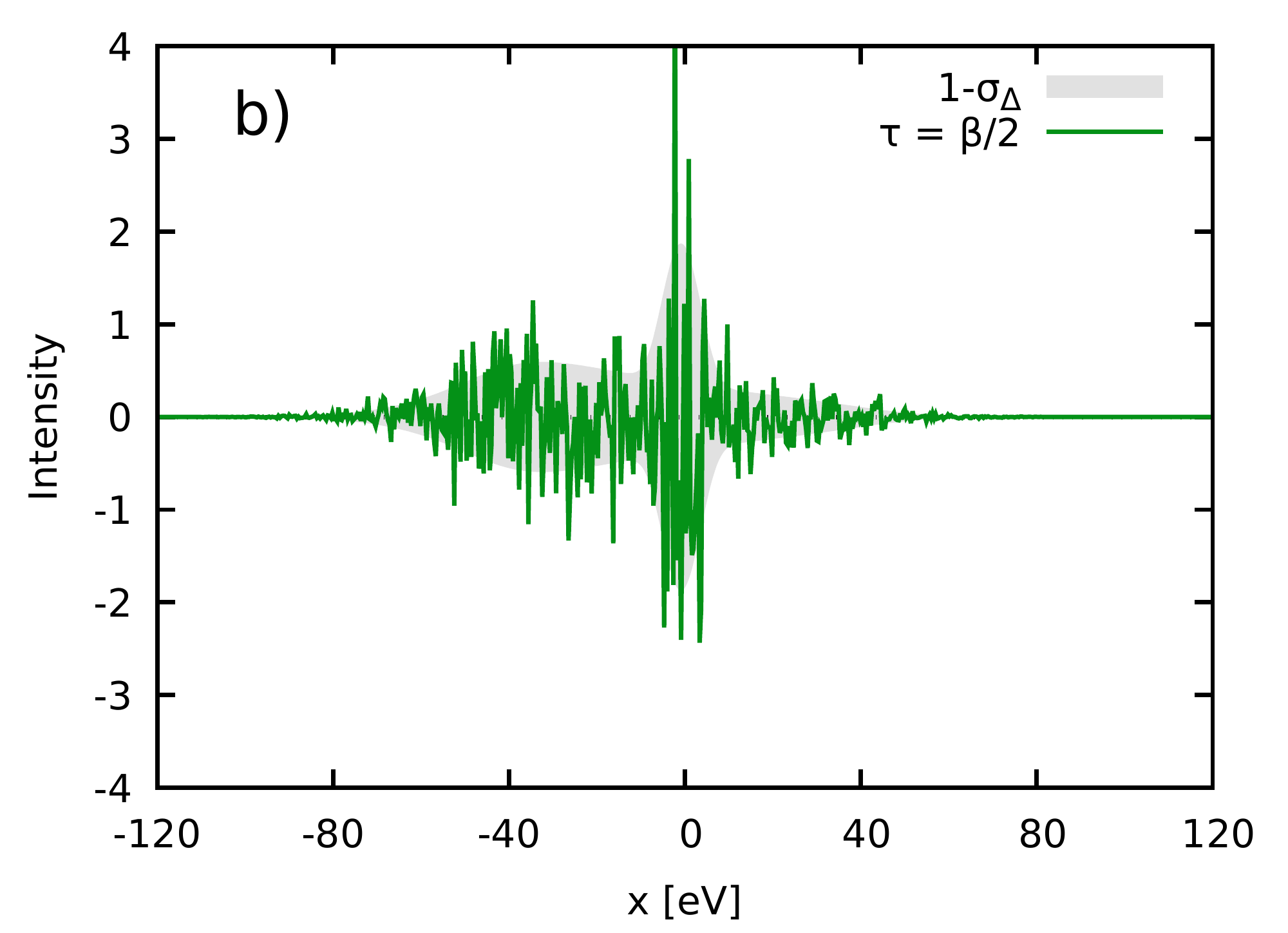}\\\vspace{-0.8cm}\includegraphics[width=0.45\textwidth]{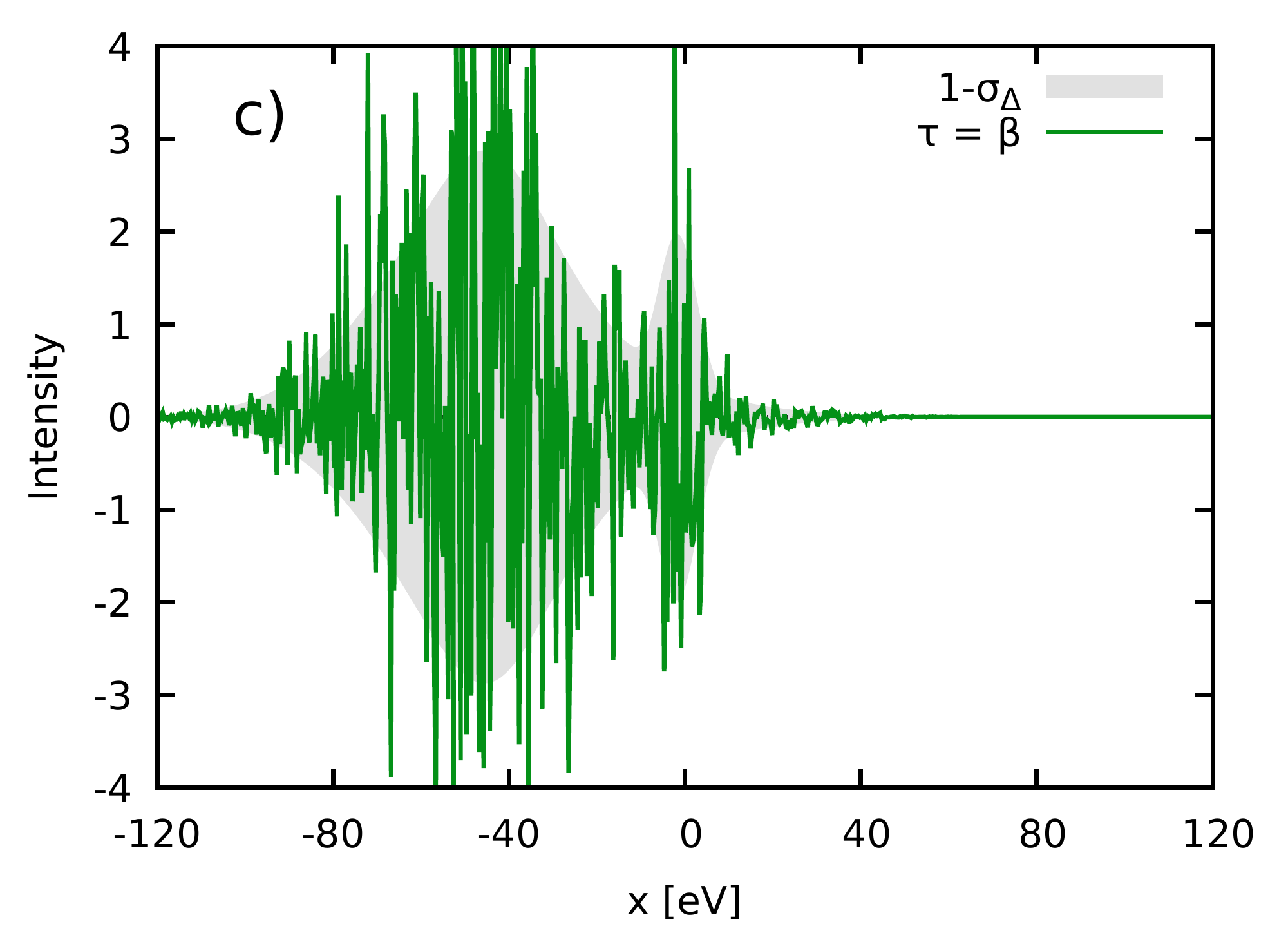}
\caption{\label{fig:sqrt_noise_contribution} Frequency-resolved contribution to the Laplace transform of Gaussian random noise of unit variance $\mathcal{L}_x\left[\xi_1(\omega)\sqrt{I(\mathbf{q},\omega)}\right]$, see Eq.~(\ref{eq:unit_noise}), for a synthetic intensity from the UEG with $r_s=2$, $\Theta=1$, and $q=q_\textnormal{F}$ convolved with a Gaussian instrument function of width $\sigma=3.33\,$eV. The shaded grey area depicts the corresponding $1\sigma_\Delta$ interval that is defined by $\sqrt{I(\mathbf{q},\omega)}$.
}
\end{figure}

\begin{figure*}\centering\includegraphics[width=0.33\textwidth]{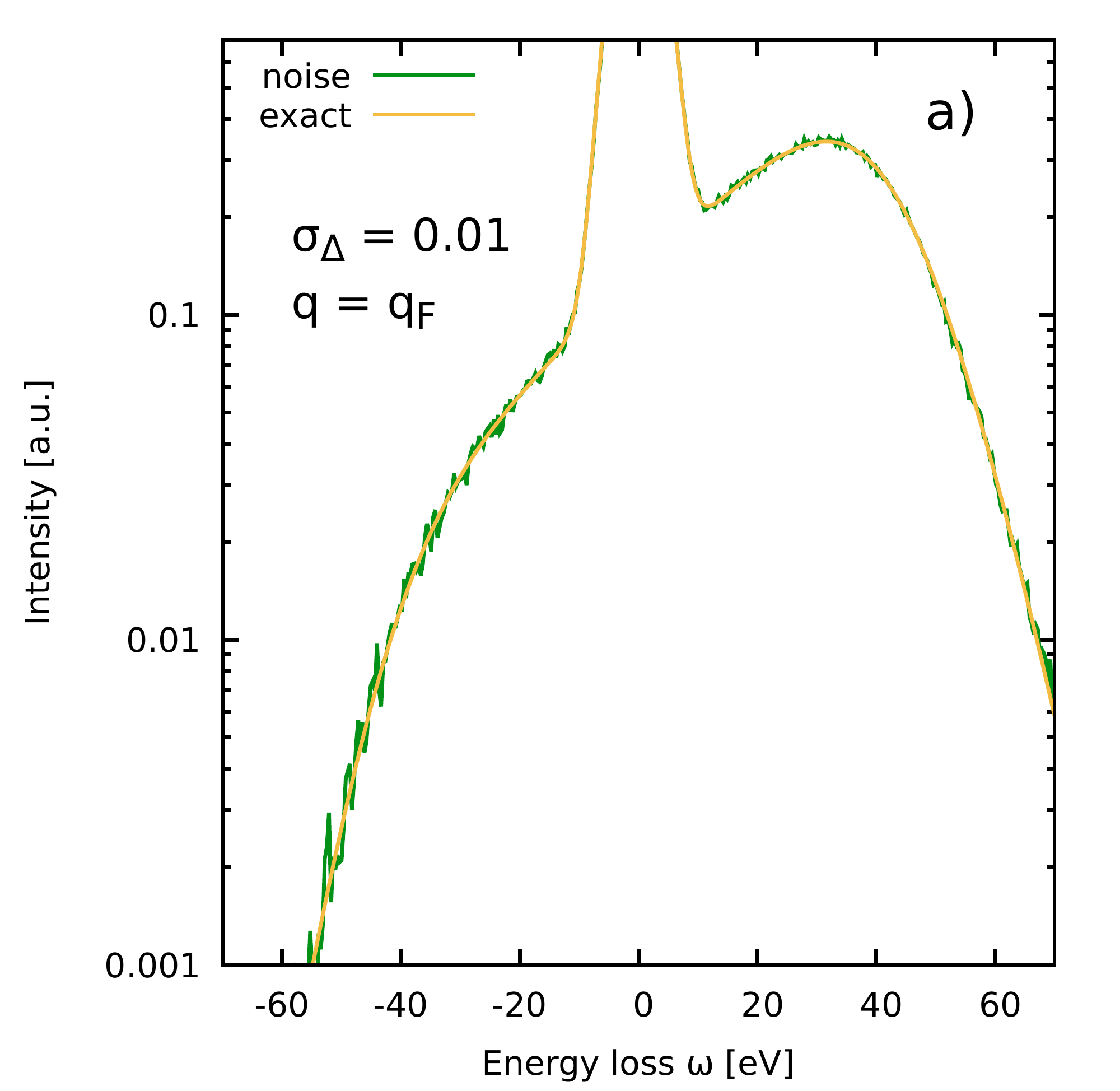}\includegraphics[width=0.33\textwidth]{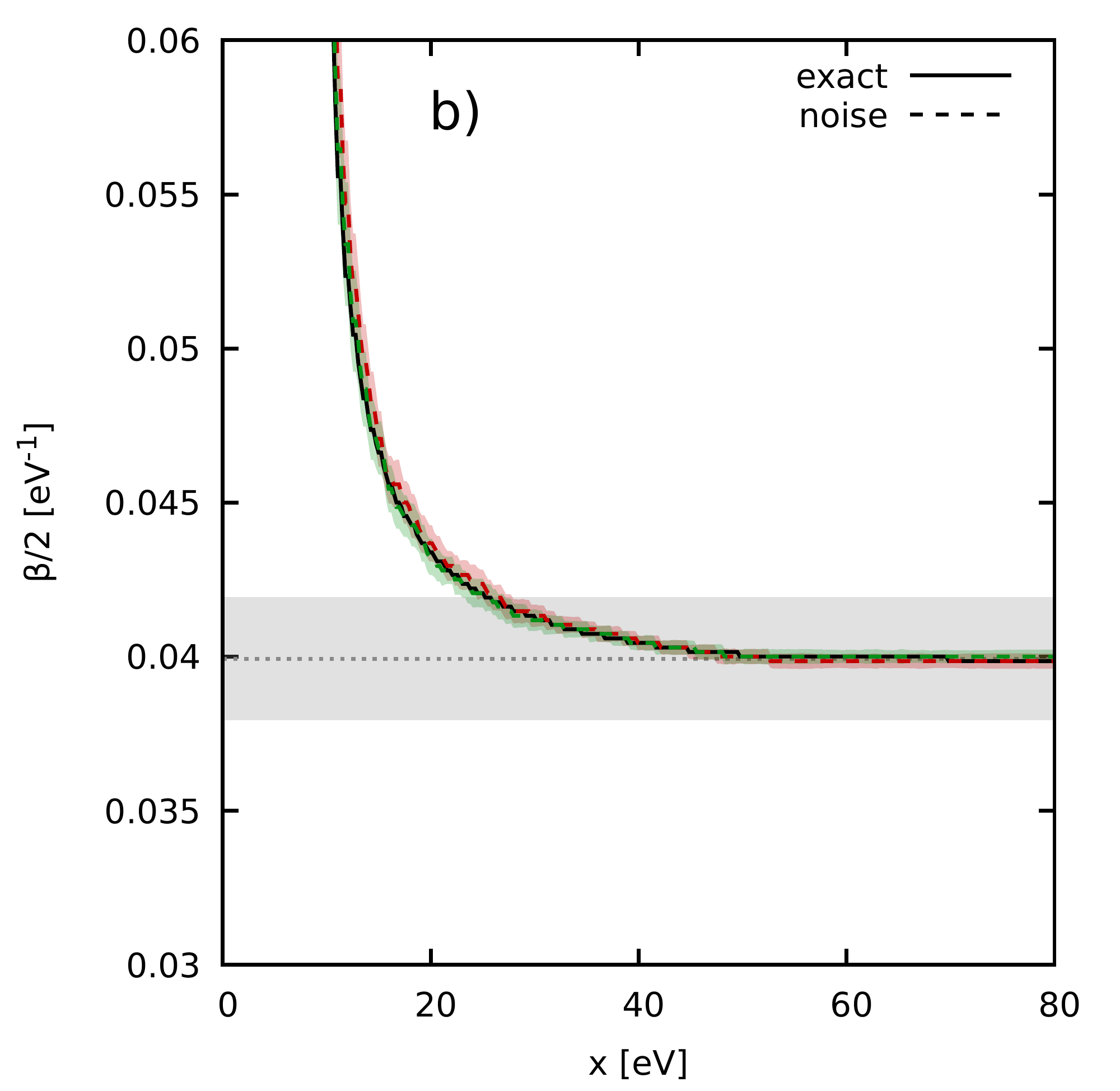}
\includegraphics[width=0.33\textwidth]{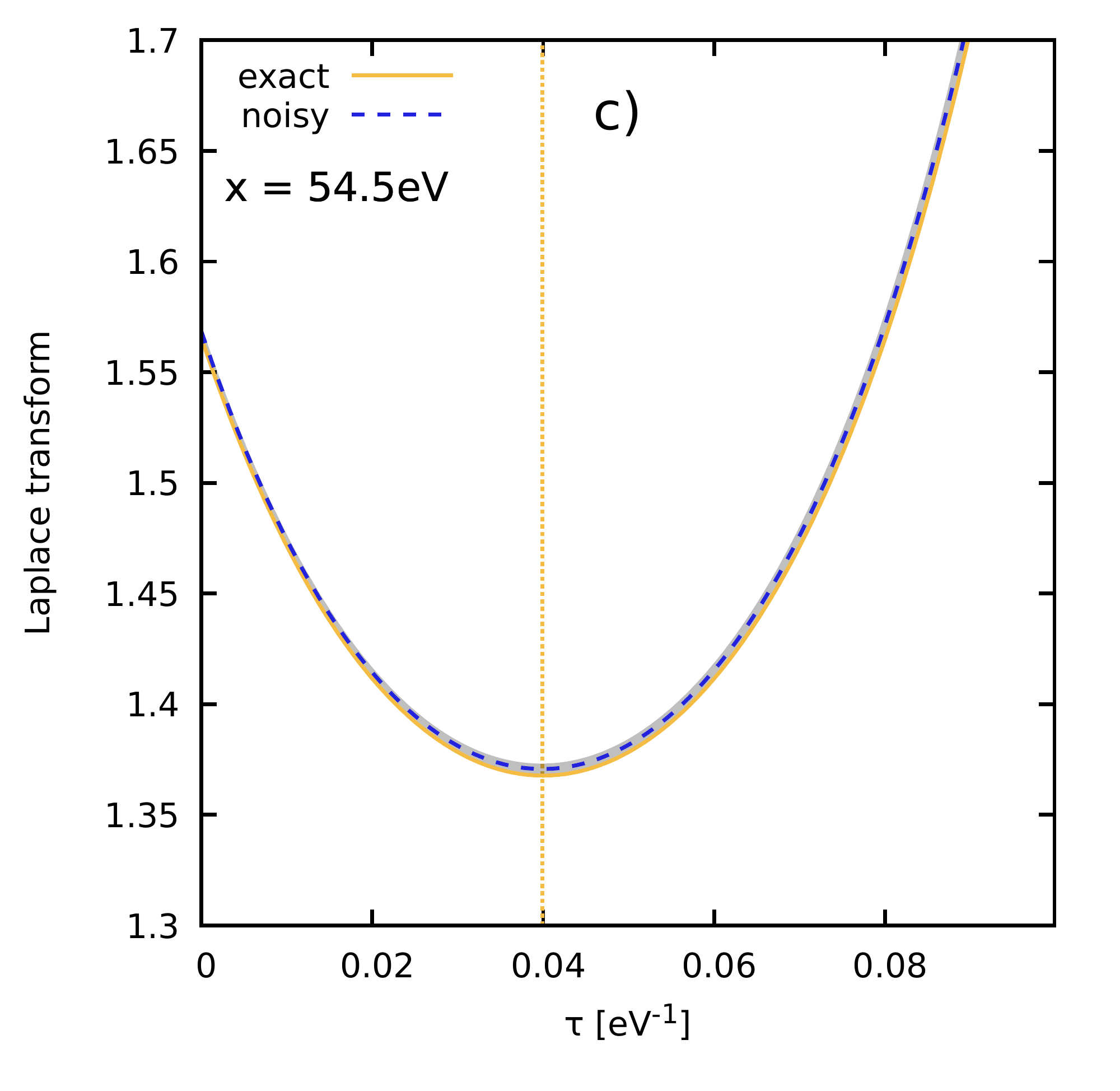}\\\includegraphics[width=0.33\textwidth]{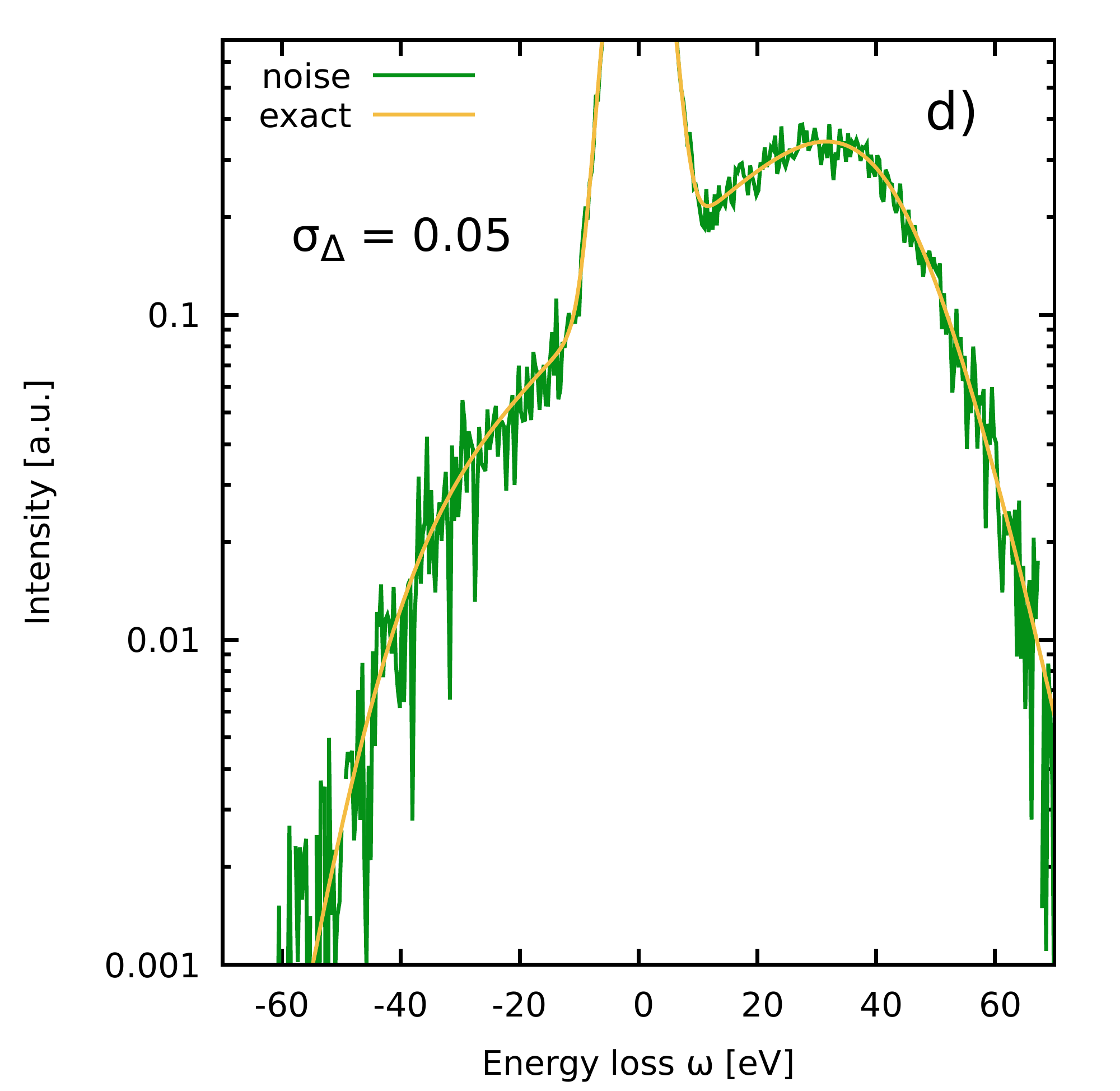}\includegraphics[width=0.33\textwidth]{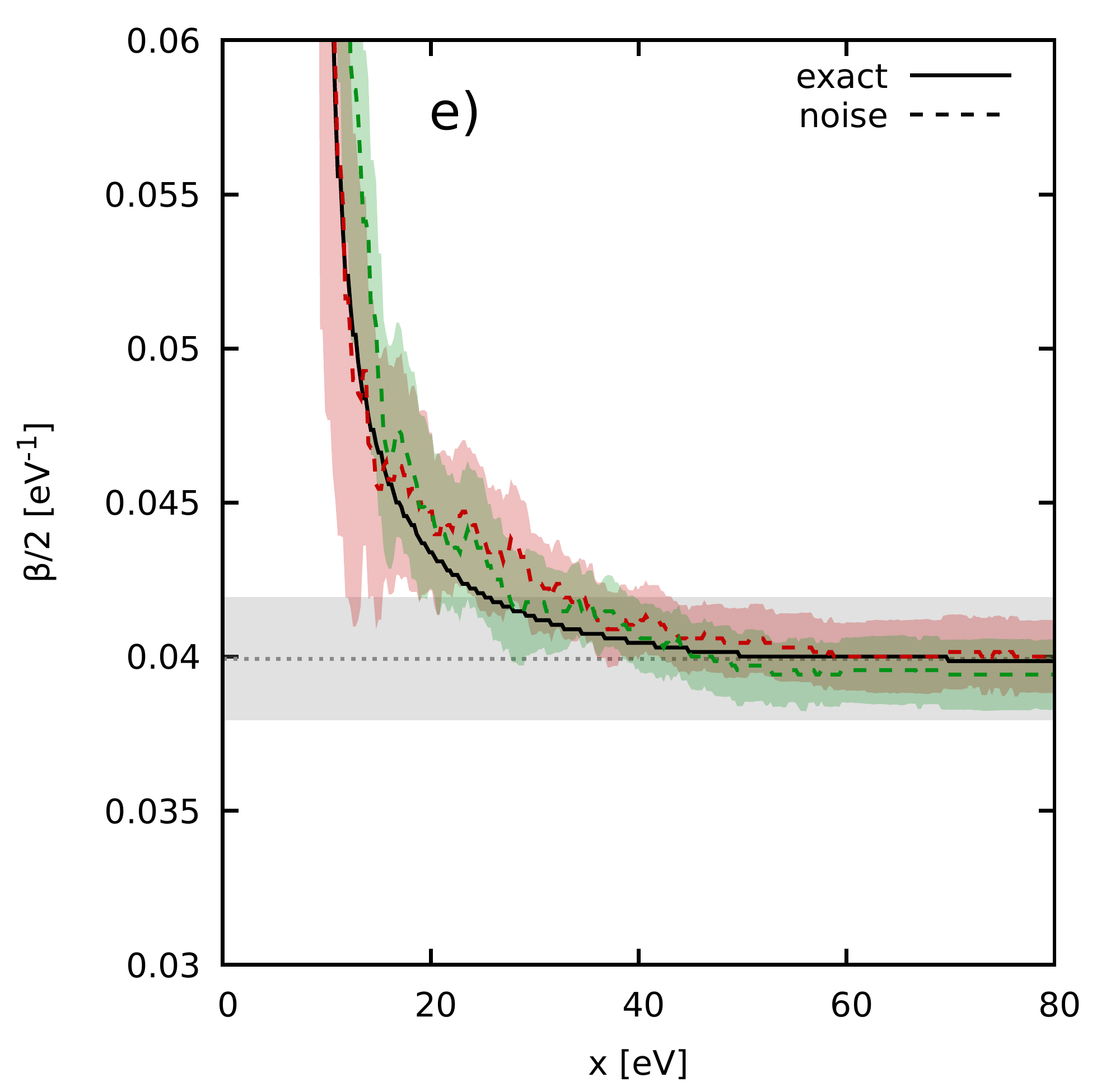}
\includegraphics[width=0.33\textwidth]{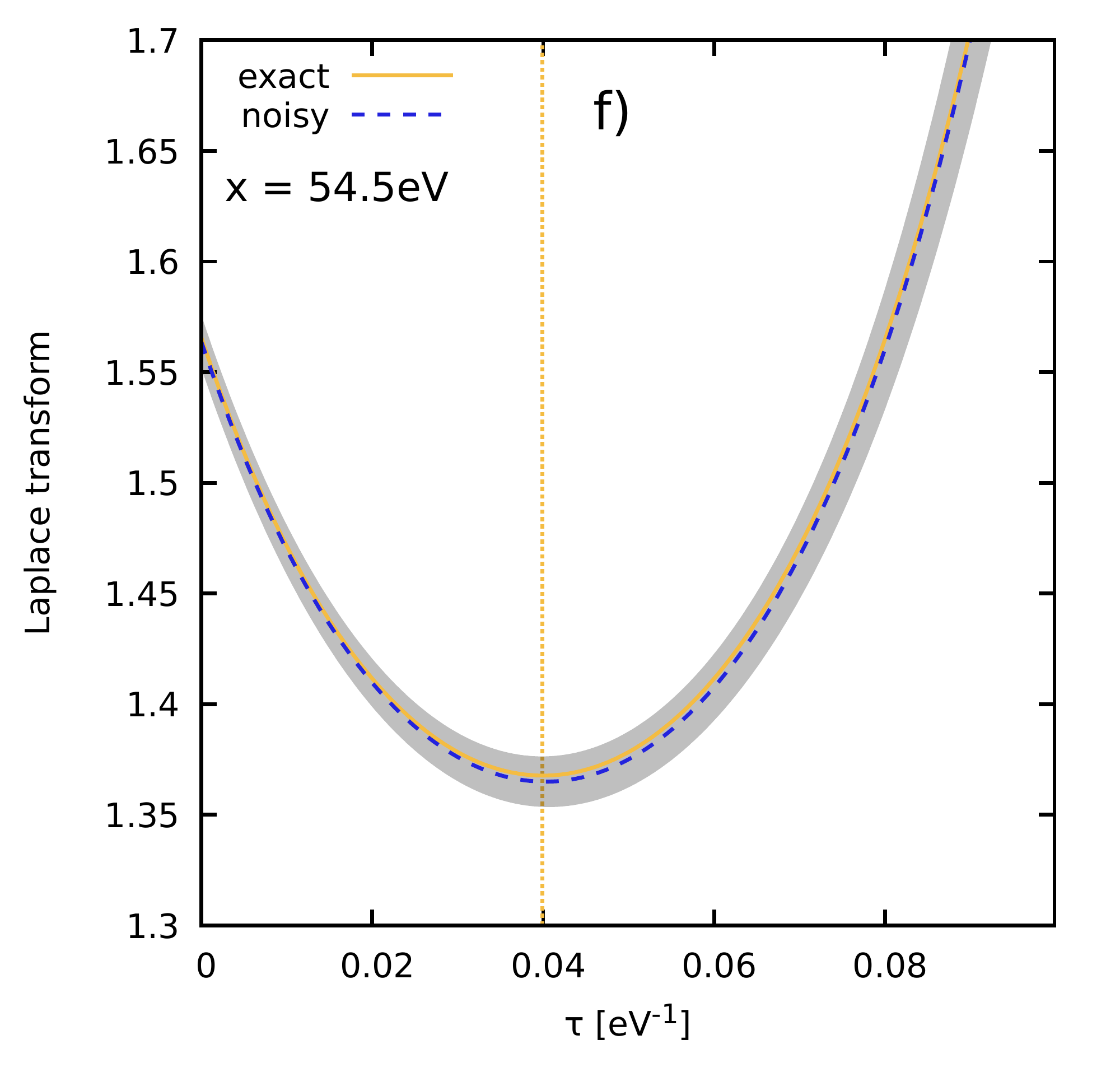}\\\includegraphics[width=0.33\textwidth]{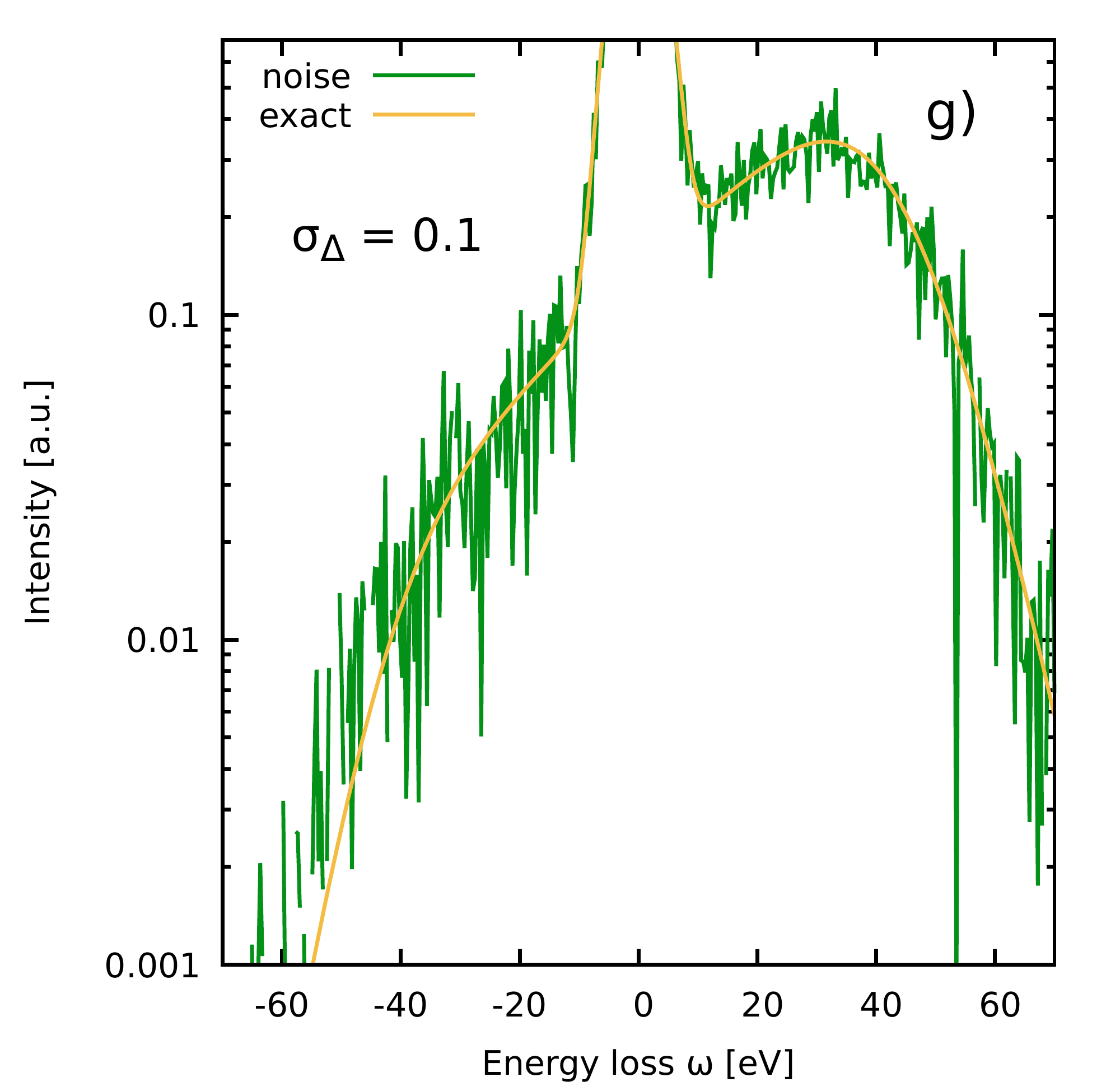}\includegraphics[width=0.33\textwidth]{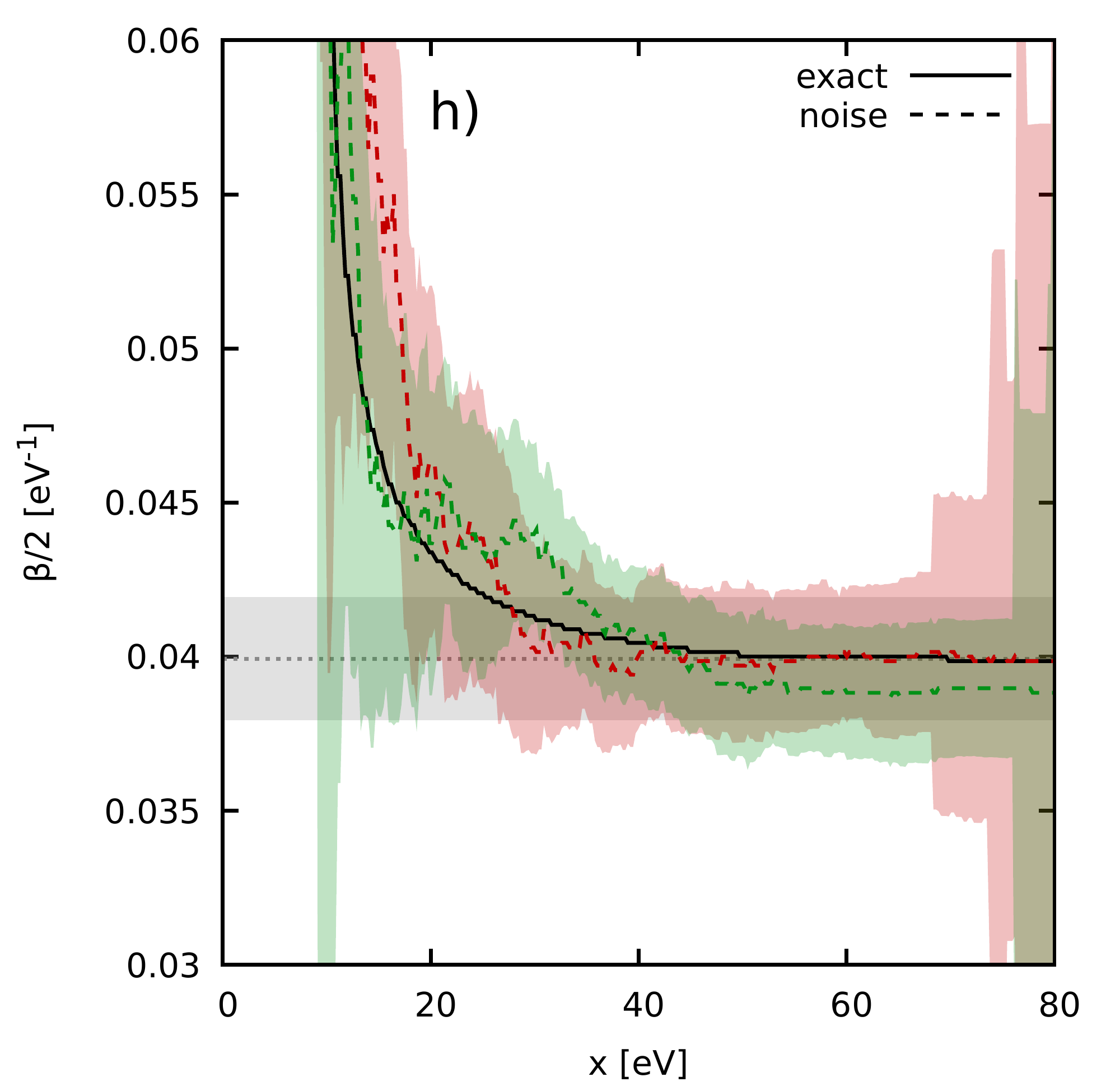}
\includegraphics[width=0.33\textwidth]{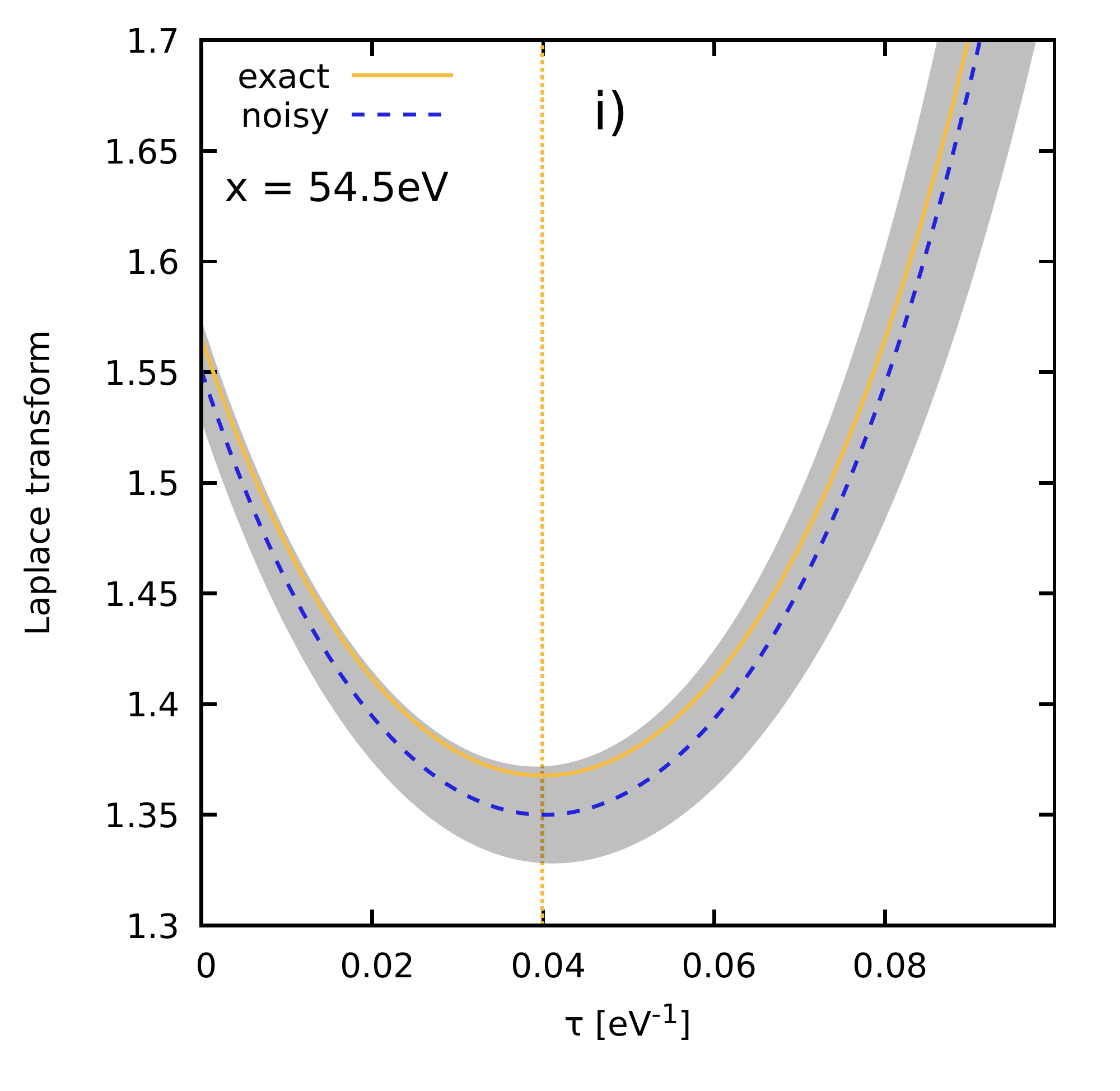}
\caption{\label{fig:sqrt_Noise0p01} 
Temperature extraction of noisy input data for the intensity of a UEG at $r_s=2$, $\Theta=1$, and $q=q_\textnormal{F}$ convolved with a Gaussian instrument function of width $\sigma=3.33\,$eV. Left: Exact (yellow) and perturbed (green) intensity; center: extraction of the temperature with respect to the integration boundary $x$, with the shaded grey area indicated an interval of $\pm5\%$ included as a reference; right: Laplace transform $F(\mathbf{q},\tau)$ with corresponding uncertainty obtained from Eq.~(\ref{eq:F_alpha}). Top: $\sigma_\Delta=0.01$, center: $\sigma_\Delta=0.05$, bottom: $\sigma_\Delta=0.1$.
}
\end{figure*} 

In Fig.~\ref{fig:sqrt_noise_contribution}, we analyze how much Gaussian random noise of unit variance contributes to the two-sided Laplace transform as a function of the frequency $\omega$ for the synthetic UEG model at $r_s=2$, $\Theta=1$, and $q=q_\textnormal{F}$ convolved with a Gaussian probe function of width $\sigma=3.33\,$eV; the shaded grey area depicts the corresponding $1\sigma_\Delta$ interval determined by $\sqrt{I(\mathbf{q},\omega)}$. Fig.~\ref{fig:sqrt_noise_contribution} a) was obtained for $\tau=0$, where most contributions are due to the positive frequency range. Conversely, Fig.~\ref{fig:sqrt_noise_contribution} b) corresponds to $\tau=\beta/2$, i.e., the location of the minimum in $\mathcal{L}\left[S(\mathbf{q},\omega)\right]$. In this case, the contribution of the noise to the Laplace transform of the intensity looks nearly identical to Fig.~\ref{fig:sqrt_noise_contribution} a), but mirrored at the $y$-axis. Finally, Fig.~\ref{fig:sqrt_noise_contribution} c) shows the same information for $\tau=\beta$. In this case, the noise in the negative frequency range is substantially increased compared to the previous two cases. In practice, it can thus be expected that we can resolve $F(\mathbf{q},\tau)$ with higher accuracy in the range of $0\leq\tau\leq\beta/2$ compared to $\tau>\beta/2$.

To illustrate the remarkable robustness of our methodology with respect to noise in the experimental data, we perturb synthetic intensities with a series of realistic noise of different pre-factors $\sigma_\Delta$ in Fig.~\ref{fig:sqrt_Noise0p01}. The top row was obtained for $\sigma_\Delta=0.01$, and the intensity itself is shown in Fig.~\ref{fig:sqrt_Noise0p01} a), with the green and yellow curves showing the perturbed and exact data, respectively.
The extraction of the (inverse) temperature from the location of the minimum in $F(\mathbf{q},\tau)$ is shown in Fig.~\ref{fig:sqrt_Noise0p01} b), where the shaded grey area indicates an interval of $\pm 5\%$ around the exact value. The solid black line shows the usual convergence with respect to the integration boundary $x$ of the exact intensity, and the dashed red and green curves have been obtained using two independent sets of random noise. 
Clearly, both curves attain the correct inverse temperature in the limit of large $x$ despite the perturbation.

An additional question of high practical importance is whether it is possible to quantify the uncertainty in the extracted temperature when the noisy experimental data for the intensity is taken as the only input.
This requires the reconstruction of the variance $\sigma_\Delta$ of the Gaussian noise in Eq.~(\ref{eq:Laplace_noise}), which can be reliably accomplished in the following way. First, we compute a smoothed intensity $I_\textnormal{smooth}(\mathbf{q},\omega)$ by averaging the experimental signal over a number of adjacent frequency points. The precise number of frequencies and the particular values of the weights are of minor importance for this idea. Then, we compute a set of corresponding noise according to
\begin{eqnarray}\label{eq:reconstruct_noise}
\Delta I_\textnormal{exp}(\mathbf{q},\omega) = I_\textnormal{exp}(\mathbf{q},\omega) - I_\textnormal{smooth}(\mathbf{q},\omega)\ . 
\end{eqnarray}
Finally, we obtain the reconstructed Gaussian noise variable $\xi_{\sigma_\Delta}(\omega)$ as
\begin{eqnarray}\label{eq:reconstruct_Gaussian}
\xi_{\sigma_\Delta}(\omega) = \frac{\Delta I_\textnormal{exp}(\mathbf{q},\omega) }{\sqrt{I_\textnormal{smooth}(\mathbf{q},\omega)}}\ .
\end{eqnarray}

\begin{figure}\centering
\includegraphics[width=0.45\textwidth]{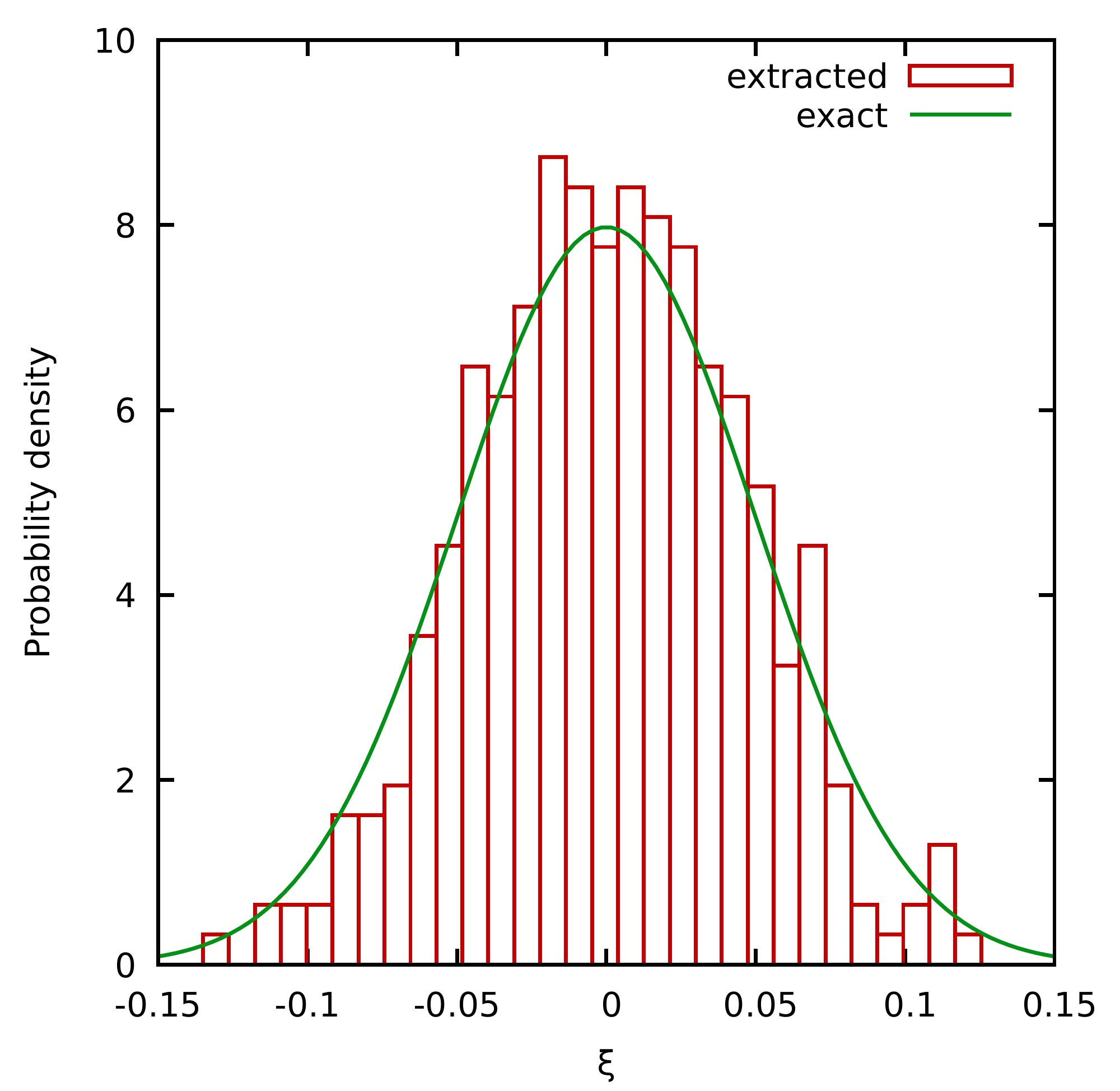}
\caption{\label{fig:hist} Red: Reconstructed noise, Eq.~(\ref{eq:reconstruct_Gaussian}); green: exact normal distribution of random noise with $\sigma_\Delta=0.05$.
}
\end{figure} 

In Fig.~\ref{fig:hist}, we show a histogram of the thus reconstructed noise (red bars), which is in excellent agreement with the true distribution $\sigma_\Delta=0.05$ (depicted in the center row of Fig.~\ref{fig:sqrt_Noise0p01}). In practice, one can then obtain the reconstructed $\sigma_\Delta$ either from a Gaussian fit of the histogram or by evaluating 
\begin{eqnarray}\label{eq:sigma_delta}
\sigma_\Delta = \left(
\frac{1}{M}\sum_{i=0}^{M-1}\xi^2_{\sigma_\Delta}(\omega_i)
\right)^{1/2}\ .
\end{eqnarray}
To estimate the corresponding uncertainty in $F(\mathbf{q},\tau)$ and in this way also in the extracted inverse temperature $\beta$, we generate a set of $K$ independent random noise samples from this distribution, resulting in a set of trial functions
\begin{eqnarray}
F^\alpha_x(\mathbf{q},\tau) =  \mathcal{L}_x\left[ I_\textnormal{exp}(\mathbf{q},\omega)\right] + \mathcal{L}_x\left[\xi^\alpha_{\sigma_\Delta}(\omega)\sqrt{I_\textnormal{smooth}(\mathbf{q},\omega)}\right]\ , \nonumber \\ \label{eq:F_alpha}
\end{eqnarray}
with $\alpha=0,\dots,K-1$. From Eq.~(\ref{eq:F_alpha}), we can directly estimate the associated variance due to the random noise in the experimental intensity both in the Laplace transform and subsequently in the extracted temperature.

\begin{figure*}\centering\includegraphics[width=0.33\textwidth]{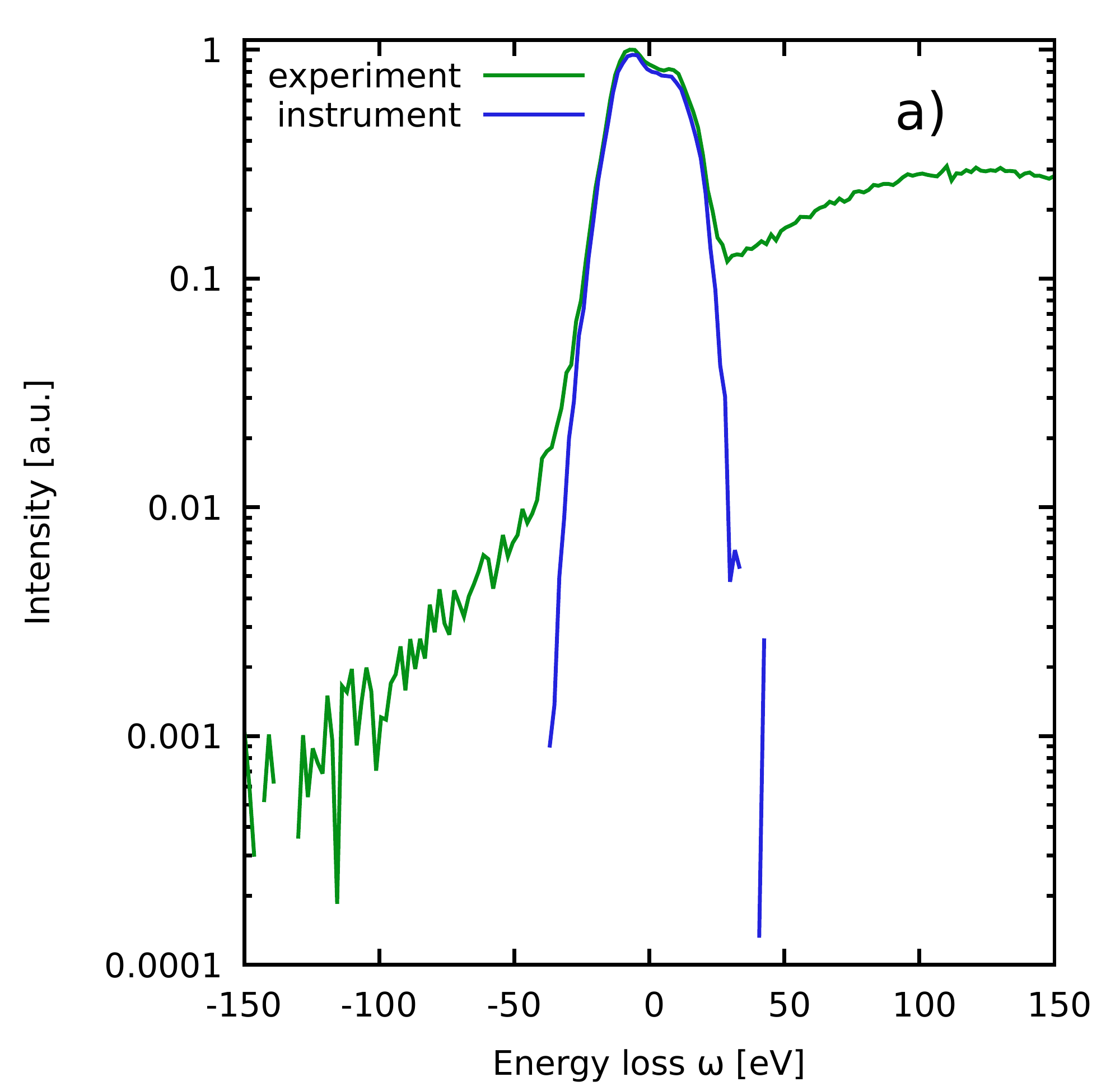}\includegraphics[width=0.33\textwidth]{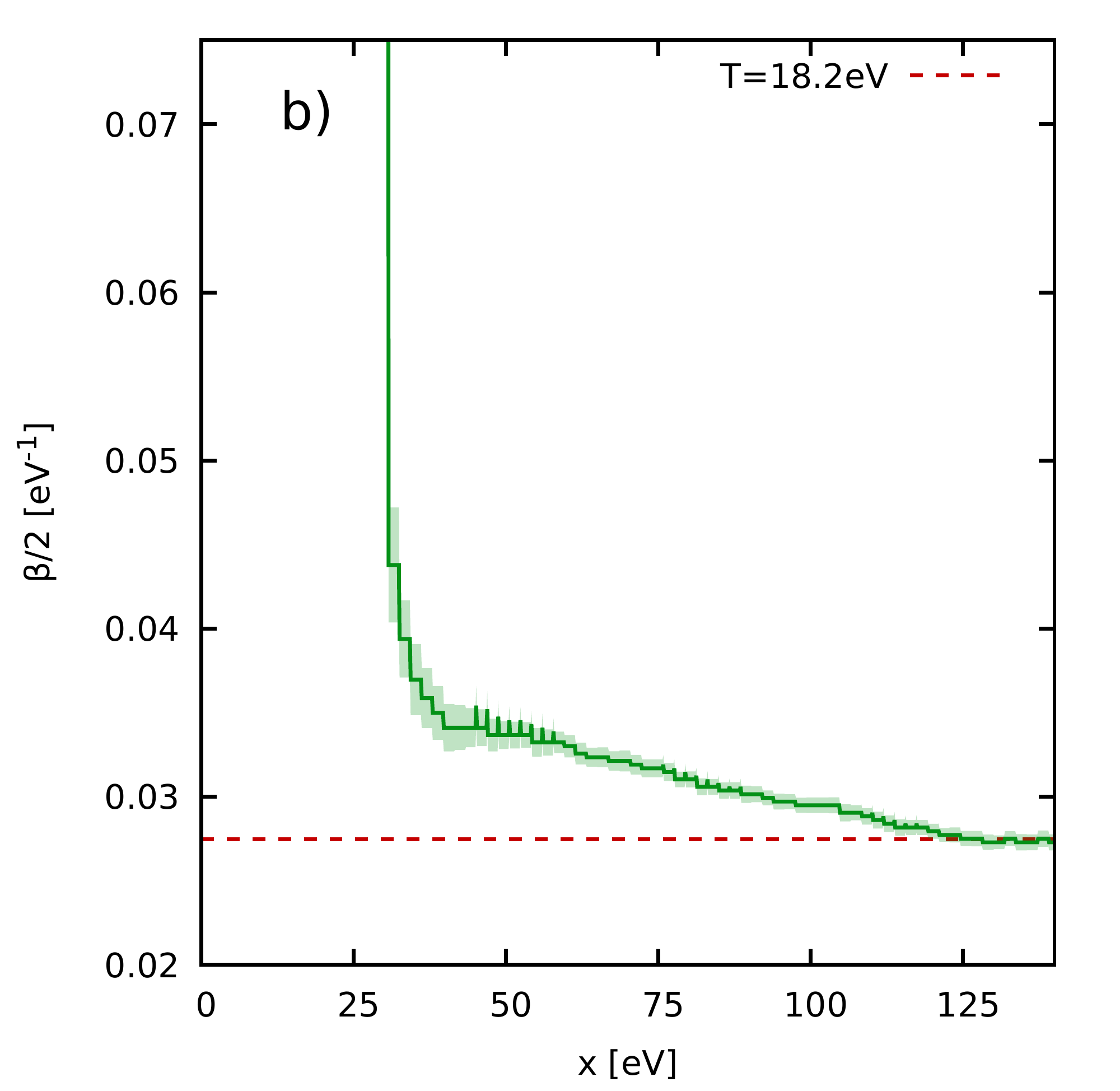}\includegraphics[width=0.33\textwidth]{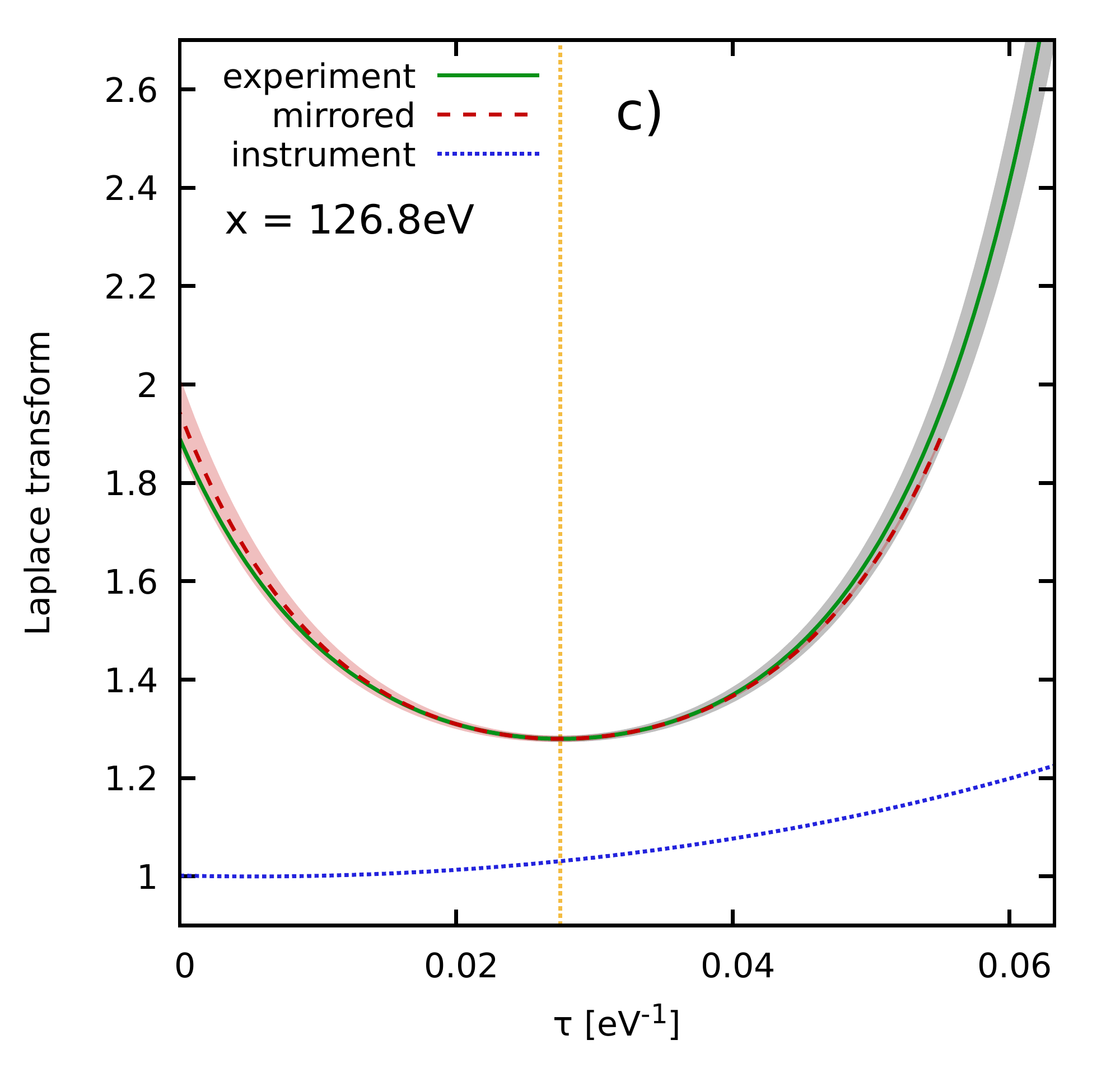}\\\includegraphics[width=0.33\textwidth]{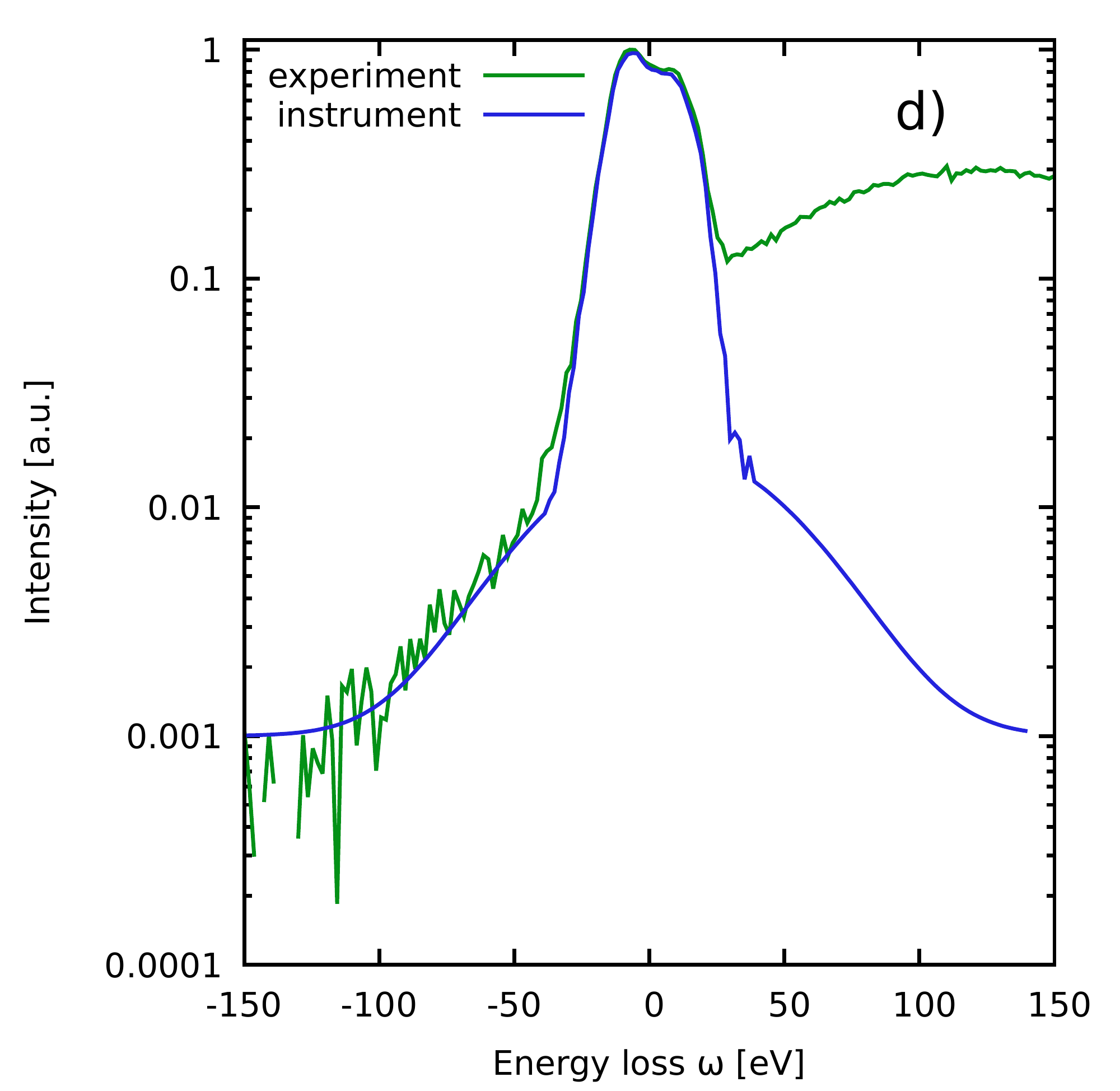}\includegraphics[width=0.33\textwidth]{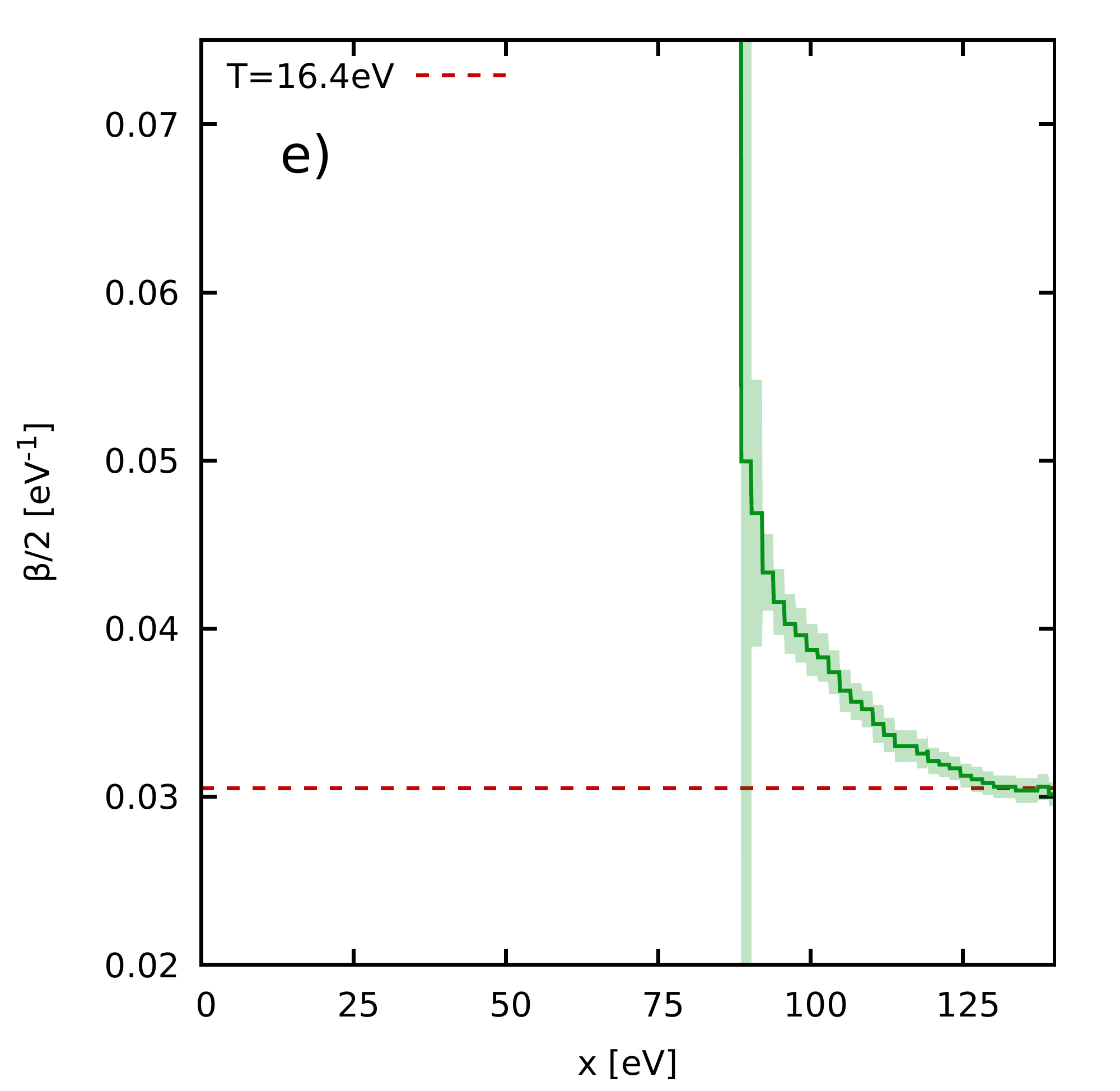}
\includegraphics[width=0.33\textwidth]{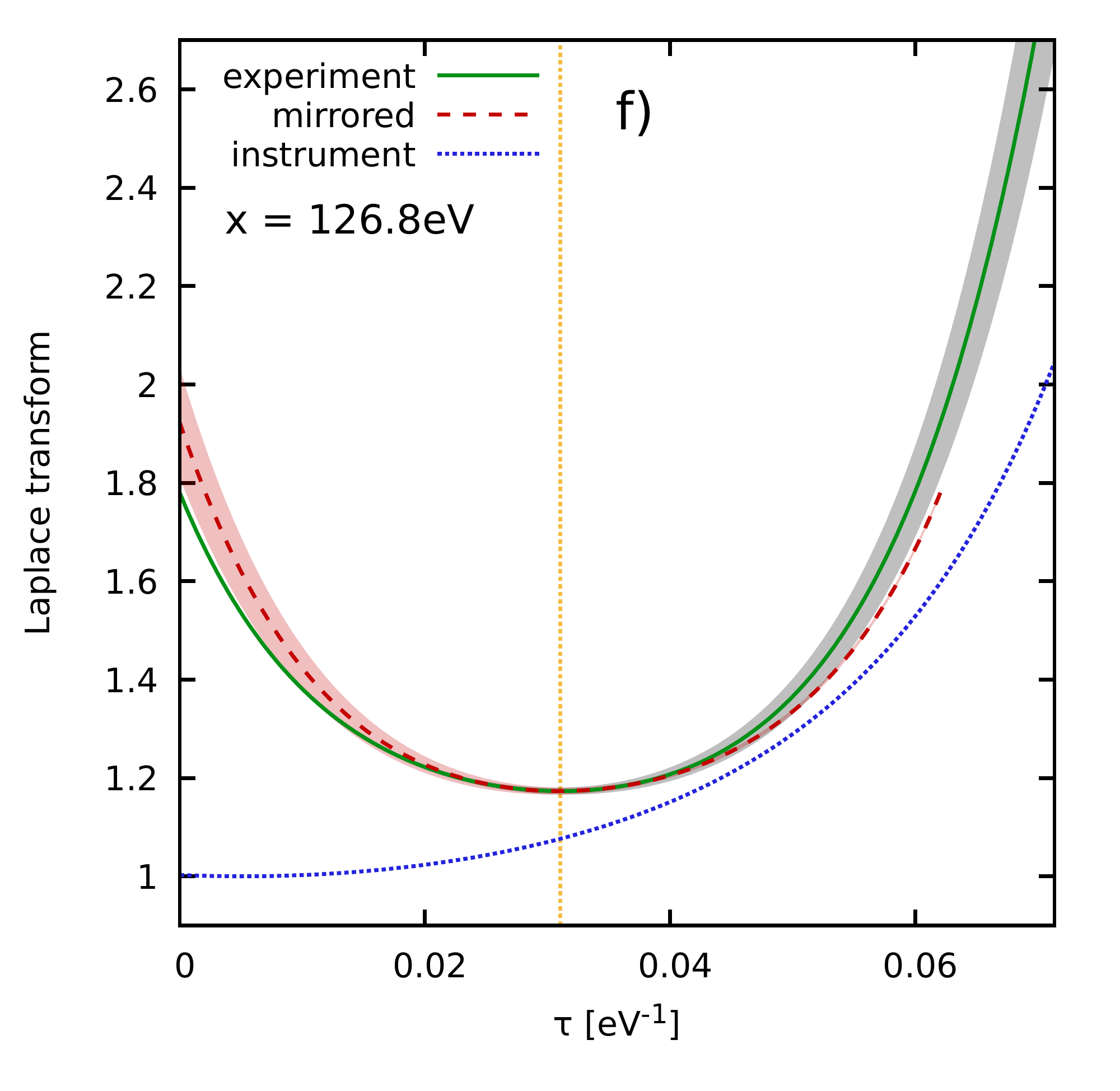}
\caption{\label{fig:sqrt_Kraus} Analysis of the graphite data collected by Kraus \emph{et al.}~\cite{kraus_xrts} for two possible instrument functions (top and bottom row). Left: Experimental XRTS intensity (green) and instrument function (blue); center: extraction of the temperature with respect to the integration boundary $x$; right: $F_x(\mathbf{q},\tau)$ for $x=126.8\,$eV.
}
\end{figure*} 

\begin{figure*}\centering
\includegraphics[width=0.4\textwidth]{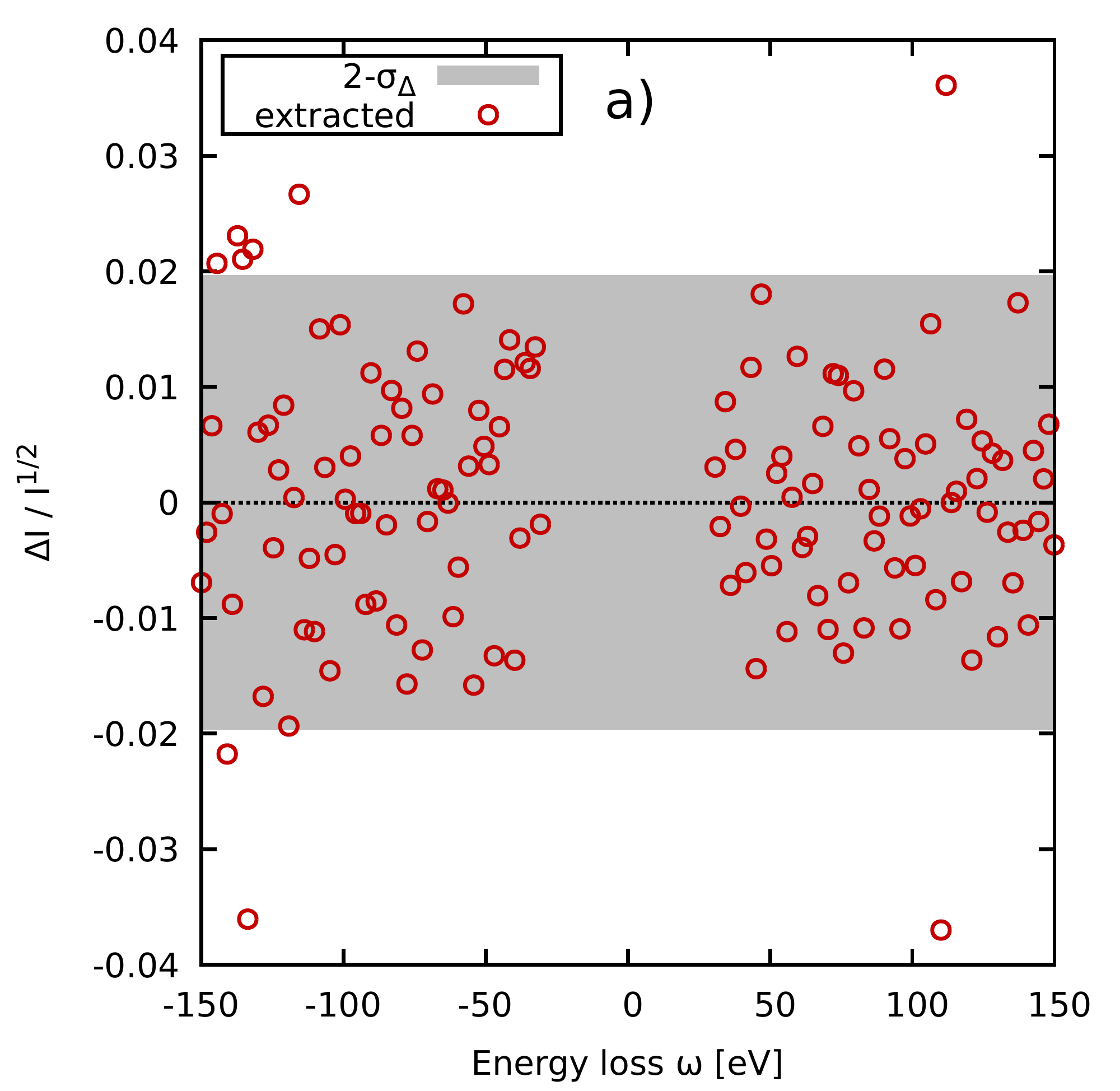}\includegraphics[width=0.4\textwidth]{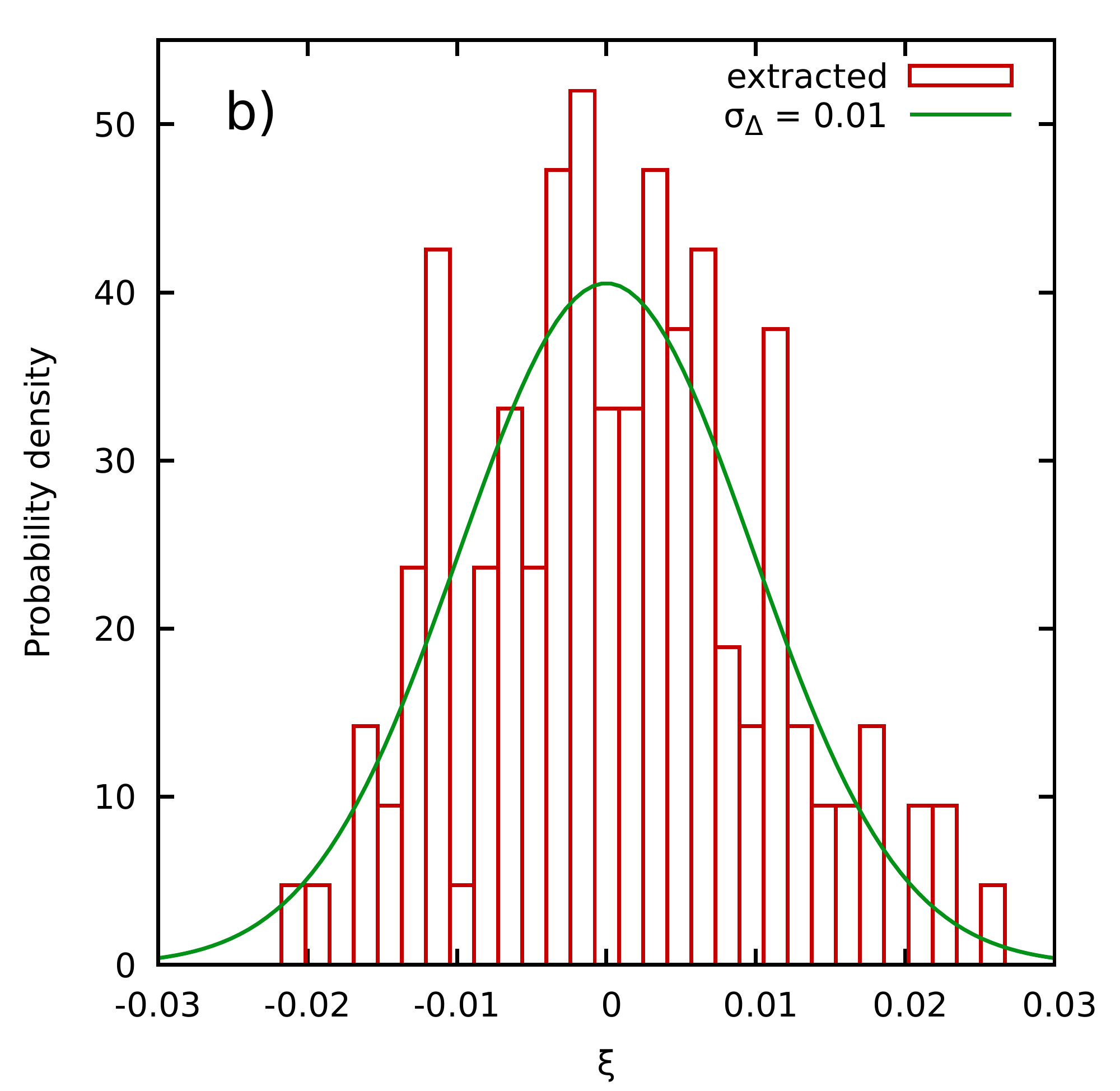}
\caption{\label{fig:hist_Kraus} Error analysis of graphite data by Kraus \emph{et al.}~\cite{kraus_xrts} shown in Fig.~\ref{fig:sqrt_Kraus}.
}
\end{figure*} 
The results for this uncertainty in $\beta$ (given as the $2\sigma$-interval) are included in Fig.~\ref{fig:sqrt_Noise0p01} as the shaded red and green areas, which, indeed, give a real measure for the fluctuation around the exact curve. 
The corresponding results for the Laplace transform $F(\mathbf{q},\tau)$ are shown in Fig.~\ref{fig:sqrt_Noise0p01} c), where the solid yellow curve shows the exact result. The dashed black curve has been obtained by taking as input the perturbed data and is close to the former, although a small yet significant deviation is observed. The associated uncertainty in $F(\mathbf{q},\tau)$ computed from Eq.~(\ref{eq:F_alpha}) has been included as the shaded grey area and nicely fits the observed difference. 

In the center row of Fig.~\ref{fig:sqrt_Noise0p01}, we repeat this analysis for a larger magnitude of the random noise, $\sigma_\Delta=0.05$. Evidently, the larger noise level is directly propagated into larger fluctuations in the extraction of the temperature shown in Fig.~\ref{fig:sqrt_Noise0p01} e). At the same time, we stress that 1) the error bars from Eq.~(\ref{eq:F_alpha}) capture these fluctuations very well and 2) that the extracted temperature is accurate to $\sim2\%$ despite the substantial noise level in the input data.
The imaginary time intermediate scattering function depicted in Fig.~\ref{fig:sqrt_Noise0p01} f) exhibits similar behavior.

Finally, we consider an even higher noise level in $\sigma_\Delta=0.1$ shown in the bottom row of Fig.~\ref{fig:sqrt_Noise0p01}.
Still, the extracted temperatures remain within $\sim4\%$ of the exact result, and our estimated uncertainty measures are accurate both for $\beta$ and $F(\mathbf{q},\tau)$.

\section{Analysis of experimental scattering data\label{sec:experiment}}
In the previous sections, we have unambiguously demonstrated the capability of our new approach for extracting the exact temperature from a scattering intensity signal in different situations. Moreover, we have shown that our method is highly robust with respect to noisy input data and have introduced a framework for the rigorous quantification of the associated uncertainty both in the temperature and in the imaginary time intermediate scattering function $F(\mathbf{q},\tau)$. In the following, we turn our attention to actual experimentally measured data and reanalyze 1) an experiment on warm dense graphite by Kraus \emph{et al.}~\cite{kraus_xrts} and 2) the pioneering investigation of plasmons in warm dense beryllium by Glenzer \emph{et al.}~\cite{Glenzer_PRL_2007}.

\subsection{Graphite\label{sec:graphite}}

In Fig.~\ref{fig:sqrt_Kraus}, we show our new analysis of the XRTS signal on isochorically heated graphite by Kraus \emph{et al.}~\cite{kraus_xrts}. In the left column, we show the measured intensity as the green curve, where accurate data is available over three orders of magnitude. Interestingly, the main source of uncertainty in this experiment is due to the somewhat unclear shape of the instrument function $R(\omega)$. Two plausible possibilities are shown as the blue curves in Fig.~\ref{fig:sqrt_Kraus} a) and Fig.~\ref{fig:sqrt_Kraus} d). 

From a practical perspective, it is very useful to start the investigation of the experimental data set by analyzing the distribution of the noise. In Fig.~\ref{fig:hist_Kraus} a), we show the corresponding results for Eq.~(\ref{eq:reconstruct_Gaussian}) as the red circles. Evidently, the deviations fluctuate around the origin, and the overall amplitude of $\xi_{\sigma_\Delta}(\omega)$ appears to be approximately constant over the entire $\omega$ range. This is a nice empirical validation of the functional form in Eq.~(\ref{eq:scattering_error}).
For completeness, we note that the elastic feature itself has been omitted from this analysis, as it would require a separate smoothing procedure, which is not needed for accurate quantification of the noise level. 
In Fig.~\ref{fig:sqrt_Kraus} b), we show the corresponding histogram as the red bars, which can be well reproduced by a Gaussian fit (green curve). Both the fit and the direct evaluation of Eq.~(\ref{eq:sigma_delta}) give a variance of $\sigma_\Delta\sim10^{-2}$, which is used to quantify the uncertainty of both the temperature and $F(\mathbf{q},\tau)$ in the following.

Going back to Fig.~\ref{fig:sqrt_Kraus} b) and d), we find that convergence with the integration boundary $x$ starts around $x=125\,$eV. We note that going beyond $x=140\,$eV does not make sense in practice, as the experimentally measured intensity vanishes within the given noise level for $\omega\lesssim-140\,$eV. From these panels, we can clearly see that using either the narrow or the broad instrument function (which we truncate at $\omega=90\,$eV as the constant asymptotes given in the original Ref.~\cite{kraus_xrts} are plainly unphysical and would lead to a diverging Laplace transform $\mathcal{L}\left[R(\omega)\right]$) has a substantial impact on the extracted temperature. Consequently, we gave our final estimate as $T=18\pm2\,$eV in our previous investigation~\cite{Dornheim_T_2022}. At the same time, we stress that the resulting uncertainty due to the somewhat unknown $R(\omega)$ is considerably smaller than in the original Ref.~\cite{kraus_xrts}, where the applied Chihara fit gave $T=21\,$eV with an uncertainty of $\sim50\%$. 

Finally, we show our estimates for $F_x(\mathbf{q},\tau)\approx F(\mathbf{q},\tau)$ for a converged integration boundary of $x=126.8\,$eV in the right column of Fig.~\ref{fig:sqrt_Kraus}. In particular, the solid green curves show our direct evaluation of Eq.~(\ref{eq:F_truncated}), and the shaded grey area indicates the corresponding uncertainty interval. In addition, we also mirror this function around $\tau=\beta/2$ (which is estimated from the minimum in the green curve), i.e., $F(\mathbf{q},\beta-\tau)$, and the results are shown as the dashed red curve, with the shaded red area indicating the correspondingly mirrored uncertainty interval. Evidently, our extracted results for $F(\mathbf{q},\tau)$ are fairly symmetric within the given uncertainty range for both instrument functions, although the degree of asymmetry [i.e., the difference between $F(\mathbf{q},\tau)$ and $F(\mathbf{q},\beta-\tau)$] is somewhat smaller for the narrower instrument function (top). 
Lastly, the dotted blue curves show the two-sided Laplace transformation of the respective $R(\omega)$. Clearly, the impact of the latter is more pronounced for the broader instrument function, as expected.

In conclusion, our present analysis is in agreement with the less controlled Chihara model-based analysis in the original Ref.~\cite{kraus_xrts} and highlights the importance of accurate characterization of the instrument function in future scattering experiments.

\begin{figure*}\centering\includegraphics[width=0.33\textwidth]{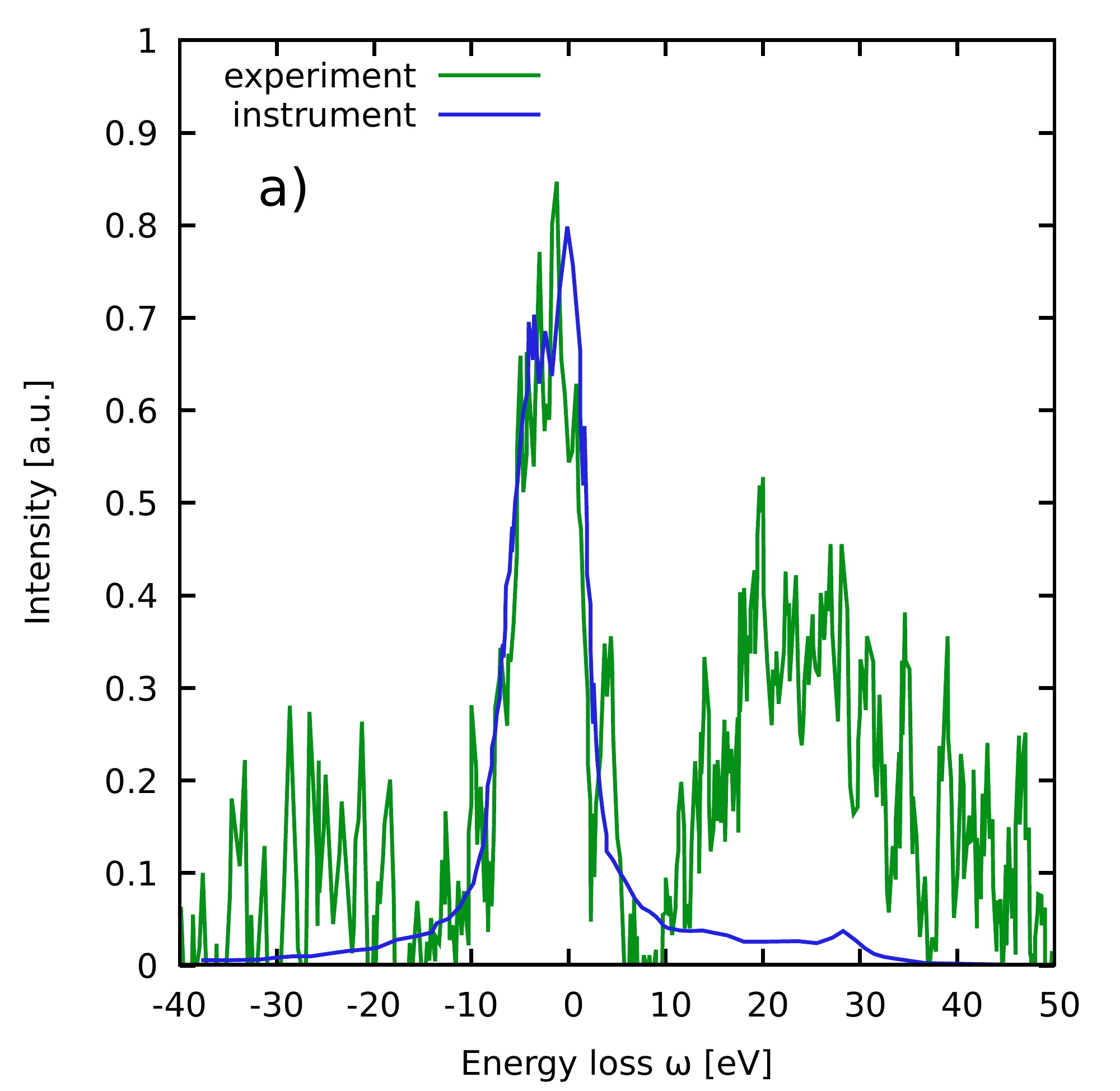}\includegraphics[width=0.33\textwidth]{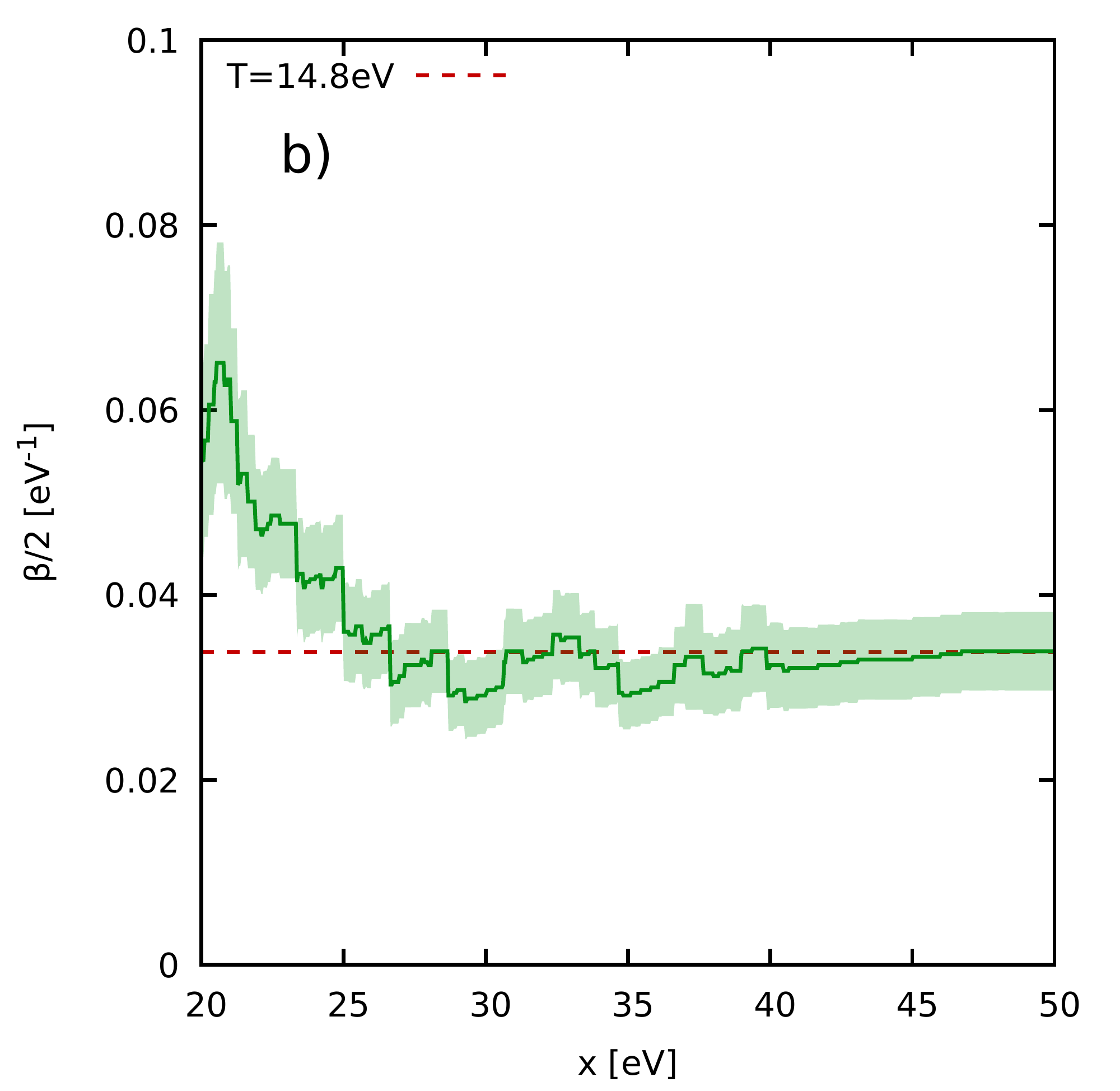}\includegraphics[width=0.33\textwidth]{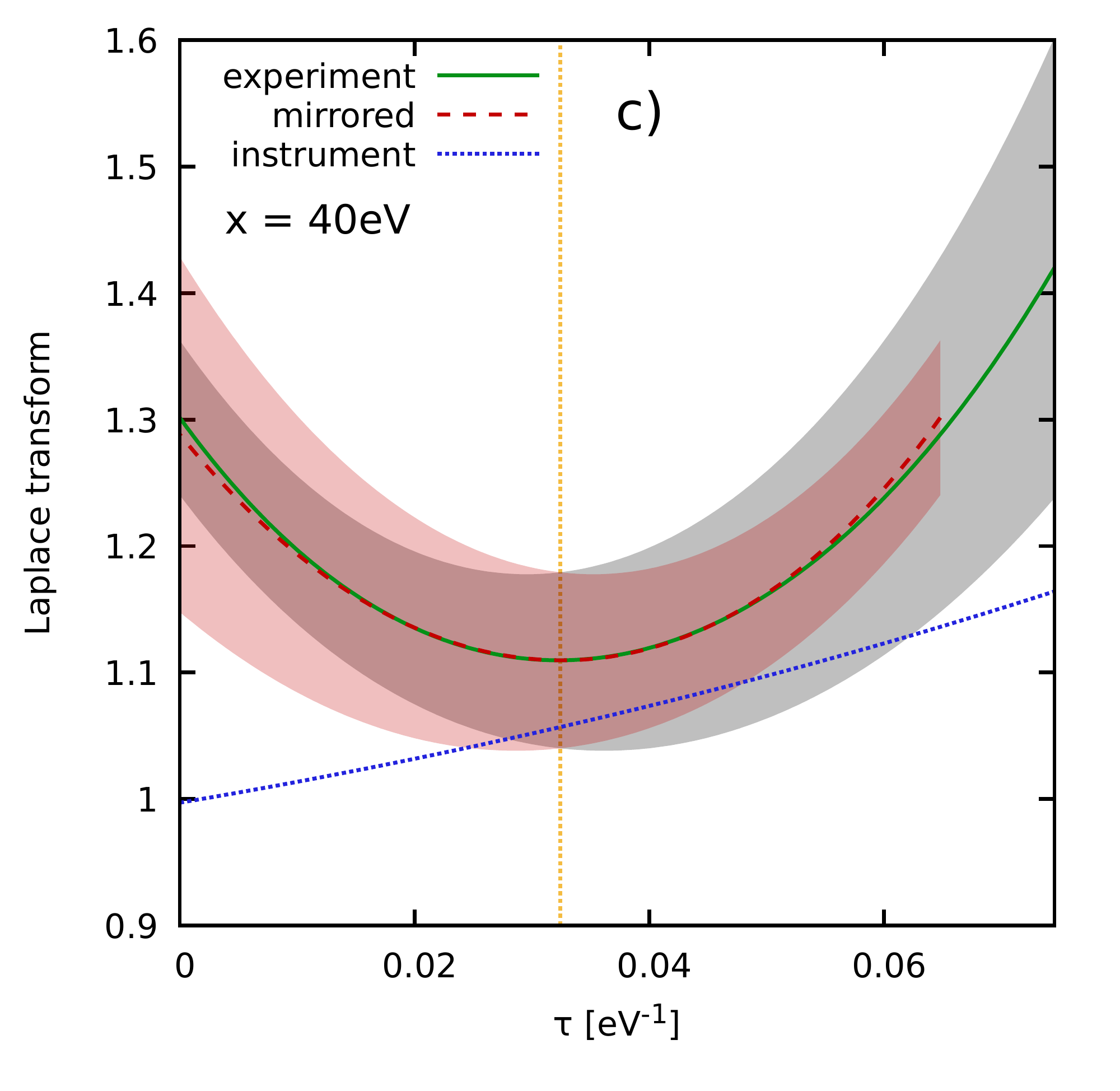}
\caption{\label{fig:sqrt_Glenzer} Temperature analysis of beryllium data by Glenzer \emph{et al.}~\cite{Glenzer_PRL_2007}. Left: Experimental XRTS intensity (green) and instrument function (blue); center: extraction of the temperature with respect to the integration boundary $x$; right: $F_x(\mathbf{q},\tau)$ for $x=40\,$eV.
}
\end{figure*} 

\begin{figure*}\centering
\includegraphics[width=0.4\textwidth]{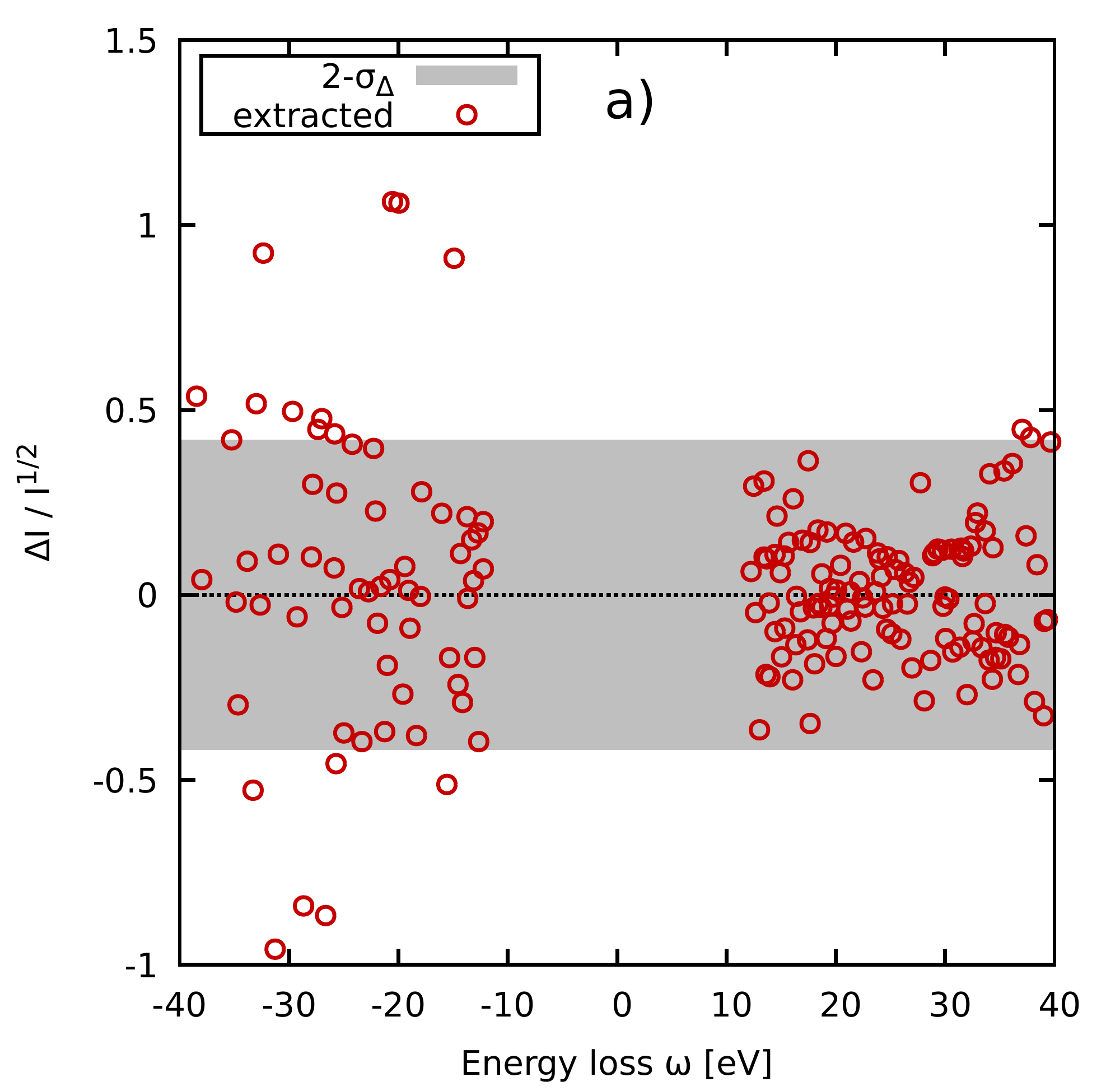}\includegraphics[width=0.4\textwidth]{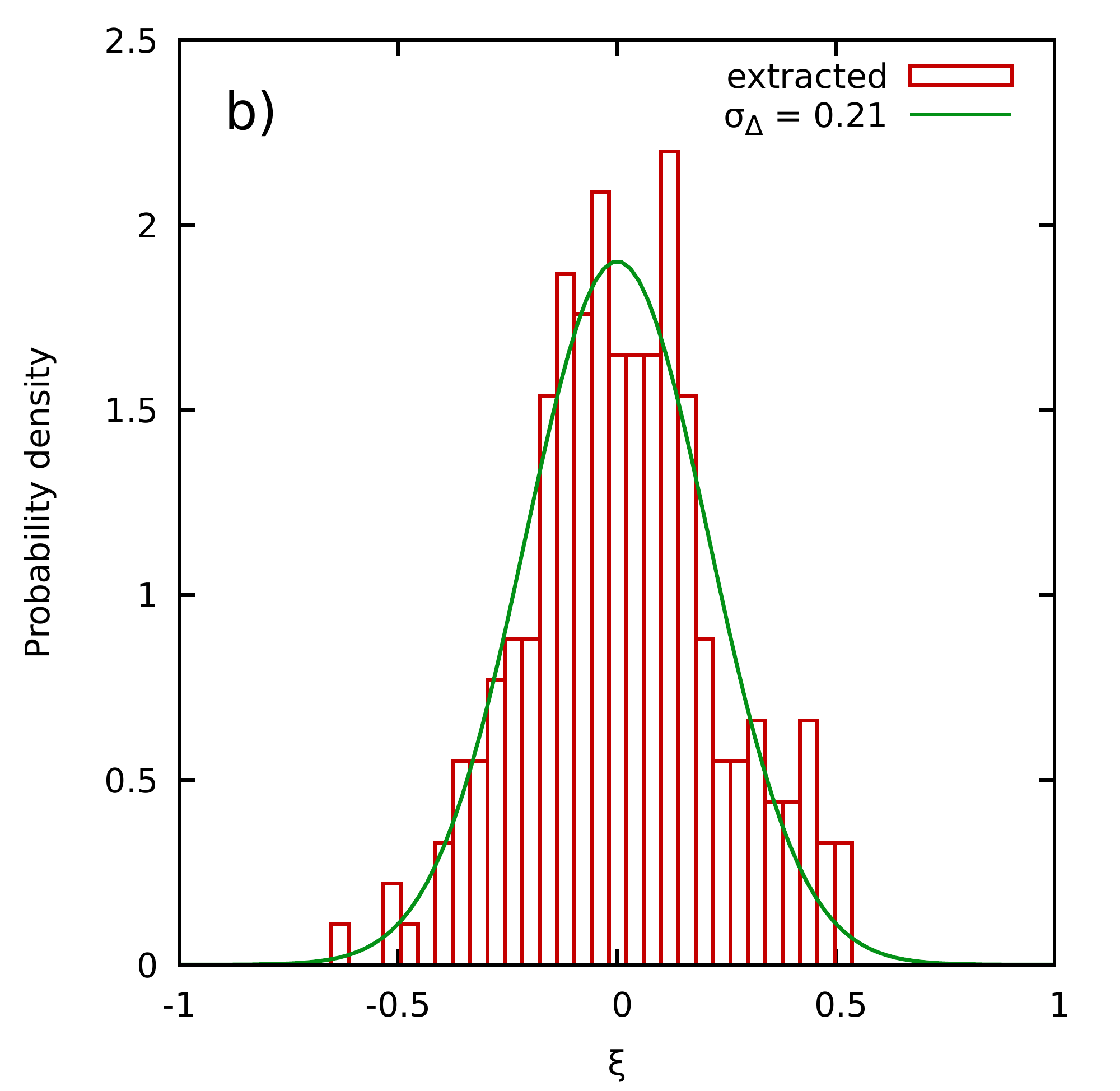}
\caption{\label{fig:hist_Glenzer} Error analysis of beryllium data by Glenzer \emph{et al.}~\cite{Glenzer_PRL_2007} shown in Fig.~\ref{fig:sqrt_Glenzer}.
}
\end{figure*}

\subsection{Beryllium\label{sec:beryllium}}
The second experimental data set that we reanalyze in the present work is the scattering experiment focusing on the plasmons in warm dense beryllium shown in Fig.~\ref{fig:sqrt_Glenzer}. 
Panel a) shows the experimentally measured XRTS intensity as the green curve, which is clearly afflicted with a much larger noise level compared to the graphite data that we have considered previously. The corresponding reconstruction of the error distribution is shown in Fig.~\ref{fig:hist_Glenzer}, where we find again good agreement with the functional form of Eq.~(\ref{eq:scattering_error}); the few outliers for $\omega<0$ are likely an artefact from a nearly vanishing intensity signal for some $\omega$, leading to increased values for $\Delta I(\mathbf{q},\omega)/\sqrt{I(\mathbf{q},\omega)}$.
Fig.~\ref{fig:hist_Glenzer} shows the corresponding histogram of $\xi_{\sigma_\Delta}(\omega)$ as the red bars, which is accurately reproduced by a Gaussian fit with a variance of $\sigma_\Delta=0.21$ (green curve).

The reconstructed distribution of the random noise is again used to rigorously quantify the uncertainty. In Fig.~\ref{fig:sqrt_Glenzer} b), we show the convergence of the extracted (inverse) temperature with the integration boundary $x$, and the shaded grey uncertainty interval very plausibly fits the observed noise in the green curve. As a final result, we obtain a temperature of $T=14.8\pm2\,$eV. This is close to, though significantly different, from $T=12\,$eV given in the original Ref.~\cite{Glenzer_PRL_2007} which was obtained from an approximate Mermin model~\cite{Mermin_model}. In the context of the present work, the main point of Fig.~\ref{fig:sqrt_Glenzer} b) is the remarkable robustness of our methodology even with respect to the considerable noise level in experimental data. 
Finally, we show our extracted imaginary time intermediate scattering function in Fig.~\ref{fig:sqrt_Glenzer} c). Evidently, the direct evaluation of Eq.~(\ref{eq:F_truncated}) (solid green) is in nearly perfect agreement with the mirrored curve (dashed red) over the entire $\tau$-range despite the given uncertainty range (shaded areas).

\section{Summary and Outook\label{sec:summary}}
In this work, we have given a detailed introduction to our new, model-free methodology for extracting the temperature of arbitrarily complex materials from XRTS measurements~\cite{Dornheim_T_2022}. In particular, we have presented an extensive analysis of synthetic scattering spectra over a wide range of wave numbers and temperatures. Evidently, the method works exceptionally well covering the range from the collective to the single-particle regimes. 
In addition, we have studied the impact of the width of the instrument function, which decisively determines the minimum temperature that can be extracted from a corresponding XRTS signal. Naturally, these findings have direct implications for the design of future experimental XRTS setups. 

Furthermore, we have introduced a framework for quantifying rigorously the impact of experimental noise both on the extracted temperature and on $F(\mathbf{q},\tau)$ itself. In practice, the method is well-behaved and highly robust even against substantial noise levels. 
As a practical demonstration, we have reanalyzed the XRTS experiments by Kraus \emph{et al.}~\cite{kraus_xrts} and Glenzer \emph{et al.}~\cite{Glenzer_PRL_2007}.

We are convinced that our new methodology will have a considerable impact on a gamut of applications related to the study of WDM including the highly active fields of inertial confinement fusion~\cite{Betti2016,Zylstra2022} and laboratory astrophysics~\cite{takabe_kuramitsu_2021}. On the one hand, we note that our method is particularly suited for modern X-ray free electron laser facilities with a high repetition rate such as LCLS~\cite{LCLS_2016}, SACLA~\cite{SACLA_2011}, and the European XFEL~\cite{Tschentscher_2017, Voigt_POP_2021}.
Specifically, its negligible computation cost will open up unprecedented possibilities for the on-the-fly interpretation of XRTS experiments. 
On the other hand, the robustness with respect to noise makes our approach also the method of choice for less advanced laser diagnostics at other facilities such as NIF~\cite{Moses_NIF}. 

Future developments might include a more rigorous analysis on the impact potential uncertainties in the characterization of the instrument function have on the extracted temperature. Furthermore, extending our framework to take into account the spatial inhomogeneity of a sample such as the fuel capsule in an ICF experiment~\cite{Chapman_POP_2014,Poole_POP_2022} seems promising. Finally, we note that $F(\mathbf{q},\tau)$ contains the same information as $S(\mathbf{q},\omega)$ and, therefore, can be used to extract physical information beyond the temperature such as quasi-particle excitation energies or even more complicated phenomena such as the roton feature~\cite{Dornheim_Nature_2022} in the strongly coupled electron liquid~\cite{dornheim2022physical}.

\begin{appendix}

\section{Deriving the approximate form of the wave number shift}
\label{app:A}

It is well known that evaluation of the dynamic properties of matter in which the particle distribution functions are isotropic in momentum space requires only the magnitude of the momentum shift variable $q = |\mathbf{q}|$. In the context of XRTS, momentum conservation within the scattering process gives the shift in the wave vector as $\mathbf{q} = \mathbf{q}_{0} - \mathbf{q}_{\text{s}}$. The magnitude then immediately follows from application of the cosine formula
\begin{align}
    \label{eq:app_mom_cons_1}
    q(q_{0},q_{s},\theta)
    = &\,
    \sqrt{q_{0}^{2} + q_{\text{s}}^{2} - 2q_{0}q_{\text{s}}\cos\theta}
    \,,
\end{align}
in which $\theta = \angle(\mathbf{q}_{0}, \mathbf{q}_{\text{s}})$ is the scattering angle defined in Fig.\,\ref{fig:XRTS}. In the case where electromagnetic dispersion can be ignored, i.e., for target electron densities far below the critical density $n_{\text{crit}}(\omega) = \varepsilon_{0}m_{e}\omega^{2}/e^{2}$, one may write the relationship between frequency and wave number as $\omega = c q$, such that \eqref{eq:app_mom_cons_1} becomes
\begin{align}
    \label{eq:app_mom_cons_2}
    q(\omega_{0},\omega,\theta)
    = &\,
    \frac{\omega_{0}}{c}\sqrt{1 + \tilde{\omega}^{2} - 2\tilde{\omega}\cos\theta}
    \,,
\end{align}
where $\tilde{\omega} = \omega_{\text{s}}/\omega_{0}$. From the consummate energy conservation: $\omega = \omega_{0} - \omega_{\text{s}}$, one has $\tilde{\omega} = (1 - \omega / \omega_{0})$. For XRTS the energy of the probe, $E_{0}=\hbar\omega_{0}$, is sufficiently high that the spectral features of the scattered spectrum occupy a dynamic range that strongly fulfills the condition $|\omega| / \omega_{0} \ll 1$ and, thus, we have both $\tilde{\omega} \approx 1$ and 
\begin{align}
    \label{eq:app_mom_cons_3}
    q(\omega_{0},\omega,\theta)
    \approx &\,
    \frac{\omega_{0}}{c}\sqrt{2(1 - \cos\theta)}
    \nonumber\\
    = &\,
    \frac{\omega_{0}}{c}\sqrt{4\sin^{2}(\theta/2)}
    \nonumber\\
    = &\,
    \frac{2\omega_{0}}{c}\sin(\theta/2)
    \equiv 
    q_{\text{approx}}(\omega_{0},\theta)
    \,,
\end{align}
which is the well-known approximate form used throughout the analysis of XRTS experiments. The fact that this relationship is only approximate and is not appropriate for all scattering experiments is well known in the optical Thomson scattering (OTS) community; it is essential to account for the change in $q$ across the spectral range in order to capture the change in the Landau damping rate on the red- and blue-shifted plasmon resonances \cite{Ross_RevSciInstrum_2010}, which strongly influence forward data fitting results. On the contrary, this is seldom discussed in the context of XRTS data analysis.

Note also that the other dynamic term in the definition of the scattered power spectrum Eq.\,\eqref{eq:power_spectrum_2} can, and indeed should, be similarly approximated as unity. These consistent simplifications allow the usage of the spectral two-sided Laplace transform of the reduced intensity $I(\mathbf{q},\omega)$ at constant $\mathbf{q}$, and further enables the latter to be robustly taken to be interpreted as the convolution defined in Eq.\,\eqref{eq:intensity}.

The restriction that the approximate form of $q$ must be well-fulfilled in order to properly use the Laplace transform indicates that further work must be undertaken in order to use the temperature diagnostic described herein on systems where the full expression is required, such as OTS experiments.

\section{\label{sec:shift}Invariance with respect to an energy shift $\omega_0$}

The combined source and instrument function of an XRTS experiment has a global maximum around an energy shift $\omega_0$, which is either determined by the XFEL energy~\cite{siegfried_review} or a backlighter emission spectrum~\cite{MacDonald_POP_2021}. Similarly, the measured intensity, too, will be centered around the same $\omega_0$. In practice, it might however not be easily possible to unambiguously resolve $\omega_0$ e.g.~due to the inevitable experimental noise. Here, we briefly demonstrate that the extracted ITCF $F(\mathbf{q},\tau)$ is, in fact, invariant with respect to $\omega_0$.
Let $\overline{R}(\omega)=R(\omega-\omega_0)$ denote a shifted source and instrument function that is centered around $\omega=0$, and $\overline{I}(\omega)$ the equally shifted intensity; the dynamic structure factor $S(\omega)$ is always centered around $\omega=0$, and the wave number $q$ is dropped from the following considerations for simplicity.

The two-sided Laplace transform of $\overline{R}(\omega)$ is given by
\begin{eqnarray}\label{eq:b1}
\mathcal{L}\left[\overline{R}(\omega)\right] &=& \int_{-\infty}^\infty \textnormal{d}\omega\ e^{-\tau\omega} \overline{R}(\omega) \\\nonumber &=& \int_{-\infty}^\infty \textnormal{d}\omega\ e^{-\tau(\omega-\omega_0)} R(\omega) \\\nonumber &=& e^{\omega_0\tau} \mathcal{L}\left[R(\omega)\right]\ ,
\end{eqnarray}
where the second line has been obtained by inserting $R(\omega)$ and shifting the integration variable by $\omega\to\omega-\omega_0$.
Similarly, we find for the convolved XRTS intensity
\begin{widetext} 
\begin{eqnarray}\label{eq:b2}
\mathcal{L}\left[ \overline{R}(\omega) \circledast S(\omega)\right] &=& \int_{-\infty}^\infty \textnormal{d}\omega\ e^{-\tau\omega} \int_{-\infty}^\infty \textnormal{d}\Omega\ S(\Omega) \overline{R}(\omega-\Omega)\\ &=& \int_{-\infty}^\infty \textnormal{d}\Omega\ S(\Omega) \int_{-\infty}^\infty \textnormal{d}\omega\ e^{-\tau(\omega-\omega_0)} R(\omega-\Omega) \nonumber = e^{\tau\omega_0} \mathcal{L}\left[R(\omega)\circledast S(\omega)\right]\ .
\end{eqnarray}
\end{widetext} 
Clearly, the exponential pre-factors in the last lines of Eqs.~(\ref{eq:b1}) and (\ref{eq:b2}) cancel when being inserted into Eq.~(\ref{eq:convolution_theorem}), which means that $F(\mathbf{q},\tau)$ is invariant with respect to an energy shift by $\omega_0$.

\end{appendix}

\section*{Data availability}

The data that support the findings of this study are available from the corresponding author upon reasonable request.

\section*{Acknowledgments}
This work was partially supported by the Center for Advanced Systems Understanding (CASUS) which is financed by Germany’s Federal Ministry of Education and Research (BMBF) and by the Saxon state government out of the State budget approved by the Saxon State Parliament. The work of Ti.~D.~was performed under the auspices of the U.S. Department of Energy by Lawrence Livermore National Laboratory under Contract No. DE-AC52-07NA27344.
The PIMC calculations for the UEG were carried out at the Norddeutscher Verbund f\"ur Hoch- und H\"ochstleistungsrechnen (HLRN) under grant shp00026, and on a Bull Cluster at the Center for Information Services and High Performance Computing (ZIH) at Technische Universit\"at Dresden.

\bibliography{bibliography}
\end{document}